\documentclass[letterpaper,11pt]{article}

\usepackage{jheppub}
\usepackage{graphicx}
\usepackage{amsmath}
\usepackage{xspace}
\usepackage[small]{subfigure}
\usepackage{afterpage}

\newcommand{\eq}[1]{eq.~\eqref{eq:#1}}
\newcommand{\eqs}[2]{eqs.~\eqref{eq:#1} and \eqref{eq:#2}}
\renewcommand{\sec}[1]{section~\ref{sec:#1}}
\newcommand{\secs}[2]{sections~\ref{sec:#1} and \ref{sec:#2}}
\newcommand{\subsec}[1]{section~\ref{subsec:#1}}
\newcommand{\app}[1]{appendix~\ref{app:#1}}
\newcommand{\fig}[1]{figure~\ref{fig:#1}}
\newcommand{\figs}[2]{figures~\ref{fig:#1} and \ref{fig:#2}}

\newcommand{\abs}[1]{\lvert#1\rvert}
\newcommand{\Abs}[1]{\bigl\lvert#1\bigr\rvert}
\newcommand{\ord}[1]{{\mathcal O}(#1)}
\newcommand{\ORd}[1]{{\mathcal O}\Bigl(#1\Bigr)}
\newcommand{\ORD}[1]{{\mathcal O}\biggl(#1\biggr)}

\newcommand{\Mae}[3]{\bigl\langle#1\bigr\rvert#2\bigr\rvert#3\bigr\rangle}
\newcommand{\MAe}[3]{\Bigl\langle#1\Bigr\rvert#2\Bigr\rvert#3\Bigr\rangle}

\newcommand{\ket}[1]{\lvert#1\rangle}

\newcommand{\df}{\mathrm{d}}
\newcommand{\img}{\mathrm{i}}

\newcommand{\Li}{\mathrm{Li}}

\newcommand{\sdt}{\!\cdot\!}
\newcommand{\tr}{\mathrm{tr}}

\newcommand{\lra}{\leftrightarrow}

\newcommand{\al}{\alpha}
\newcommand{\bt}{\beta}

\newcommand{\Ga}{\Gamma}
\newcommand{\de}{\delta}
\newcommand{\De}{\Delta}
\newcommand{\eps}{\epsilon}
\newcommand{\ve}{\varepsilon}

\newcommand{\si}{\sigma}
\newcommand{\w}{\omega}

\newcommand{\Tau}{\mathcal{T}}

\newcommand{\cB}{{\mathcal B}}

\newcommand{\cI}{{\mathcal I}}
\newcommand{\cL}{{\mathcal L}}

\newcommand{\hp}{\hat{p}}

\newcommand{\bn}{\bar{n}}
\newcommand{\bnP}{\overline {\mathcal P}}

\newcommand{\bnslash}{\bar{n}\!\!\!\slash}
\newcommand{\pslash}{p\!\!\!\slash}

\newcommand{\qslash}{q\!\!\!\slash}

\newcommand{\GeV}{\,\mathrm{GeV}}
\newcommand{\TeV}{\,\mathrm{TeV}}

\newcommand{\nn}{\nonumber}

\newcommand{\lqcd}{\Lambda_\mathrm{QCD}}

\newcommand{\lb}{ {\tilde{b}} }

\newcommand{\cm}{\mathrm{cm}}          
\newcommand{\soft}{\mathrm{soft}}      
\newcommand{\sing}{\mathrm{s}}         
\newcommand{\ns}{\mathrm{ns}}          
\newcommand{\res}{\mathrm{res}}        

\newcommand{\FO}{\mathrm{FO}}
\newcommand{\LO}{\mathrm{LO}}
\newcommand{\NLO}{\mathrm{NLO}}
\newcommand{\NNLO}{\mathrm{NNLO}}
\newcommand{\NNLL}{\mathrm{NNLL}}

\newcommand{\MSbar}{$\overline{\text{MS}}$\xspace}
\newcommand{\Ecm}{E_\mathrm{cm}}

\newcommand{\Disc}{\mathrm{Disc}}
\newcommand{\cut}{\mathrm{cut}}

\newcommand{\bare}{\mathrm{bare}}
\newcommand{\cusp}{\mathrm{cusp}}
\renewcommand{\max}{\mathrm{max}}

\newcommand{\jet}{\mathrm{jet}}

\newcommand{\tB}{\widetilde{B}}

\newcommand{\zero}{{(0)}}
\newcommand{\one}{{(1)}}
\newcommand{\two}{{(2)}}

\newcommand{\op}{{\mathcal{O}}}
\newcommand{\oq}{{\mathcal{Q}}}

\newcommand{\Pythia}{\textsc{Pythia}\xspace}
\newcommand{\Herwig}{\textsc{Herwig}\xspace}
\newcommand{\MCNLO}{\textsc{MC@NLO}\xspace}
\newcommand{\POWHEG}{\textsc{POWHEG}\xspace}
\newcommand{\Fehip}{\textsc{FEHiP}\xspace}
\newcommand{\FastJet}{\textsc{FastJet}\xspace}

\newcommand{\Tcm}{\mathcal{T}_\mathrm{cm}}
\newcommand{\taucm}{\tau}
\newcommand{\TB}{\mathcal{T}_B}
\newcommand{\tauB}{\tau_B}
\newcommand{\Tcmc}{\mathcal{T}_\mathrm{cm}^\mathrm{cut}}

\allowdisplaybreaks[3]

\title{Higgs Production with a Central Jet Veto at NNLL+NNLO}

\author[a]{Carola F.~Berger,}
\author[a]{Claudio Marcantonini,}
\author[a,b]{Iain W.~Stewart,}
\author[a]{Frank J.~Tackmann,}
\author[a,c]{and Wouter J.~Waalewijn}

\affiliation[a]{Center for Theoretical Physics, Massachusetts Institute of Technology, Cambridge, MA~02139, U.S.A.}
\affiliation[b]{Center for the Fundamental Laws of Nature, Harvard University, Cambridge, MA~02138, U.S.A.}
\affiliation[c]{Department of Physics, University of California at San Diego, La Jolla, CA~92093, U.S.A.}

\emailAdd{cfberger@mit.edu}
\emailAdd{cmarcant@mit.edu}
\emailAdd{iains@mit.edu}
\emailAdd{frank@mit.edu}
\emailAdd{wouterw@physics.ucsd.edu}

\abstract{ A major ingredient in Higgs searches at the Tevatron and LHC is the
  elimination of backgrounds with jets.  In current $H\to WW\to \ell\nu\ell\nu$
  searches, jet algorithms are used to veto central jets to obtain a $0$-jet
  sample, which is then analyzed to discover the Higgs signal.  Imposing this
  tight jet veto induces large double logarithms which significantly modify the
  Higgs production cross section. These jet-veto logarithms are presently
  only accounted for at fixed order or with the
  leading-logarithmic summation from parton-shower Monte Carlos. Here we
  consider Higgs production with an inclusive event-shape variable for the jet
  veto, namely beam thrust $\Tcm$, which has a close correspondence with a
  traditional $p_T$ jet veto. $\Tcm$ allows us to systematically sum the
  large jet-veto logarithms to higher orders and to provide better estimates for
  theoretical uncertainties.  We present results for the $0$-jet Higgs
  production cross section from gluon fusion at next-to-next-to-leading-logarithmic
  order (NNLL), fully incorporating fixed-order results at next-to-next-to-leading
  order (NNLO).  At this order the
  scale uncertainty is $15-20\%$, depending on the cut, implying that a
  larger scale uncertainty should be used in current Tevatron bounds on the
  Higgs.
}


\begin{document}

{\flushright MIT--CTP 4122\\December 20, 2010\\[-9ex]}
\maketitle

\pagebreak
\section{Introduction}
\label{sec:Intro}

The discovery of the Higgs boson is a major goal of the Large Hadron Collider
(LHC) and current analyses at the Tevatron. The decay $H\to WW^*$ is the
dominant channel for Higgs masses $m_H\gtrsim 130\GeV$. Hence, the $H\to
W^+W^-\to \ell^+\nu \ell^- \bar\nu$ channel has strong discovery potential and
plays a very important role for early searches that are statistically limited.  It
is the dominant channel in the current Tevatron exclusion
limit~\cite{Aaltonen:2010yv}.  The presence of the final-state neutrinos does
not allow the reconstruction of the Higgs invariant mass, and hence sideband
methods cannot be used for this channel to determine the backgrounds directly
from data. At the LHC and Tevatron, $t\bar t \to W^+W^-b\bar b$ events
constitute a large background, dominating the signal by a factor of $10$ to $40$
depending on the Higgs mass and center-of-mass energy.  Requiring a minimum
missing energy is not effective against this background since it also contains
two neutrinos. To eliminate the huge background from top-quark decays one
imposes a stringent jet veto to define a $0$-jet sample for the search, where
one only allows soft jets with $p_T^\jet\le p_T^\cut$. The latest ATLAS study~\cite{Aad:2009wy} vetoes any jets
with transverse momentum $p_T^\jet \geq 20\GeV$ and pseudorapidity
$\abs{\eta^\jet}\leq 4.8$, which reduces the $t\bar t$ background to a
negligible level.  The latest CMS study~\cite{CMSnoteWW} rejects all events that
have jets with $p_T^\jet \gtrsim 25\GeV$ and $\abs{\eta^\jet} \leq 2.5$, which
reduces this background by a factor of $\sim 40$. After the jet veto, the main irreducible
background stems from the direct production channel $pp\to W^+W^-$, which at
this point still dominates the signal by a factor of about $4:1$. The final
discrimination against this and other backgrounds is achieved by exploiting
several kinematic variables~\cite{Dittmar:1996ss}.

The Tevatron Higgs searches analyze their data using a jet algorithm and Monte
Carlo to implement a jet veto and divide the data into $0$-jet, $1$-jet, and
$\geq 2$-jet samples for all jets with $p_T^\jet \geq 15 \GeV$ and $\abs{\eta^\jet} \leq
2.4-2.5$~\cite{Aaltonen:2010yv, Aaltonen:2010cm, Abazov:2010ct}.  For $m_H
\gtrsim 130\GeV$ the sensitivity is completely dominated by the $0$-jet and
$1$-jet samples in $H\to WW$. At lower Higgs masses, the $WH$, $ZH$, and
vector-boson-fusion production channels with higher jet multiplicities are
included to increase sensitivity. With the latest update from ICHEP
2010~\cite{:2010ar}, the Tevatron excludes a range of Higgs masses $m_H =
158-175\GeV$ at $95\%$ confidence level.  For these exclusion limits it is
important to have a good theoretical understanding of the jet production cross
sections and a reliable estimate of theory uncertainties separately for each jet
bin, as emphasized in ref.~\cite{Anastasiou:2009bt}.  The theory uncertainties
in the Higgs production cross section were investigated recently in
refs.~\cite{Baglio:2010um, Demartin:2010er, Baglio:2010yf}. For their $0$-jet bin,
the Tevatron analyses use an uncertainty of 7\%, which is taken from the fixed
next-to-next-to-leading order (NNLO) analysis of the $0$-jet bin in
ref.~\cite{Anastasiou:2009bt}.  With our resummed next-to-next-to-leading
logarithmic order (NNLL) plus NNLO calculation of a $0$-jet cross section we
will see that the perturbative uncertainties are actually larger, $\simeq 20\%$,
due to the presence of large logarithms that are not accounted for in the
fixed-order analysis.

Theoretically, the inclusive Higgs production cross section has been studied
extensively in the literature and is known to NNLO~\cite{Dawson:1990zj,
  Djouadi:1991tka, Spira:1995rr, Harlander:2002wh, Anastasiou:2002yz,
  Ravindran:2003um, Pak:2009dg, Harlander:2009my} and including NLO electroweak
corrections~\cite{Aglietti:2004nj, Actis:2008ug, Anastasiou:2008tj} (for reviews and additional
references see e.g.\ refs.~\cite{Djouadi:2005gi, Boughezal:2009fw}).  However,
Higgs production in a $0$-jet sample differs substantially from inclusive Higgs
production. In particular, the jet veto induces large double logarithms
$\alpha_s^n \ln^m(p_T^\cut/m_H)$ with $m\leq 2n$ that are not present in the
inclusive cross section, and also induces a sizable dependence on the choice of
jet algorithm used to define the veto (see e.g. ref.~\cite{Anastasiou:2008ik}).
Theoretical studies of the jet veto are available in fixed-order calculations at
NNLO~\cite{Catani:2001cr, Anastasiou:2004xq, Davatz:2006ut}, and include
additional kinematic selection cuts~\cite{Anastasiou:2007mz, Anastasiou:2008ik,
  Grazzini:2008tf, Anastasiou:2009bt} (see also ref.~\cite{Berger:2010nc}).

Currently, the only method available to experiments to incorporate the effect of the jet veto
and the accompanying large logarithms beyond fixed order is to use parton-shower Monte Carlos, such
as \MCNLO~\cite{Frixione:2002ik, Frixione:2003ei}, \POWHEG~\cite{Alioli:2008tz, Hamilton:2009za},
\Pythia~\cite{Sjostrand:2006za, Sjostrand:2007gs}, and \Herwig~\cite{Corcella:2000bw, Corcella:2002jc}.
This allows one to take into account the dependence of the $0$-jet sample on the choice of jet algorithm, but
for the large logarithms it limits the accuracy to the leading-logarithmic
summation provided by the parton shower. The
comparison~\cite{Davatz:2006ut, Anastasiou:2008ik, Anastasiou:2009bt} of the
results at fixed NLO with those from \MCNLO, \Herwig, and
\Pythia (the latter two reweighted to the total NLO cross section), shows
differences of $20-30\%$, cf. tables 4 and 1 of refs.~\cite{Anastasiou:2008ik,
  Anastasiou:2009bt} respectively. This shows the importance of resumming the
phase-space logarithms caused by the jet veto.  Furthermore, the \Herwig and
\Pythia parton-level results obtained in ref.~\cite{Anastasiou:2009bt} differ by
about $15\%$, which is an indication that subleading phase-space logarithms are
relevant.

Theoretically, one can also study the Higgs production as a function of the
Higgs transverse momentum, $p_T^H$, both in fixed-order perturbation theory for
large $p_T^H$~\cite{deFlorian:1999zd, Ravindran:2002dc, Glosser:2002gm,
  Anastasiou:2005qj} and with a resummation of logarithms of $p_T^H$ at small
$p_T^H$~\cite{Collins:1984kg, Balazs:2000wv, Berger:2002ut, Bozzi:2003jy,
  Kulesza:2003wn, Idilbi:2005er, Bozzi:2005wk, Mantry:2009qz}. A further method
is the so-called joint factorization~\cite{Kulesza:2003wn, Laenen:2000ij}, which
allows one to simultaneously resum logarithms at threshold and small $p_T^H$ by
introducing $p_T$-dependent PDFs.  For $H\to W^+W^-\to \ell^+\nu\ell^-\bar\nu$
the missing neutrino momenta make a direct measurement of small $p_T^H$
impossible.  Instead the NNLL resummed $p_T^H$ spectrum~\cite{deFlorian:2009hc}
is used to reweight the \Pythia Higgs spectrum in the Tevatron search, which is
important for estimating the efficiency of selection cuts.  The study of
$p_T$-resummation is also motivated by the fact that the jet veto automatically
forces $p_T^H$ to be small, see e.g.\ refs.~\cite{Davatz:2004zg,
  Anastasiou:2008ik, Anastasiou:2009bt}.  However, the logarithms at small
$p_T^H$ summed at NNLL differ from those induced by the jet veto. Thus studies
of the small-$p_T^H$ spectrum can only provide a qualitative template for the
effect of the jet veto.\footnote{On the other hand, the hadronic $E_T$ spectrum
  could be considered for a central jet veto, and the resummation at small $E_T$
  was carried out in ref.~\cite{Papaefstathiou:2010bw} at NLL order.}

In this paper we explore a jet veto in $pp\to H X$ at the LHC and $p\bar p\to
HX$ at the Tevatron using an inclusive kinematic variable called beam
thrust~\cite{Stewart:2009yx}.  Beam thrust does not require a jet algorithm and
is well-suited for carrying out higher-order logarithmic resummation. It allows
us to directly predict a $0$-jet Higgs production cross section using
factorization techniques without relying on parton showers or hadronization
models from Monte Carlo.  We will present results for both the differential
beam-thrust spectrum as well as the integrated $pp\to H +0j$ cross section with
a cut on beam thrust working at NNLL and including the NNLO
corrections.\footnote{Our NNLL resummation is in the jet-veto variable, and is
  not the same as NNLL threshold resummations for the total
  cross section~\cite{Catani:2003zt, Moch:2005ky, Idilbi:2005ni,
Laenen:2005uz, Ravindran:2006cg, Ahrens:2010rs}.}
With the large logarithms under control, we are also able to perform a realistic
assessment of the perturbative theory uncertainties.  Since a factorization
theorem exists for the beam thrust spectrum we are also able to rigorously
account for the leading effect of nonperturbative hadronization corrections.  A
final advantage of beam thrust is that the cross section for the dominant
irreducible background, $pp\to WW + 0j$, can be computed with precisely the same
jet veto and similar precision, which we leave to future work.

While $H\to WW$ provides the most obvious motivation for studying the effect of
jet vetoes, one can also consider the case of $H\to \gamma\gamma$.  Here, the
Higgs signal appears as a small bump in the $\gamma\gamma$ invariant mass
spectrum on top of a smooth but overwhelming QCD background. The signal and
background are separated from each other by a combined fit to both.  The main
reducible backgrounds are $pp\to jj$ and $pp\to j\gamma$, while the irreducible
background comes from QCD diphoton production, $pp\to \gamma\gamma$.
Experimentally, it is still advantageous to separate the data into $0$-jet,
$1$-jet, and $\geq 2$-jet samples because in each sample the background has a
different shape, which helps to gain sensitivity in the fit.  However, this
separation introduces the same theoretical issues as for the jet veto in $H\to
WW$.  Beam thrust provides a continuous measure of the $0$-jettiness of an
event.  Hence, instead of using separate jet samples it may be useful to perform
a combined fit to the beam thrust and $\gamma\gamma$ invariant mass spectra.
The theoretical formulas presented here can be used to study $H\to\gamma\gamma$,
and we briefly comment on this, however we choose to focus on $H\to WW$.

In $H\to WW$, where missing energy plays an important
role, the appropriate version of beam thrust is defined in the hadronic
center-of-mass frame~\cite{Stewart:2009yx, Stewart:2010tn} by
\begin{equation} \label{eq:TauBcm}
\taucm = \frac{\Tcm}{m_H}
\,,\qquad
\Tcm
= \sum_k\, \abs{\vec p_{kT}}\, e^{-\abs{\eta_k}}
= \sum_k\, (E_k - \abs{p_k^z})
\,.\end{equation}
The central jet veto using beam thrust is implemented by requiring $\Tcm
\ll m_H$, or equivalently $\taucm \ll 1$.  Since the mass of the Higgs,
$m_H$, is unknown, for our analysis the dimension-one variable $\Tcm$ is
more convenient than the dimensionless $\taucm$.  The sum over $k$ in
\eq{TauBcm} runs over all particles in the final state, excluding the signal
leptons from the $W$ decays. Here $\vec{p}_{kT}$ and $\eta_k$ are the measured
transverse momentum and rapidity of particle $k$ with respect to the beam axis
(taken to be the $z$ axis).%
\footnote{Just as for jet algorithms, experimentally the sum will be over
  pseudo-particles constructed from calorimeter clusters and possibly
  supplemented by tracking information.  Using information from the tracking
  systems is important to reduce the impact of pile-up as it allows to distinguish
  particles originating from the primary hard interaction from those due to
  secondary minimum-bias interactions.} For simplicity we assume all particles
to be massless.

To see that a cut on $\Tcm \ll m_H$ vetoes central jets, first note that the
absolute value in \eq{TauBcm} divides all particles $k$ into two hemispheres
$\eta_k,p_k^z > 0$ and $\eta_k, p_k^z < 0$.  We can now distinguish between
energetic particles with $E_k \sim m_H$ and soft particles with $E_k \ll m_H$.
The latter only give small contributions to $\Tcm$.  Energetic particles moving
in the forward direction have $E_k - \abs{p_k^z} \ll m_H$, so they also
contribute only small amounts.  In particular, unmeasured particles beyond the
rapidity reach of the detector are exponentially suppressed, $\abs{\vec{p}_{kT}}
e^{-\abs{\eta_k}} \approx 2E_k e^{-2\abs{\eta_k}}$, and give negligible
contributions. On the other hand, energetic particles in the central region have
$E_k - \abs{p_k^z} \sim E_k \sim m_H$ and give a large contribution.  Therefore,
a cut $\Tcm \le \Tcmc \ll m_H$ provides an inclusive veto on central energetic
jets without requiring a jet algorithm.

\begin{figure}[t]
\includegraphics[width=0.49\textwidth]{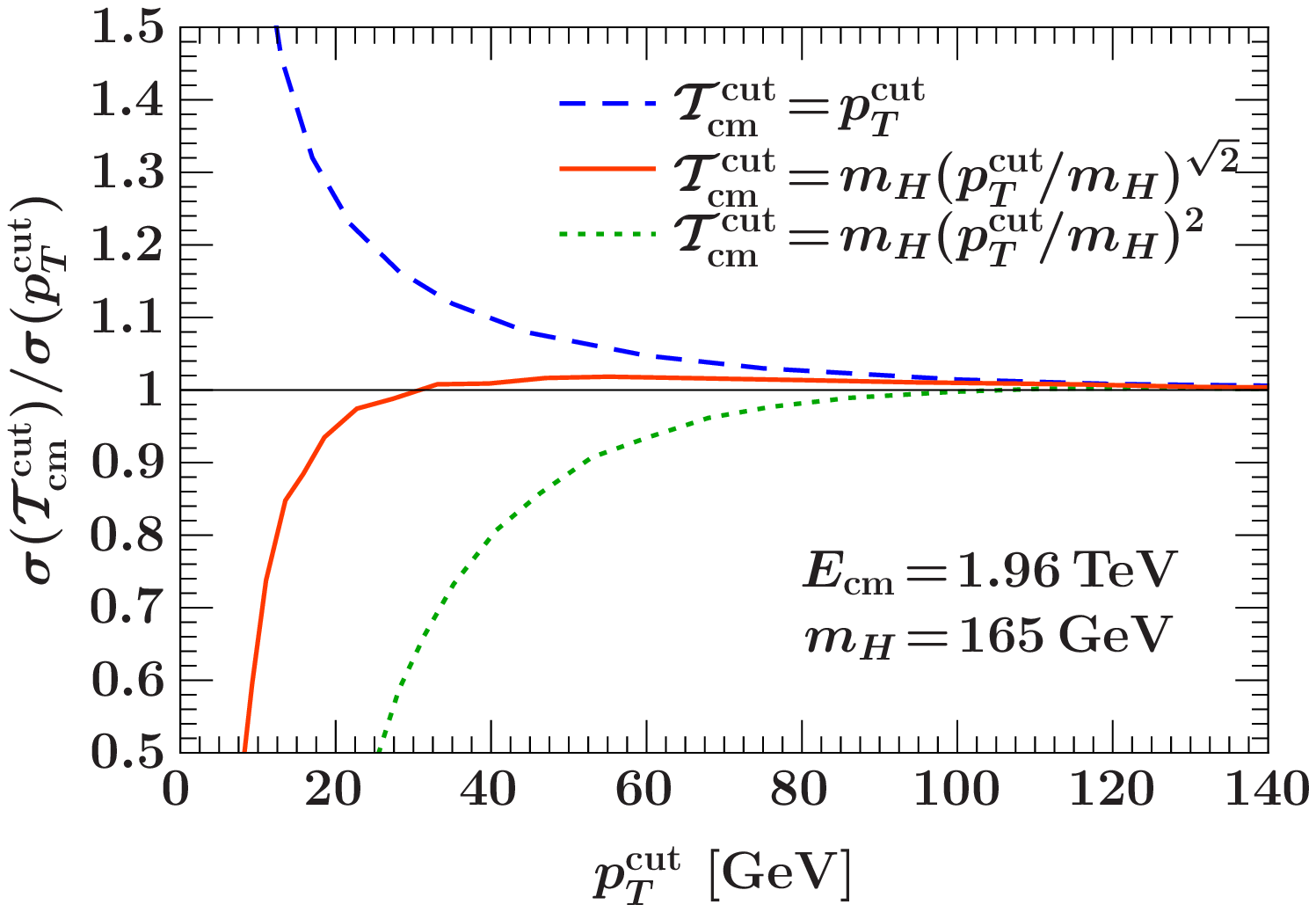}%
\hfill\includegraphics[width=0.49\textwidth]{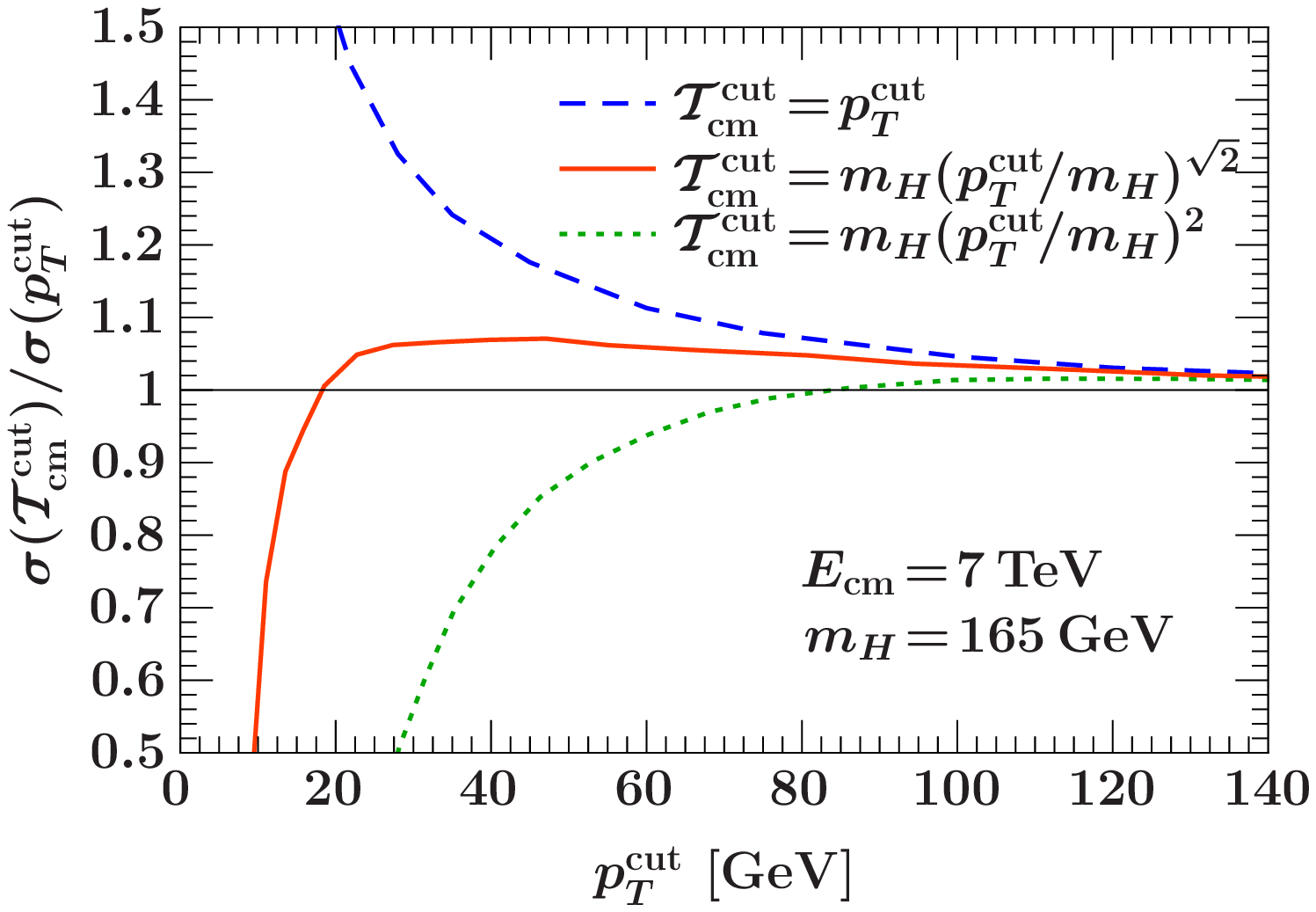}%
\vspace{-0.5ex}
\caption{Comparison of different relations between $p_T^\cut$ and $\Tcmc$ for
  the NNLO cross section, where the left panel is for the Tevatron and the right
  panel is for the LHC. The relation $\Tcmc \simeq m_H(p_T^\cut/m_H)^{\sqrt{2}}$
  yields the same leading large logarithm at $\ord{\alpha_s}$ and also the best
  overall agreement at NNLO. Here we used MSTW2008 NNLO PDFs~\cite{Martin:2009iq}
  and evaluate the cross section at $\mu = m_H$.}
\label{fig:pTvsTauB}
\end{figure}

An important question is how the $0$-jet cross section $\sigma(\Tcmc)$ with the
jet-veto cut $\Tcm\leq \Tcmc$ compares to the more standard $\sigma(p_T^\cut)$
with a traditional jet-veto cut on the maximum $p_T$ of the jets,
$p_T^{\rm jet}\le p_T^\cut$. To relate the uncertainties due to the large logarithms
in these two cross sections, we can compare their leading double-logarithmic terms
at $\ord{\alpha_s}$ using our computation and the one in ref.~\cite{Catani:2001cr}:
\begin{align} \label{eq:LLcut}
 \sigma(\Tcmc) & \propto \Bigl( 1 - \frac{\alpha_s C_A}{\pi}
 \ln^2 \frac{\Tcmc}{m_H} + \dotsb \Bigr)
 \,,
& \sigma(p_T^\cut) & \propto \Bigl( 1 - \frac{2 \alpha_s C_A}{\pi}
 \ln^2 \frac{p_T^\cut}{m_H} + \dotsb \Bigr)
 \, .
\end{align}
To obtain agreement for the leading-logarithmic terms in \eq{LLcut} the correct
correspondence between the two variables is
\begin{align} \label{eq:TauBtopT}
  \Tcmc \simeq m_H \biggl(\frac{p_T^\cut}{m_H}\biggr)^{\sqrt{2}}
\,.\end{align}
We can check the accuracy of this relation for the two jet vetoes at NNLO
numerically using the \Fehip program~\cite{Anastasiou:2004xq, Anastasiou:2005qj}
This fixed-order comparison contains not only leading
logarithms, but also subleading logarithms and non-logarithmic terms. The ratio of the NNLO
cross sections using different trial relations for the correspondence between
$\Tcmc$ and $p_T^\cut$ are shown in \fig{pTvsTauB}. With the relation in
\eq{TauBtopT} the NNLO cross sections differ by $\leq 2\%$ at the Tevatron, and
$\leq 7\%$ at the LHC, throughout the range of interesting cuts. If we multiply
the prefactor in \eq{TauBtopT} by a factor of $1/2$ or $2$ then the agreement is
substantially worse, close to that of the dotted and dashed curves in
\fig{pTvsTauB}. This confirms that \eq{TauBtopT} provides a realistic
estimate for the correspondence. However, it does not directly test the
correspondence for the cross sections with resummation at NLL order or beyond.

\begin{figure}[t]
\includegraphics[width=0.49\textwidth]{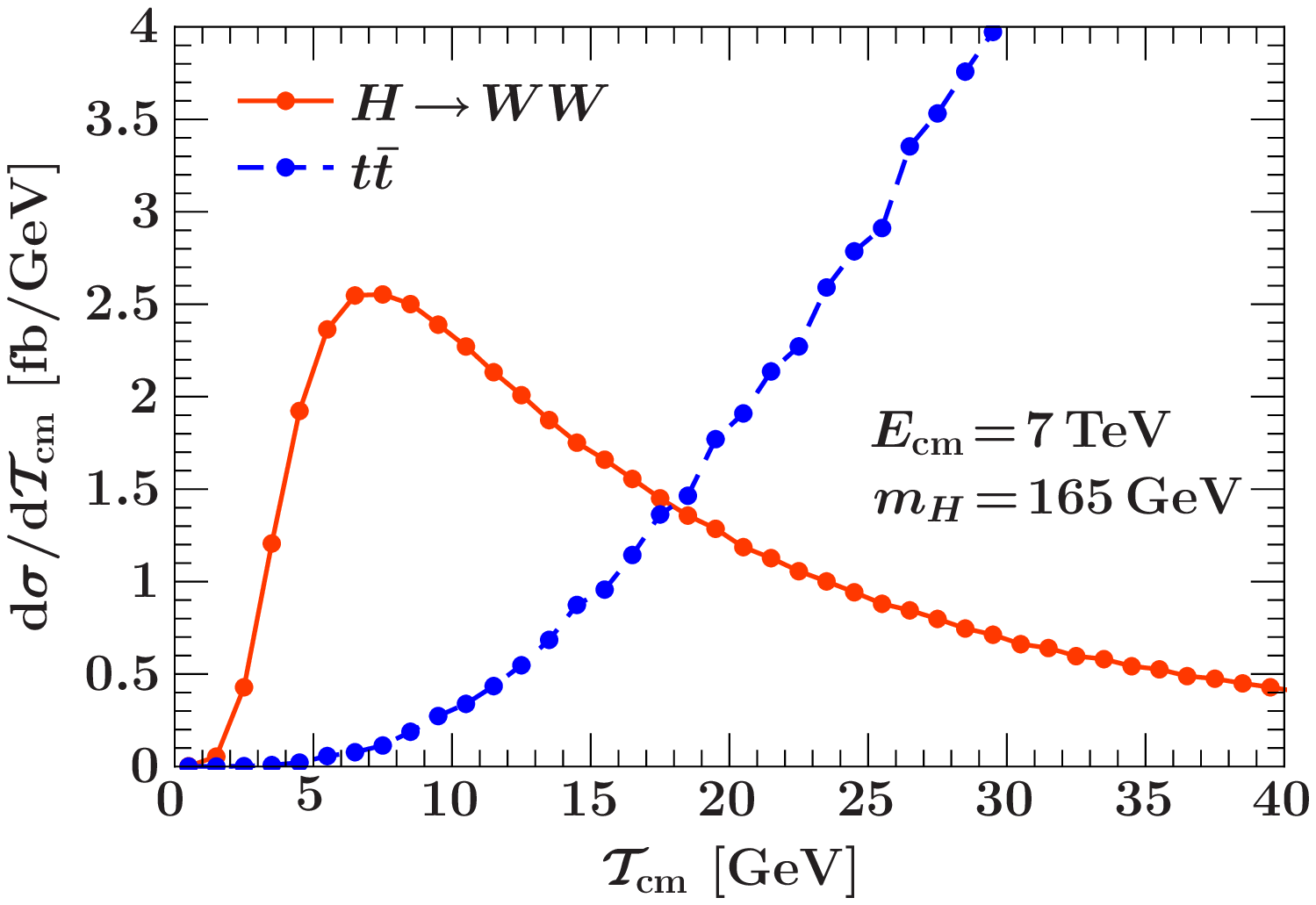}
\hfill%
\includegraphics[width=0.49\textwidth]{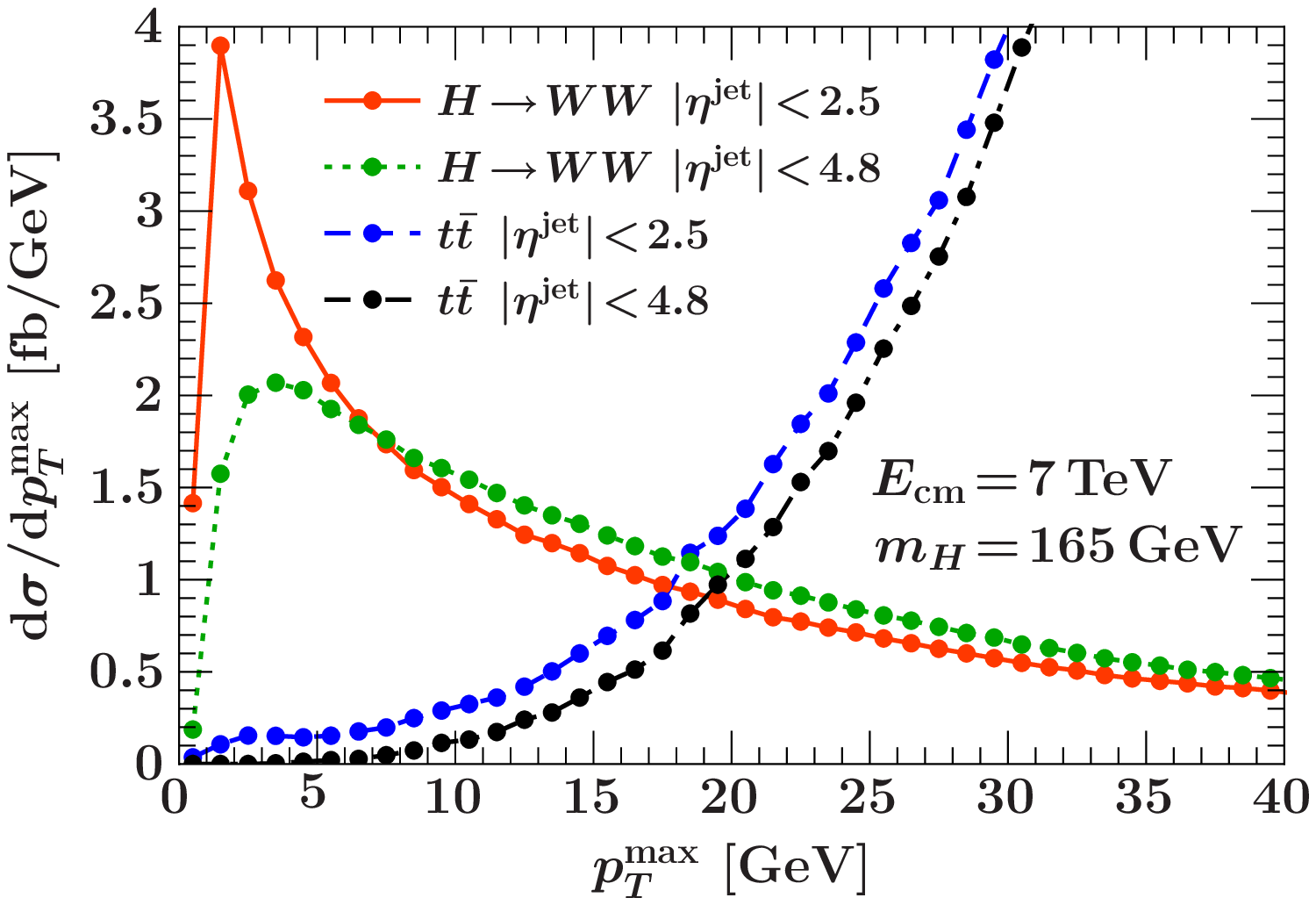}
\vspace{-0.5ex}
\caption{Comparison of the Higgs signal and $t\bar t$ background using \Pythia.
  The differential spectrum in $\Tcm$ is shown on the left, and in
  $p_T^\mathrm{max}$, the $p_T$ of the hardest jet, on the right.  For the jet
  algorithm we use the anti-$k_t$ algorithm with $R = 0.4$, only considering
  jets with $\abs{\eta^\jet} < 2.5$ or $\abs{\eta^\jet} < 4.8$.}
\label{fig:Pythiafigs}
\end{figure}

To illustrate the relative size of the $H\to WW$ signal compared to the $t\bar
t\to WWb\bar b$ background as a function of either $\Tcm$ or the $p_T$ of the
hardest jet, $p_T^\mathrm{max}$, we use \Pythia 8~\cite{Sjostrand:2007gs}
to simulate $gg\to H \to WW$ for $m_H = 165\GeV$ and $t\bar
t \to WWb\bar b$ events. In both cases we turn off multiple interactions in
\Pythia, since the corresponding uncertainty is hard to estimate without
dedicated LHC tunes. Following the selection cuts from ATLAS in
ref.~\cite{Aad:2009wy} we force one $W$ to decay into an electron and one into a
muon. We then require both leptons to have $p_T > 15\GeV$ and $\abs{\eta} <
2.5$. For the dilepton invariant mass we require $12\GeV < m_{\ell\ell} <
300\GeV$, and for the missing transverse momentum, $p_T^\mathrm{miss} > 30
\GeV$. We have not attempted to implement any lepton isolation criteria since
they should have a similar effect on the Higgs signal and $t\bar t$ background.
For the $p_T$ jet veto we define jets using the anti-$k_t$
algorithm~\cite{Cacciari:2008gp} with $R = 0.4$ implemented in the \FastJet
package~\cite{fastjet}. The results for the differential cross section in $\Tcm$
and $p_T^\mathrm{max}$ after the above cuts are shown in \fig{Pythiafigs}, where
the normalization corresponds to the total cross sections $\sigma_{gg\to H} = 8
\, {\rm pb}$ and $\sigma_{t\bar t} = 163\,{\rm pb}$ (see e.g.\
ref.~\cite{Kidonakis:2010dk}).  Note that the above selection cuts have no
effect on the shape of the Higgs signal and a small $5-20\%$ effect on the shape
of the $t\bar t$ background. In this simulation a signal to background ratio of
one is achieved with cuts $\Tcm < 31\GeV$, $p_T^\mathrm{max} < 32\GeV$ for
$\abs{\eta} < 2.5$, and $p_T^\mathrm{max} < 33\GeV$ for $\abs{\eta} < 4.8$.  It
will be very interesting to see the performance of $\Tcm$ in a full experimental
analysis including a $b$-jet veto from $b$-tagging which will further improve
the suppression of $t\to Wb$ decays with only small effects on the Higgs signal.

We have also tested the correspondence between the $\Tcm$ and $p_T^\cut$
variables using partonic \Pythia 8 Higgs samples for the LHC at $7\TeV$.
The cut $p_T < p_T^\cut$ is applied for $R=0.4$ anti-$k_T$ jets with
rapidities $\abs{\eta} < \eta^\cut$. For $\eta^\cut=4.8$ the variable
correspondence is roughly midway between $\Tcm=p_T^\cut$ and the relation in
\eq{TauBtopT}, whereas for $\eta^\cut = 2.5$ the correspondence is closer
to $\Tcm=p_T^\cut$. For the Tevatron the correspondence is also closer to
$\Tcm=p_T^\cut$ with less dependence on $\eta^\cut$. To estimate the impact
of our results on the uncertainties for the $p_T^\cut$ jet veto we will consider
the range between \eq{TauBtopT} and $\Tcm=p_T^\cut$. Further discussion on
how to apply our results to the experimental analyses using reweighted
Monte Carlo samples is left to \sec{conclusions}.

Including the resummation of large logarithms for $\Tcm\ll m_H$, the production
cross section from gluon fusion, $gg\to H$, is given by the factorization
theorem~\cite{Stewart:2009yx}
\begin{align} \label{eq:TauBcm_fact}
\frac{\df\sigma}{\df \Tcm}
&= \sigma_0\, H_{gg}(m_t, m_H^2, \mu)  \int\!\df Y \int\!\df t_a\, \df t_b\,
B_g(t_a, x_a, \mu)\, B_g(t_b, x_b, \mu)
\nn\\ &\quad \times
S_B^{gg}\Bigl(\Tcm - \frac{e^{-Y} t_a + e^Y t_b}{m_H}, \mu\Bigr)
+ \frac{\df\sigma^\ns}{\df \Tcm}
\,,\end{align}
where
\begin{equation}
x_a = \frac{m_H}{\Ecm}\,e^{Y}
\,,\qquad
x_b = \frac{m_H}{\Ecm}\,e^{-Y}
\,,\qquad
\sigma_0 = \frac{\sqrt{2} G_F\, m_H^2}{576 \pi \Ecm^2}
\,,\end{equation}
$\Ecm$ is the total center-of-mass energy, and $Y$ is the rapidity of the
Higgs.\footnote{For $H\to \gamma\gamma$ the Higgs rapidity $Y$ is measurable.
  With no additional jets in the event it provides the boost of the partonic
  hard collision relative to the hadronic center-of-mass frame. In this case one
  can account for this boost in the definition of beam thrust, $\TB =
  \sum_k\, \abs{\vec p_{kT}} \, e^{-\abs{\eta_k - Y}}$, which effectively
  defines beam thrust in the partonic center-of-mass frame.  Just as for
  $\Tcm$, a jet veto is obtained by imposing $\TB \ll m_H$.
  The factorization theorem for the gluon-fusion production cross section for
  $\TB \ll m_H$ is the same as in \eq{TauBcm_fact} but with $Y$ set to zero
  inside $S_B^{gg}$~\cite{Stewart:2009yx}.  The difference between
  $\df\sigma/\Tcm$ and $\df\sigma/\df\TB$ first appears at NLO and NNLL
  and is numerically small, at the 4\% level.}
The limits on the $Y$ integration are $\ln(m_H/\Ecm) \leq Y \leq -
\ln(m_H/\Ecm)$.

In this paper we focus our attention on the Higgs production cross section. The
leptonic decay of the Higgs does not alter the factorization structure for the summation of
large logarithms in the first term in \eq{TauBcm_fact}, where it can be included
straightforwardly as was done in ref.~\cite{Stewart:2009yx} for
the simpler case of $pp\to Z/\gamma\to \ell^+\ell^-$. Its effect on the second
term can be more involved. Including the Higgs decay is of course important in
practical applications, which use additional leptonic variables to discriminate
against the $pp\to WW$ background. A further investigation of these effects
using factorization is left for future work.

By using a cut on $\Tcm \leq \Tcmc$ to implement the jet veto, the resulting
large double logarithms in the $0$-jet cross section have the form $\alpha_s^n
\ln^m(\Tcmc/m_H)$ with $m \leq 2n$.  Measuring $\Tcm$ introduces two new energy
scales into the problem. In addition to the hard-interaction scale $\mu_H\simeq
m_H$, one is now sensitive to an intermediate beam scale $\mu_B^2 \simeq \Tcm
m_H$ and a soft scale $\mu_S \simeq \Tcm$. In the first term in
\eq{TauBcm_fact}, the physics at each of these scales is factorized into
separate hard, beam, and soft functions, $H_{gg}$, $B_g$, $S_B^{gg}$,
which are briefly discussed below. The veto induced logarithms
are systematically summed using this factorized result for the singular terms in
the cross section.  These functions and the nonsingular cross section
components, $\df\sigma^\ns / \df \Tcm$, are discussed in detail in \sec{calc}.
The full expression in \eq{TauBcm_fact} applies for any value of $\Tcm$, and
reduces to the fixed-order result when $\Tcm\simeq m_H$.

When $\Tcm \ll m_H$ the absence of additional hard jets in the final state
implies that the dominant corrections appearing at $\mu_H$ are hard virtual
corrections, which are described by the hard function, $H_{gg}(m_t, m_H^2,
\mu_H)$. It contains the virtual top-quark loop that generates the effective
$ggH$ vertex plus the effects of any additionally exchanged hard virtual gluons.
The jet veto explicitly restricts the energetic initial-state radiation (ISR)
emitted by the incoming gluon to be collinear to the proton direction. As a
result, the energetic ISR cannot be described by the evolution of the standard
parton distribution functions (PDFs), which would treat it fully inclusively. In
this situation, as discussed in detail in refs.~\cite{Stewart:2009yx,
  Stewart:2010qs}, the initial state containing the colliding gluon is described
by a gluon beam function, $B_g(t, x, \mu_B)$, which depends on the momentum
fraction $x$ and spacelike virtuality $-t < 0$ of the gluon annihilated in the
hard interaction. The beam function can be computed as~\cite{Fleming:2006cd,
  Stewart:2010qs}
\begin{equation} \label{eq:Bg_OPE}
B_g(t,x,\mu_B)
= \sum_{j = \{g,q,\bar{q}\}} \int_x^1 \! \frac{\df \xi}{\xi}\, \cI_{gj}\Bigl(t,\frac{x}{\xi},\mu_B \Bigr) f_j(\xi, \mu_B)
\,.\end{equation}
Here, $f_j(\xi, \mu_B)$ is the standard PDF describing the probability to find a
parton $j$ with light-cone momentum fraction $\xi$ in the proton, which is
probed at the beam scale $\mu_B$. The virtual and real collinear ISR emitted by
the parton $j$ builds up an incoming jet and is described by the perturbative
coefficients $\cI_{gj}(t, x/\xi, \mu_B)$. At tree level, a gluon from the proton
directly enters the hard interaction without emitting any radiation, so $B_g(t,
x, \mu_B) = \delta(t) f_g(x, \mu_B)$. Beyond tree level, real emissions decrease
the parton's momentum fraction to $x \leq \xi$ and push it off shell with $-t <
0$.  $\Tcm$ for small values is given by
\begin{equation} \label{eq:TauB_sep}
\Tcm = \frac{e^{-Y} t_a}{m_H} + \frac{e^{Y} t_b}{m_H}
   + \Tcm^\soft + \mathcal{O}(\Tcm^{\,2})
\,,\end{equation}
where the $t_a$- and $t_b$-dependent terms are the total contributions from
forward and backward collinear ISR. Here $\Tcm^\soft$ is the total contribution
from soft radiation and is determined by the beam-thrust soft function
$S_B^{gg}(\Tcm^\soft, \mu_S)$. Neither $\Tcm^\soft$ nor $t_{a,b}$ are physical
observables that can be measured separately. Hence, in \eq{TauBcm_fact} we
integrate over $t_a$ and $t_b$ subject to the constraint in \eq{TauB_sep}, where
the integration limits are determined by $t_{a,b} \geq 0$ and $\Tcm^\soft \geq
0$.

In \sec{calc} we describe all the ingredients required for our calculation of
the 0-jet Higgs production cross section from gluon fusion at NNLL+NNLO.  The
hard, beam, and soft functions are discussed in sections~\ref{subsec:hard},
\ref{subsec:beam}, and \ref{subsec:soft}, respectively. In
sections~\ref{subsec:nonsingular} and \ref{subsec:NNLLNNLO} we describe how we
add the nonsingular NNLO corrections, which are terms not contained in the NNLL
result. The treatment of running renormalization scales is described in
\subsec{profiles}, the impact of $\pi^2$ summation and PDF choices in
\subsec{pi2pdf}, and the size of hadronization corrections in
\subsec{Nonperturbative}.  Details of the calculations are relegated to
appendices. (In \app{calculation} we calculate the one-loop matching of the
gluon beam function onto gluon and quark PDFs, and verify at one loop that the
IR divergences of the gluon beam function match those of the gluon PDF. In
\app{rge} we present analytic fixed-order results for the hard and beam
functions with terms up to NNLO, as well as results for the singular NLO and
NNLO beam thrust cross section.)  In \sec{results} we present our results for
the Higgs production cross section as a function of beam thrust up to NNLL+NNLO
order.  In \subsec{convergence} we study the convergence of our resummed
predictions. In \subsec{fixedorder} we compare our resummed to the fixed-order
predictions, and our main results for the theoretical scale uncertainties are
presented in figs.~\ref{fig:compNNLO_Tev} and \ref{fig:compNNLO_LHC}. The origin
of the large $K$-factors for Higgs production is discussed in \subsec{Kfactor}.
Section~\ref{sec:conclusions} contains our conclusions and outlook, including
comments on the implications of our results for the current Tevatron Higgs
limits.  Readers not interested in technical details should focus their reading
on the introduction to \sec{calc} (skipping its subsections), and then read
\secs{results}{conclusions}.

\section{Components of the Calculation}
\label{sec:calc}

The differential cross section for $\Tcm$ in \eq{TauBcm_fact} can be separated into a singular and nonsingular piece
\begin{equation} \label{eq:TauBcm_sns}
\frac{\df \si}{\df \Tcm} = \biggl(\frac{\df \si^\sing}{\df \Tcm} + \frac{\df \si^\ns}{\df \Tcm}\biggr) \biggl[1 + \ORD{\frac{\lqcd}{m_H}} \biggr]
\,.\end{equation}
Including the renormalization group running of the hard, beam, and soft functions, we have
\begin{align} \label{eq:TauBcm_run}
\frac{\df\sigma^\sing}{\df \Tcm}
&= \sigma_0 H_{gg}(m_t, m_H^2, \mu_H)\, U_H(m_H^2, \mu_H, \mu) \int\!\df Y \int\!\df t_a\,\df t_b
\\ &\quad \times
\int\!\df t_a'\, B_g(t_a - t_a', x_a, \mu_B)\, U^g_B(t_a', \mu_B, \mu)
\int\!\df t_b'\, B_g(t_b - t_b', x_b, \mu_B)\, U^g_B(t_b', \mu_B, \mu)
\nn\\\nn &\quad \times
\int\!\df k\, S_B^{gg}\Bigl(\Tcm - \frac{e^{-Y} t_a + e^{Y} t_b}{m_H} - k, \mu_S \Bigr)\, U_S(k, \mu_S, \mu)
\,.\end{align}
Equation~\eqref{eq:TauBcm_run} is valid to all orders in
perturbation theory and is derived in ref.~\cite{Stewart:2009yx} using the formalism
of soft-collinear effective theory (SCET)~\cite{Bauer:2000ew, Bauer:2000yr,
  Bauer:2001ct, Bauer:2001yt, Bauer:2002nz}. In addition we will consider the cumulant,
\begin{align}
  \sigma(\Tcmc) = \int_0^{\Tcmc} \!
  \df \Tcm\, \frac{\df\sigma}{\df \Tcm}
\,,\end{align}
which gives the cross section with the jet-veto cut $\Tcm < \Tcmc$. For $\sigma(\Tcmc)$
the relevant scales are $\mu_H \simeq m_H$, $\mu_B^2 \simeq \Tcmc m_H$, and $\mu_S \simeq \Tcmc$.

\begin{table}[t]
  \centering
  \begin{tabular}{l | c c c c c c}
  \hline \hline
  & matching (singular) & nonsingular & $\gamma_x$ & $\Gamma_\cusp$ & $\beta$ & PDF \\ \hline
  LO & LO & LO & - & - & $1$-loop & LO \\
  NLO & NLO & NLO & - & - & $2$-loop & NLO \\
  NNLO & NNLO & NNLO & - & - & $3$-loop & NNLO \\ \hline
  LL & LO & - & - & $1$-loop & $1$-loop & LO \\
  NLL & LO & - & $1$-loop & $2$-loop & $2$-loop & LO \\
  NNLL & NLO & - & $2$-loop & $3$-loop & $3$-loop & NLO \\ 
  NLL$'$+NLO & NLO & NLO & $1$-loop & $2$-loop & $2$-loop & NLO \\
  NNLL+NNLO & (N)NLO & NNLO & $2$-loop & $3$-loop & $3$-loop & NNLO \\ \hline
  NNLL$'$+NNLO & NNLO & NNLO & $2$-loop & $3$-loop & $3$-loop & NNLO \\
  N$^3$LL+NNLO & NNLO & NNLO & $3$-loop & $4$-loop & $4$-loop & NNLO \\
  \hline\hline
  \end{tabular}
  \caption{The order counting we use in fixed-order and resummed perturbation theory. The
    last two rows are beyond the level of our calculations here, but are
    discussed in the text.}
\label{tab:counting}
\end{table}

Letting $v-\img 0$ be the Fourier conjugate variable to $\taucm = \Tcm/m_H$, the Fourier-transformed singular cross section exponentiates and has the form
\begin{equation} \label{eq:logseries}
\ln\frac{\df\sigma^\sing}{\df v} \sim \ln v\, (\alpha_s \ln v)^k + (\alpha_s \ln v)^k + \alpha_s (\alpha_s \ln v)^k + \dotsb
\,,\end{equation}
where $k \geq 1$. The three sets of terms represent the LL, NLL, and NNLL
corrections, respectively. As usual for problems involving Sudakov double
logarithms, the summation happens in the exponent of the cross section, which
sums a much larger set of terms compared to counting the leading logarithms in
the cross section.
To sum the terms in \eq{logseries} to all orders in $\alpha_s$, the
hard function, $H_{gg}$, beam functions, $B_g$, and soft function, $S_B^{gg}$,
in \eq{TauBcm_run} are each evaluated at their natural scales, $\mu_H \simeq
m_H$, $\mu_B \simeq \sqrt{\Tcm m_H}$, $\mu_S \simeq \Tcm$, and are then evolved
to the common scale $\mu$ by their respective renormalization group evolution
factors $U_H$, $U_B^g$, and $U_S$ to sum the series of large logarithms. In
table~\ref{tab:counting} we show various orders in resummed perturbation theory
and the corresponding accuracy needed for the matching (i.e.\ the fixed-order
results for the hard, beam, and soft functions) and anomalous dimensions
($\gamma_x$, $\Gamma_{\rm cusp}$) that enter the singular corrections.  To NNLL
order we require the NLO fixed-order corrections for $H_{gg}$, $B_g$, and
$S_B^{gg}$, as well as the two-loop non-cusp and three-loop cusp anomalous
dimensions in the evolution factors, and the three-loop running of $\alpha_s$.

The nonsingular contributions, $\df\sigma^\ns/\df\Tcm$ in \eq{TauBcm_sns}, are
$\ord{\Tcm/m_H}$ suppressed relative to the resummed contribution,
$\df\sigma^\sing/\df\Tcm$. They become important at large $\Tcm$ and are
required to ensure that the resummed results also reproduce the fixed-order
cross section at a given order.

For the various combinations in table~\ref{tab:counting} we show the order at
which nonsingular corrections are included, which for consistency agrees with
the order for the singular matching corrections. For example, to include the
fixed NLO corrections in the NLL result requires including both the singular and
nonsingular NLO terms, which we denote as NLL$'$+NLO. Similarly at one higher
order we would obtain NNLL$'$+NNLO. The prime in both cases refers to the fact
that the matching corrections in the resummed result are included at one higher
order than what would be necessary for the resummation only.  The complete
NNLO matching corrections for the beam and soft functions, which we would
need at NNLL$'$ and N$^3$LL, are not available at present. Instead, for our
final result, which we denote as NNLL+NNLO, we only include the $\mu$-dependent
NNLO terms in $H_{gg}$, $B_g$, and $S_B^{gg}$, which we compute using the
two-loop RGEs. The remaining $\mu$-independent NNLO terms are added in addition to the
nonsingular NNLO terms, as discussed in \subsec{NNLLNNLO}, such that the
fixed-order expansion of our final result always reproduces the complete NNLO
expression.

In the following sections~\ref{subsec:hard} to~\ref{subsec:soft}, the hard, beam,
and soft function are discussed in turn, including expressions for their
fixed-order corrections as well as their NNLL evolution. The one-loop
results for the hard and soft function are easily obtained from known results.
The one-loop calculation for the gluon beam function is performed in \app{calc}.
The anomalous dimensions are all known and given in \app{rgeapp}.  The basic
SCET ingredients relevant to our context are reviewed in
refs.~\cite{Stewart:2009yx, Stewart:2010qs}.  To obtain numerical results for
the cross section, we use the identities from App.~B of
ref.~\cite{Ligeti:2008ac} to evaluate the required convolutions of the various
plus distributions in the fixed-order results and evolution kernels.  In
\subsec{nonsingular} we discuss how to extract the nonsingular contributions at
NLO and NNLO, and in \subsec{NNLLNNLO} how these are combined with the resummed
singular result to give our final result valid to NNLL+NNLO.

The scale $\mu$ in \eq{TauBcm_run} is an arbitrary auxiliary scale and the cross section is
manifestly independent of it at each order in resummed perturbation theory.
This fact can be used to eliminate one of the evolution factors by setting $\mu$
equal to one of $\mu_H$, $\mu_B$, or $\mu_S$. The relevant factorization scales
in the resummed result at which a fixed-order perturbative series is evaluated
are the three scales $\mu_H$, $\mu_B$, and $\mu_S$. Hence, their dependence
only cancels out up to the order one is
working, and the residual dependence on these scales can be used to provide an
improved estimate of theoretical uncertainties from higher orders in
perturbation theory.  The choice of scales used for our central value and to
estimate the perturbative uncertainties is discussed in \subsec{profiles}.
Finally, in \subsec{pi2pdf} we briefly discuss the effect that the $\pi^2$ summation
and the order of the used PDFs have on our results.

\subsection{Hard Virtual Corrections}
\label{subsec:hard}

The hard function contains hard virtual corrections at the scale of order $m_H$,
including the virtual top-quark loop that generates the effective $ggH$ vertex. It is obtained by
matching the full theory onto the effective Hamiltonian in SCET
\begin{equation} \label{eq:Heff}
\mathcal{H}_\mathrm{eff} = \frac{H}{12\pi v} \sum_{n_1, n_2}\int\!\df\w_1\df\w_2\, C_{ggH}(m_t, 2\lb_1\cdot\lb_2)\,
(2\lb_1\cdot \lb_2) g_{\mu\nu}\, \cB_{n_1,-\w_1\perp}^{\mu c} \cB_{n_2,-\w_2\perp}^{\nu c}
\,.\end{equation}
Here, $H$ denotes the physical Higgs field and $v = (\sqrt{2}G_F)^{-1/2} = 246
\GeV$ the Higgs vacuum expectation value. The $\cB_{n,\w}^\mu$ fields are gauge
invariant fields in SCET that describe energetic gluons with large momentum
$\lb_i = \w_i n_i/2$, where $n_i$ are unit light-cone vectors, $n_i^2 = 0$. The
matching coefficient $C_{ggH}$ depends on the top-quark mass and the invariant
mass $2\lb_1 \cdot \lb_2$ of the two gluons. For the case we are interested in
we have $2\lb_1\cdot \lb_2 = q^2$, where $q$ is the total momentum of the Higgs,
i.e.\ of the $WW$ or $\gamma\gamma$ pair.  In addition to the operator shown in
\eq{Heff}, there are also operators where the Higgs couples to two collinear
quark fields. The tree-level matching onto these operators is proportional to the
light quark mass, $m_q$, and are numerically very small.  There are potentially larger matching
contributions from QCD loops where the Higgs couples to a top quark, but these
are also $m_q/m_H$ suppressed due to helicity conservation. Hence, we neglect
these collinear quark operators in our analysis.

The hard function is defined as
\begin{equation}
H_{gg}(m_t, q^2, \mu) = \Abs{C_{ggH}(m_t, q^2, \mu)}^2
\,.\end{equation}
It is evaluated at $q^2 = m_H^2$ in \eq{TauBcm_run} because we consider the production of an on-shell Higgs.
(Including the decay of the Higgs, the cross section differential in $q^2$ is proportional to $\sigma_0 L H_{gg}(m_t, q^2, \mu)$, where $L$ contains the squared Higgs propagator and decay matrix element. In the narrow width approximation $L$ reduces to $L = \delta(q^2 - m_H^2) \mathrm{Br}$, where $\mathrm{Br}$ is the appropriate Higgs branching ratio, e.g.\ $\mathrm{Br}(H\to WW)$ or $\mathrm{Br}(H\to \gamma\gamma)$.)

By matching onto \eq{Heff} we integrate out all degrees of freedom above the
scale $\mu_H$, which are the heavy top quark as well as gluons and light quarks
with offshellness above $\mu_H$. This can be done in either one or two steps. In
the one-step matching used here we integrate out both the top quark and hard
off-shell modes at the same time. This allows us to keep the full dependence on
$m_H^2/m_t^2$. In pure dimensional regularization with \MSbar the matching
coefficient $C_{ggH}(m_t, q^2, \mu_H)$ is given by the infrared-finite part of
the full $m_t$-dependent $ggH$ form factor, which is known analytically at NLO
(corresponding to two loops)~\cite{Harlander:2005rq, Anastasiou:2006hc} and in
an expansion in $q^2/m_t^2$ at NNLO (three loops)~\cite{Harlander:2009bw,
  Pak:2009bx}.

We write the Wilson coefficient as
\begin{align} \label{eq:CggH}
C_{ggH}(m_t, q^2, \mu_H)
&= \alpha_s(\mu_H) F^\zero\Bigl(\frac{q^2}{4m_t^2}\Bigr)
\biggl\{1 + \frac{\alpha_s(\mu_H)}{4\pi}
\biggl[ C^\one\Bigl(\frac{-q^2 - \img 0}{\mu_H^2} \Bigr)
+ F^\one\Bigl(\frac{q^2}{4m_t^2}\Bigr) \biggr]
\nn \\ & \quad
+ \frac{\alpha_s^2(\mu_H)}{(4\pi)^2}
\biggl[ C^\two\Bigl(\frac{-q^2 - \img 0}{\mu_H^2}, \frac{q^2}{4m_t^2} \Bigr)
+ F^\two\Bigl(\frac{q^2}{4m_t^2}\Bigr) \biggr]
\biggr\}
\,,\end{align}
where $F^\zero(0) = 1$. At NNLL we need the NLO coefficients
\begin{equation}
C^\one(x_H) = C_A \Bigl(-\ln^2 x_H + \frac{\pi^2}{6} \Bigr)
\,,\qquad
F^\one(0) = 5C_A - 3C_F
\,.\end{equation}
The dependence of $F^\zero(z)$ and $F^\one(z)$ on $z = q^2/(4m_t^2)$, which
encodes the $m_t$ dependence, is given in \eq{Ftop}.  At NNLL+NNLO we also need
to include the NNLO terms that depend logarithmically on the hard scale $\mu_H$,
which follow from the two-loop RGE of the Wilson coefficient (see \eq{C_RGE}),
and are given by
\begin{align}
C^\two(x_H, z)
&= \frac{1}{2} C_A^2 \ln^4 x_H + \frac{1}{3} C_A \beta_0 \ln^3 x_H +
  C_A\Bigl[\Bigl(-\frac{4}{3} + \frac{\pi^2}{6} \Bigr) C_A  - \frac{5}{3}\beta_0 - F^\one(z) \Bigr] \ln^2 x_H
\nn \\ & \quad
+ \Bigl[\Bigl(\frac{59}{9} - 2\zeta_3 \Bigr) C_A^2  + \Bigl(\frac{19}{9}-\frac{\pi^2}{3}\Bigr) C_A \beta_0 - F^\one(z) \beta_0 \Bigr] \ln x_H
\,.\end{align}
The remaining $\mu_H$-independent NNLO terms are contained in $F^\two(z)$. Although these are known in an expansion in $z$, we do not include them, since the corresponding $\mu$-independent NNLO terms are not known for the beam and soft functions.

To minimize the
large logarithms in $C_{ggH}$ we should evaluate \eq{CggH} at the hard scale
$\mu_H$ with $\abs{\mu_H^2} \sim q^2 \sim m_t^2$. For the simplest choice
$\mu_H^2 = q^2$ the double logarithms of $-q^2/\mu_H^2$ are not minimized since
they give rise to additional $\pi^2$ terms from the analytic continuation of the
form factor from spacelike to timelike argument, $\ln^2(-1-\img0) = -\pi^2$,
which causes rather large perturbative corrections. These $\pi^2$ terms can be
summed along with the double logarithms by taking $\mu_H = -\img \sqrt{q^2}$ or
in our case $\mu_H = -\img m_H$~\cite{Parisi:1979xd, Sterman:1986aj,
  Magnea:1990zb, Eynck:2003fn}. For Higgs production this method was applied in
refs.~\cite{Ahrens:2008qu, Ahrens:2008nc}, where it was shown to improve the
perturbative convergence of the hard matching coefficient.
Starting at NNLO, the expansion of $C_{ggH}$ contains single logarithms $\ln(m_t^2/\mu_H^2)$,
which in \eq{CggH} are contained as $\ln x_H$ in $C^\two$ with a compensating $-\ln(-4z-\img 0)$
in $F^\two(z)$, which are not large since $m_H/m_t \simeq 1$. In \eq{CggH}, $\alpha_s(\mu_H)$ is defined
for $n_f = 5$ flavors. When written in terms of $\alpha_s(\mu_H)$ with $n_f = 6$
flavors similar $\ln(m_t^2/mu_H^2)$ terms would already appear at NLO. The additional terms
that are induced by using an imaginary scale in these logarithms are small, because
the imaginary part of $\alpha_s(-\img m_H)$ is much smaller than its real part.

The alternative two-step matching is briefly discussed in \app{hardapp}, where we
compare results with the literature. In this case, one first integrates out the
top quark at the scale $m_t$ and then matches QCD onto SCET at the slightly
lower scale $\mu_H \simeq m_H$. This allows one to sum the logarithms of
$m_H/m_t$ at the expense of neglecting $m_H^2/m_t^2$ corrections.  Since
parametrically $m_H/m_t \simeq 1$, we use the one-step matching above.
Note that we do not include electroweak corrections whose predominant effect
(of order $5\%$) is on the normalization of the cross section through
the hard function~\cite{Aglietti:2004nj, Actis:2008ug, Anastasiou:2008tj,
Chiu:2008vv, Chiu:2009mg, Ahrens:2010rs}.

Given the hard matching coefficient at the scale $\mu_H$ we use its renormalization group evolution to obtain it at any other scale $\mu$,
\begin{equation} \label{eq:Hrun}
H_{gg}(m_t, q^2, \mu) = H_{gg}(m_t, q^2, \mu_H)\, U_H(q^2, \mu_H, \mu)
\,,\end{equation}
where the evolution factor is given by
\begin{align}
U_H(q^2, \mu_H, \mu)
&= \Bigl\lvert e^{K_H(\mu_H, \mu)} \Bigl(\frac{- q^2 - \img 0}{\mu_H^2}\Bigr)^{\eta_H(\mu_H, \mu)} \Bigr\rvert^2
\,,\nn \\
K_H(\mu_H,\mu) &= -2 K^g_\Gamma(\mu_H,\mu) + K_{\gamma_H^g}(\mu_H,\mu)
\,, \qquad
\eta_H(\mu_H,\mu) = \eta_\Gamma^g(\mu_H,\mu)
\,,\end{align}
and the functions $K_\Gamma^g(\mu_H, \mu)$, $\eta_\Gamma^g(\mu_H, \mu)$, and $K_\gamma(\mu_H, \mu)$ are given in \app{rgeapp}. They vanish for $\mu = \mu_H$ and therefore $U_H(q^2, \mu_H, \mu_H) = 1$, consistent with \eq{Hrun}.

\subsection{Gluon Beam Function}
\label{subsec:beam}

The gluon beam function can be computed in an operator product expansion (OPE) in terms of standard gluon and quark PDFs (see \app{beamdef} for more details),
\begin{equation} \label{eq:Bg_OPE2}
B_g(t,x,\mu_B)
= \sum_{j = \{g,q,\bar{q}\}} \int_x^1 \! \frac{\df \xi}{\xi}\, \cI_{gj}\Bigl(t,\frac{x}{\xi},\mu_B \Bigr) f_j(\xi, \mu_B)
  \biggl[1 + \ORd{\frac{\lqcd^2}{t}}\biggr]
\,.\end{equation}
In ref.~\cite{Fleming:2006cd} the $\cI_{gg}$ matching coefficient was computed
at one loop in moment space.  The $\cI_{gq}$ and $\cI_{g\bar q}$ coefficients in
the sum over $j$ in \eq{Bg_OPE2} describe the case where a quark or antiquark is
taken out of the proton, it radiates a gluon which participates in the hard
collision, and the quark or antiquark then continues into the final state. These
mixing contributions start at one loop. Our one-loop calculation of $\cI_{gj}$
for $j = \{g, q, \bar{q}\}$, which are needed for the gluon beam function
in the $0$-jet Higgs cross section at NNLL, is given in some detail in \app{calculation}
and follows the analogous computation of the quark beam function in
ref.~\cite{Stewart:2010qs}.

We write the matching coefficients for the gluon beam function as
\begin{align} \label{eq:Ig}
\cI_{gg}(t,z,\mu_B)
&= \delta(t)\,\delta(1-z)
   + \frac{\alpha_s(\mu_B)}{4\pi}\, \cI_{gg}^\one(t,z,\mu_B)
   + \frac{\alpha_s^2(\mu_B)}{(4\pi)^2}\, \cI_{gg}^\two(t,z,\mu_B)
\,, \nn \\
\cI_{gq}(t,z,\mu_B)
&= \frac{\alpha_s(\mu_B)}{4\pi}\, \cI_{gq}^\one(t,z,\mu_B)
   + \frac{\alpha_s^2(\mu_B)}{(4\pi)^2}\, \cI_{gq}^\two(t,z,\mu_B)
\,.\end{align}
Our calculation in \app{calculation} yields the one-loop coefficients
\begin{align} \label{eq:Ig_results}
\cI^\one_{gg}(t,z,\mu_B)
&= 2C_A\, \theta(z)
  \biggl\{ \frac{2}{\mu_B^2} \cL_1\Bigl(\frac{t}{\mu_B^2}\Bigr) \delta(1-z)
  + \frac{1}{\mu_B^2} \cL_0\Bigl(\frac{t}{\mu_B^2}\Bigr) P_{gg}(z)
  + \delta(t)\, \cI_{gg}^{(1,\delta)}(z) \biggr\}
\,, \nn \\
\cI^\one_{gq}(t,z,\mu_B)
&= 2C_F\, \theta(z)
   \biggl\{\frac{1}{\mu_B^2} \cL_0\Bigl(\frac{t}{\mu_B^2}\Bigr) P_{gq}(z)
   + \delta(t)\, \cI_{gq}^{(1,\delta)}(z) \biggr\}
\,,\end{align}
where
\begin{align} \label{eq:Igdel_results}
\cI_{gg}^{(1,\delta)}(z)
&= \cL_1(1-z) \frac{2(1-z + z^2)^2}{z} - P_{gg}(z) \ln z - \frac{\pi^2}{6} \delta(1-z)
\,, \nn \\
\cI_{gq}^{(1,\delta)}(z)
&= P_{gq}(z)\ln \frac{1-z}{z} + \theta(1-z) z
\,.\end{align}
Here $P_{gg}(z)$ and $P_{gq}(z)$ are the $g\to gg$ and $q\to gq$ splitting
functions given in \eq{split}, and the $\cL_n(x)$ denote the standard plus distributions,
\begin{align} \label{eq:cL}
  \cL_n(x) = \bigg[\frac{\theta(x)\ln^n\!x}{x}\bigg]_+
\,,\end{align}
defined in \eq{plusdef}. From \eq{Ig_results} we see that the proper scale to
evaluate \eq{Bg_OPE2} is $\mu_B^2 \simeq t \simeq \Tcm m_H$. For our final
NNLL+NNLO result we also need the $\mu_B$-dependent terms of the two-loop
coefficients, contained in $\cI_{gg}^\two$ and $\cI_{gq}^\two$. They can be
computed from the two-loop RGE of the $\cI_{gj}$ (see \eq{Igj_RGE}), which follows from the
two-loop RGEs of the beam function and the PDFs. Our results for these
coefficients are given in \app{beamapp}.

Our result for $\cI_{gg}^\one$ is converted to moment space in \eq{hC}, and
except for a $\pi^2$ term, agrees with the corresponding moment space result
given in eq.~(68) of ref.~\cite{Fleming:2006cd}. Another comparison can be made
by considering the correspondence with the $p_T$-dependent gluon beam function
from ref.~\cite{Mantry:2009qz}, which is given in impact parameter space $y_T$
as $B_g(t_n,x,y_T,\mu)=\cI_{gg}(t_n,x/\xi,y_T,\mu)\otimes f_g(\xi,\mu)$.  Taking
the $y_T \to 0$ limit of their bare result should yield agreement with our bare
beam function.  In principle the renormalization could change in this limit, but
their results indicate that this is not the case. Translating their variable
$t_n$ into our variables, $t_n = t/z$, the $\lim_{y_T\to
  0}\cI_{gg}(t_n,z,y_T,\mu)$ from ref.~\cite{Mantry:2009qz} agrees with our
result in \eq{Ig_results}. In ref.~\cite{Mantry:2010mk} the authors changed
their variable definition from $t_n$ to our $t = t_n z$.\footnote{The translated
  result for $\cI_{gg}$ quoted in ref.~\cite{Mantry:2010mk} has a typo, it is
  missing a term $-\delta(t) P_{gg}(z) \ln z$ induced by rescaling $(1/\mu^2)
  \cL_0(t_n/\mu^2) P_{gg}(z)$. We thank S.~Mantry and F.~Petriello for
  confirming this.} Ref.~\cite{Mantry:2010mk} also calculates the other
$\cI_{ij}$ coefficients at one loop.  Our result for the mixing contribution
$\cI_{gq}^\one$ disagrees with the $y_T\to 0$ limit of their $\cI_{gq}$. In
particular, their constant term is $\cI_{gq}^{(1,\delta)}(z) = - P_{gq}(z)
\ln[(1-z)/z] + 2(1-z)/z$ which disagrees with ours in \eq{Igdel_results}.  We
have also compared the results of ref.~\cite{Mantry:2010mk} for the quark beam
function for $y_T\to 0$ with our earlier results in refs.~\cite{Stewart:2009yx,
  Stewart:2010qs}.  The coefficient $\cI_{qq}^\one$ agrees, but the mixing term
$\cI_{qg}^\one$ also disagrees.  The $\lim_{y_T\to 0} \cI_{qg}^\one$ result in
ref.~\cite{Mantry:2010mk} is missing a term $-\delta(t) P_{qg}(z)$ that is
present in refs.~\cite{Stewart:2009yx, Stewart:2010qs}.

Given the beam function at the scale $\mu_B$ from \eq{Bg_OPE2}, we can evaluate it at any other scale using its renormalization group evolution~\cite{Stewart:2010qs}
\begin{equation} \label{eq:Bgrun}
B_g(t, x, \mu) = \int\!\df t'\, B_g(t - t', x, \mu_B)\, U_B(t', \mu_B, \mu)
\,,\end{equation}
with the evolution kernel
\begin{align} \label{eq:U_def}
U_B(t, \mu_B, \mu) &= \frac{e^{K_B-\gamma_E\, \eta_B}}{\Gamma(1+\eta_B)}
\, \biggl[\frac{\eta_B}{\mu_B^2} \cL^{\eta_B} \Bigl(\frac{t}{\mu_B^2}\Bigr)
+ \delta(t) \biggr]
\,,\nn\\
K_B(\mu_B,\mu) &= 4 K^g_\Gamma(\mu_B,\mu) + K_{\gamma_B^g}(\mu_B,\mu)
\,, \qquad
\eta_B(\mu_B,\mu) = -2\eta^g_{\Gamma}(\mu_B,\mu)
\,.\end{align}
The plus distribution $\cL^\eta(x) = [\theta(x)/x^{1-\eta}]_+$ is defined in \eq{plusdef}, and the functions $K_\Gamma^g(\mu_B, \mu)$, $\eta_\Gamma^g(\mu_B, \mu)$, and $K_\gamma(\mu_B, \mu)$ are given in \app{rgeapp}. Note that for $\mu = \mu_B$ we have $U_B(t, \mu_B, \mu_B) = \delta(t)$, which is consistent with \eq{Bgrun}.

\begin{figure}[t]
\includegraphics[width=0.485\textwidth]{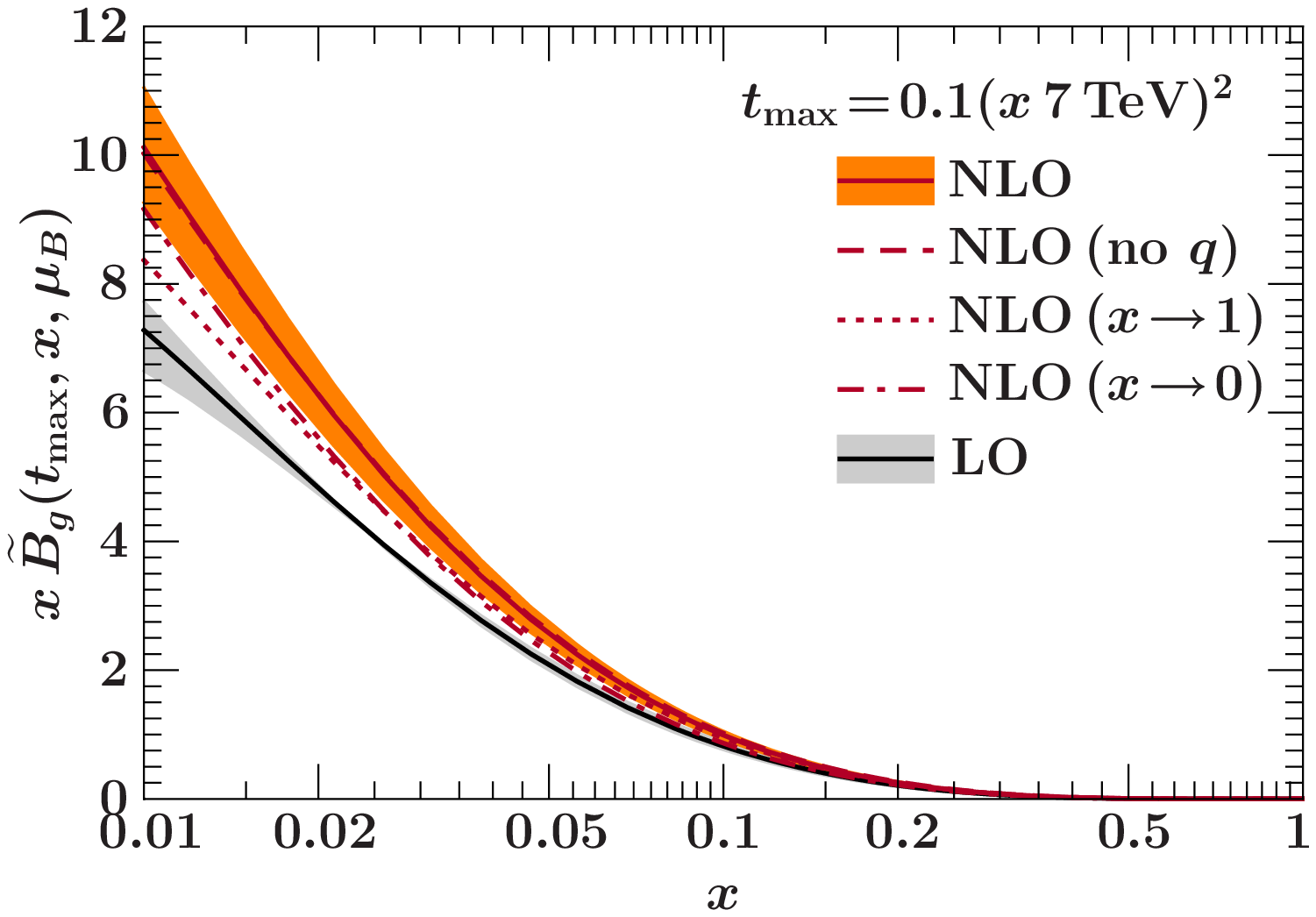}%
\hfill%
\includegraphics[width=0.495\textwidth]{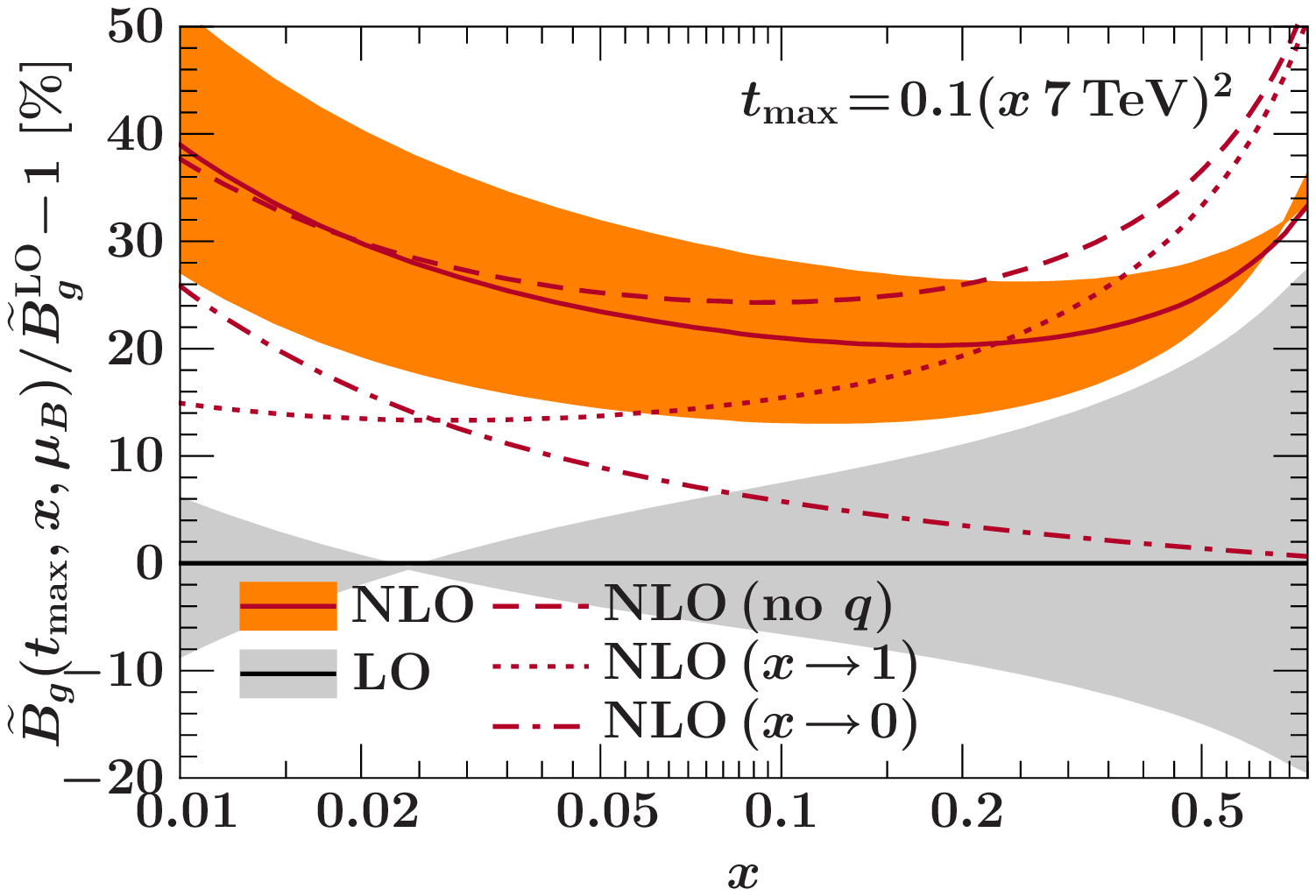}%
\vspace{-0.5ex}
\caption{The gluon beam function integrated up to $t_\max = 0.1 (x\, 7 \TeV)^2$.
  The left plot shows $x\tB_g(t_\max, x, \mu_B)$. The right plot shows all
  results relative to the LO result. The solid lines show the LO and NLO results
  with the perturbative uncertainties shown by the bands. The dashed, dotted,
  and dot-dashed lines show the NLO result without quark contribution, in the
  large $x$ limit, and the small $x$ limit, respectively. See the text for
  further details.}
\label{fig:Bg_muB}
\end{figure}

To illustrate our results for the gluon beam function we define its integral over $t \leq t_\max$,
\begin{equation} \label{eq:tB_def}
\tB_g(t_\max, x, \mu_B) = \int\! \df t\, B_g(t, x, \mu_B)\,\theta(t_\max - t)
\,.\end{equation}
In \fig{Bg_muB} we plot $\tB_g(t_\max, x, \mu_B)$ for a representative fixed
value of $t_\max = 0.1 (x\, 7\TeV)^2$. (Similar plots for the quark and
antiquark beam functions can be found in refs.~\cite{Stewart:2009yx,
  Stewart:2010qs}.) The left panel shows $x \tB_g(t_\max, x, \mu_B)$. The right
panel shows the relative corrections to the LO result $\tB_g^\mathrm{LO}(t_\max,
x, \mu_B) = f_g(x, \mu_B)$. We use MSTW2008 NLO PDFs~\cite{Martin:2009iq} with
their $\alpha_s(m_Z) = 0.12018$ and two-loop, five-flavor running for
$\alpha_s$. The bands show the perturbative uncertainties from varying the
matching scale $\mu_B$. Since at the scale $\mu_B$ there are no large logarithms
in the beam function, the $\mu_B$ variation can be used as an
indicator of higher-order perturbative uncertainties. At LO
the only scale variation is that of the PDF and the minimum and maximum scale
variation are obtained by $\mu_B = \{\sqrt{t_\max}/2,2\sqrt{t_\max}\}$ with
$\mu_B = \sqrt{t_\max}$ the central value. At NLO the maximum variation does not
occur at the endpoints of the range $\sqrt{t_\max}/2 \leq \mu_B \leq
2\sqrt{t_\max}$, but rather for approximately $\mu_B =
\{0.8\sqrt{t_\max},2.0\sqrt{t_\max}\}$ with the central value at $\mu_B =
1.5\sqrt{t_\max}$. The $\alpha_s$ corrections to the gluon beam function are
quite large, between $20\%$ to $40\%$, which is significantly larger than the
$\sim 10\%$ corrections to the quark beam function. The main reason for this is
the larger color factor for gluons than quarks.

The size of the various perturbative contributions to the beam
function is illustrated in~\fig{Bg_muB}. The dashed line shows the result obtained from
$\cI_{gg}$, without adding the mixing contribution $\cI_{gq}$, using the same
central value $\mu_B = 1.5\sqrt{t_\max}$. The mixing contributions are only
relevant above $x \gtrsim 0.2$, and are suppressed at small $x$, because of
their smaller color factor compared to $\cI_{gg}$ and the dominance of the gluon
PDF at small $x$.  This means they will be numerically small for a light Higgs.

The dotted line in~\fig{Bg_muB} shows the result in the threshold limit (again
for $\mu_B = 1.5\sqrt{t_\max}$), where we drop $\cI_{gq}$ and in addition only
keep the terms in $\cI_{gg}$ that are singular as $z \to 1$, and which are
expected to become dominant as $x\to 1$ in \eq{Bg_OPE2},
\begin{align} \label{eq:I_thres}
 \cI^{z\to 1}_{gg}(t,z,\mu_B) & = \delta(t)\delta(1-z)
+ \frac{\alpha_s(\mu_B)}{2\pi}\, C_A\, \theta(z)
  \biggl\{ \frac{2}{\mu_B^2} \cL_1\Bigl(\frac{t}{\mu_B^2}\Bigr) \delta(1-z)
\nn \\ &\qquad
  + \frac{2}{\mu_B^2} \cL_0\Bigl(\frac{t}{\mu_B^2}\Bigr) \cL_0(1-z)
  + \delta(t)\Bigl[ 2 \cL_1(1-z) - \frac{\pi^2}{6} \delta(1-z) \Bigr] \biggr\}
\,.\end{align}
The dotted line indeed approaches the dashed line for $x \gtrsim 0.2$. However, since in this region the mixing contributions become important, the threshold result does not provide a good approximation to the full result (solid line) anywhere.

Finally, the dot-dashed line in~\fig{Bg_muB} shows the result only keeping the
terms singular as $z\to 0$ (but including tree level),
\begin{align} \label{eq:I_small_x}
\cI^{z\to 0}_{gg}(t,z,\mu_B)
&= \delta(t)\delta(1-z) + \frac{\alpha_s(\mu_B)}{\pi}\, C_A\,
  \theta(z) \theta(1-z) \biggl[ \frac{1}{\mu_B^2} \cL_0\Bigl(\frac{t}{\mu_B^2}\Bigr) \frac{1}{z}
  - \delta(t) \frac{\ln z}{z} \biggr]
\,, \nn \\
\cI^{z\to 0}_{gq}(t,z,\mu_B)
&= \frac{\alpha_s(\mu_B)}{\pi}\, C_F\, \theta(z) \theta(1-z)
  \biggl[ \frac{1}{\mu_B^2} \cL_0\Bigl(\frac{t}{\mu_B^2}\Bigr) \frac{1}{z}
  - \delta(t) \frac{\ln z}{z} \biggr]
\,,\end{align}
which one might expect to dominate for $x\to 0$. Since these only have
single-logarithmic $\mu$ dependence, their central value is obtained for $\mu_B
= \sqrt{t_\max}$. This contribution indeed grows towards smaller
$x$, and makes up more than half of the total contribution at $x=0.01$, but does not
yet dominate.

\subsection{Soft Function}
\label{subsec:soft}

The soft function $S_B^{gg}(k, \mu_S)$ appearing in \eq{TauBcm_run} is defined
by the vacuum matrix element of a product of Wilson lines. For $k \simeq \Tcm
\gg \lqcd$, it can be computed in perturbation theory. We write the perturbative
soft function as
\begin{equation}
S^{gg}_{\rm pert}(k,\mu_S) = \delta(k) + \frac{\alpha_s(\mu_S)}{\pi}\, C_A\, S^\one_{gg}(k, \mu_S)
+ \frac{\alpha_s^2(\mu_S)}{\pi^2}\, C_A\, S_{gg}^{\two}(k, \mu_S) + \ORd{\frac{\alpha_s^3}{k}}
\,,\end{equation}
where the one- and two-loop coefficients are
\begin{align} \label{eq:Scoeff12}
S_{gg}^\one(k, \mu_S) &= -\frac{4}{\mu_S} \cL_1\Bigl(\frac{k}{\mu_S}\Bigr) + \frac{\pi^2}{12}\, \delta(k)
\,,\nn\\
S_{gg}^\two(k, \mu_S) &=
  8 C_A\, \frac{1}{\mu_S} \cL_3 \Bigl(\frac{k}{\mu_S}\Bigr)
  + \beta_0\, \frac{1}{\mu_S} \cL_2 \Bigl(\frac{k}{\mu_S}\Bigr)
  - \Bigl[\Bigl( \frac{4}{3} + \frac{8\pi^2}{3} \Bigr) C_A + \frac{5}{3} \beta_0 \Bigr] \frac{1}{\mu_S} \cL_1 \Bigl(\frac{k}{\mu_S}\Bigr)
\nn \\ & \quad
  + \Bigl[\Bigl(\frac{8}{9} + \frac{25}{2} \zeta_3\Bigr)C_A + \Bigl(\frac{7}{9} - \frac{\pi^2}{12}\Bigr) \beta_0\Bigr]
 \frac{1}{\mu_S} \cL_0 \Bigl(\frac{k}{\mu_S}\Bigr)
  + S^{(2,\delta)}_{gg} \delta(k)
\,,\end{align}
In ref.~\cite{Stewart:2009yx} the quark beam-thrust soft function was obtained
from the one-loop hemisphere soft function for outgoing
jets~\cite{Schwartz:2007ib, Fleming:2007xt}. The gluon beam-thrust soft function
has Wilson lines in the adjoint rather than fundamental representation and at
one loop $S^{gg}_{{\rm pert}}$ is obtained from the quark result by simply
replacing $C_F$ by $C_A$.  The $\mu_S$-dependent terms needed at NNLL+NNLO are
shown in \eq{Scoeff12} and were obtained by perturbatively solving the two-loop
RGE of the soft function (see \eq{S_RGE}). The determination of the
$\mu_S$-independent constant term, $S^{(2,\delta)}_{gg} \delta(k)$, requires the
two-loop calculation of the soft function.

The RG evolution of the soft function has the same structure as that of the
beam function,
\begin{equation} \label{eq:SBrun}
S_B^{gg}(k,\mu) =  \int\! \df k'\, S_B^{gg}(k - k',\mu_S)\, U_S(k',\mu_S,\mu)
\,,\end{equation}
with the evolution kernel
\begin{align}
U_S(k, \mu_S, \mu) & = \frac{e^{K_S -\gamma_E\, \eta_S}}{\Gamma(1 + \eta_S)}\,
\biggl[\frac{\eta_S}{\mu_S} \cL^{\eta_S} \Bigl( \frac{k}{\mu_S} \Bigr) + \delta(k) \biggr]
\,, \nn \\
K_S(\mu_S,\mu) &= -4K_\Gamma^g(\mu_S,\mu) + K_{\gamma_S^g}(\mu_S,\mu)
\,, \qquad
\eta_S(\mu_S,\mu) = 4\eta_\Gamma^g(\mu_S,\mu)
\,,\end{align}
where $\cL^\eta(x) = [\theta(x)/x^{1-\eta}]_+$ is defined in \eq{plusdef}, and
$K_\Gamma^g(\mu_S, \mu)$, $\eta_\Gamma^g(\mu_S, \mu)$, and $K_\gamma(\mu_S,
\mu)$ are given in \app{rgeapp}.

The nonperturbative corrections can be modeled and included using the methods of
refs.~\cite{Hoang:2007vb, Ligeti:2008ac}. The perturbative component
$S_\mathrm{pert}^{gg}$ and nonperturbative component $F^{gg}$ of the soft
function can be factorized as
\begin{equation} \label{eq:SF}
  S_{B}^{gg}(k,\mu_S) = \int\! \df k'\,  S_\mathrm{pert}^{gg}(k-k',\mu_S)\, F^{gg}(k') \,.
\end{equation}
At small $\Tcm \sim \lqcd$ the nonperturbative corrections to
the soft function are important.  When the spectrum is dominated by perturbative
momenta with $\Tcm \gg\lqcd$ \eq{SF} can be expanded in an OPE as
\begin{align}\label{eq:OPE}
  S_B^{gg}(k,\mu_S) = S^{gg}_\mathrm{pert}(k,\mu_S) - 2\Omega_1^{gg}\,
  \frac{\df S^{gg}_\mathrm{pert}(k,\mu_S)}{\df k} +\ORd{\frac{\lqcd^2}{k^3}}
\,,\end{align}
where the leading power correction is determined by the dimension-one
nonperturbative parameter $\Omega_1^{gg}= \int\!\df k'\, (k'/2) F^{gg}(k')$
which is parametrically $\ord{\lqcd}$. The positivity of $F^{gg}(k)$ implies
that $\Omega_1^{gg}>0$, so the factorization in \eq{SF} predicts the sign of the
correction caused by the nonperturbative effects. We will see that this simple
OPE result with one nonperturbative parameter $\Omega_1^{gg}$ gives an accurate
description of the nonperturbative effects in the $\Tcm$ spectra for the entire
region we are interested in, which includes the peak in the distribution.  The
OPE in \eq{OPE} implies that the leading nonperturbative effects can be computed
as an additive correction to the spectrum
\begin{equation} \label{eq:OPEdsdT}
\frac{\df\sigma^\sing}{\df\Tcm}
= \frac{\df\sigma^\sing_\mathrm{pert}}{\df\Tcm} - 2\Omega_1^{gg} \frac{\df^2\sigma^\sing_\mathrm{pert}}{\df\Tcm^2}
\,,
\end{equation}
and likewise for the cumulant
\begin{align} \label{eq:OPEsigc}
\sigma^\sing(\Tcmc) = \sigma^\sing_\mathrm{pert}(\Tcmc)  - 2\Omega_1^{gg} \frac{\df}{\df\Tcmc} \sigma^\sing_\mathrm{pert}(\Tcmc)
  \,.
\end{align}
To first order in the OPE expansion this is equivalent to a shift in the
variable used to evaluate the perturbative spectrum, $\Tcm \to \Tcm - 2
\Omega_1^{gg}$, or cumulant, $\Tcmc\to \Tcmc-2\Omega_1^{gg}$.  For the cumulant
the nonperturbative corrections always reduce the cross section, whereas the
distribution is reduced before the peak and increased in the tail region.  Since
the nonsingular terms in the cross section are an order of magnitude smaller
than the singular terms we can also replace $\sigma^\sing$ by $\sigma$, that is
include the nonsingular $\df\sigma^\ns/\df\Tcm$ in \eq{OPEdsdT}.  For
simplicity, we will use the purely perturbative result in most of our numerical
analysis. However, in \subsec{Nonperturbative} we will use \eqs{SF}{OPEdsdT} to
analyze the effect of nonperturbative corrections on our predictions.

\subsection{Nonsingular Contributions}
\label{subsec:nonsingular}

In this section we discuss how we incorporate the nonsingular contributions to the
cross section using fixed-order perturbation theory.  For the two beam thrust
cross sections considered in this paper, the full cross section in
fixed-order perturbation theory can be written as
\begin{align} \label{eq:sigmaFO}
\frac{\df \sigma}{\df \taucm \df Y}
&= \sigma_0\, \alpha_s^2(\mu) \Bigl\lvert F^\zero\Bigl(\frac{m_H^2}{4m_t^2}\Bigr)\Bigr\rvert^2
   \nn\\ & \quad\times
   \int\! \frac{\df \xi_a}{\xi_a}\, \frac{\df \xi_b}{\xi_b}\, \sum_{i,j}
   C_{ij}\Bigl(\frac{x_a}{\xi_a},\frac{x_b}{\xi_b},\taucm, Y, \mu, m_H, m_t\Bigr)\, f_i(\xi_a,\mu)\, f_j(\xi_b,\mu)
\,,\end{align}
where $i,j={g,q,\bar q}$ sum over parton types, and $\tau = \Tcm/m_H$.\footnote{For $H\to\gamma\gamma$
  where the boost between the partonic and hadronic center-of-mass frames is
  accounted for with $\tauB=\TB/m_H$, the appropriate replacements in
  \eq{sigmaFO} are to take $\taucm\to \tauB$, and in the fourth argument of
  $C_{ij}$ to set $Y=0$.}  To simplify the notation in the following we will
suppress the $m_H$ and $m_t$ dependence of the coefficients $C_{ij}$. The
contributions to the $C_{ij}$ can be separated into singular and nonsingular
parts,
\begin{equation} \label{eq:Cijdecomp}
C_{ij}(z_a,z_b,\tau, Y,\mu)
= C_{ij}^\sing(z_a,z_b,\tau, Y,\mu) + C_{ij}^\ns(z_a,z_b,\tau, Y,\mu)
\,,\end{equation}
where the singular terms scale as $\sim 1/\tau$ modulo logarithms and can be
written as
\begin{equation} \label{eq:Cijsing}
C_{ij}^\sing(z_a,z_b,\tau, Y,\mu) = C_{ij}^{-1} (z_a,z_b,Y,\mu)\,\delta(\tau) +
\sum_{k\geq 0} C_{ij}^k (z_a,z_b,Y,\mu)\, \cL_k(\tau)
\,,\end{equation}
where $\cL_k(\tau) = [\theta(\tau)(\ln^k\!\tau)/\tau]_+$ is defined in
\eq{plusdef}.  The resummed result for the cross section in \eq{TauBcm_run} sums
the singular contributions at small $\tau$ to all orders, counting
$(\alpha_s\ln\tau)\sim 1$.  The nonsingular contributions, $C_{ij}^\ns$, are
suppressed relative to the singular ones by $\ord{\tau}$ and it suffices to
determine them in fixed-order perturbation theory.  Hence, we can obtain them by
simply subtracting the fixed-order expansion of the singular result from the
full fixed-order result,
\begin{equation}
\frac{\df\sigma^{\ns,\mathrm{FO}}}{\df \tau}
= \frac{\df\sigma^\mathrm{FO}}{\df \tau}
- \frac{\df\sigma^{\sing,\mathrm{FO}}}{\df \tau}
\,,\end{equation}
and analogously for the cumulant
\begin{align}
\sigma^{\ns,\FO}(\tau^\cut) = \sigma^\FO(\tau^\cut)-\sigma^{\sing,\FO}(\tau^\cut)
\,.\end{align}
We must take care to use the same PDFs and renormalization scale $\mu$ for both
the $\sigma^\FO$ and $\sigma^{\sing,\FO}$ terms.  In our analysis we obtain
$\sigma^\NLO(\tau^\cut)$ and $\sigma^\NNLO(\tau^\cut)$ numerically using the
publicly available \Fehip program~\cite{Anastasiou:2004xq, Anastasiou:2005qj}, which allows one to obtain the
fixed-order NNLO Higgs production cross section for generic phase-space cuts.

\begin{figure}[t]
\includegraphics[width=0.485\textwidth]{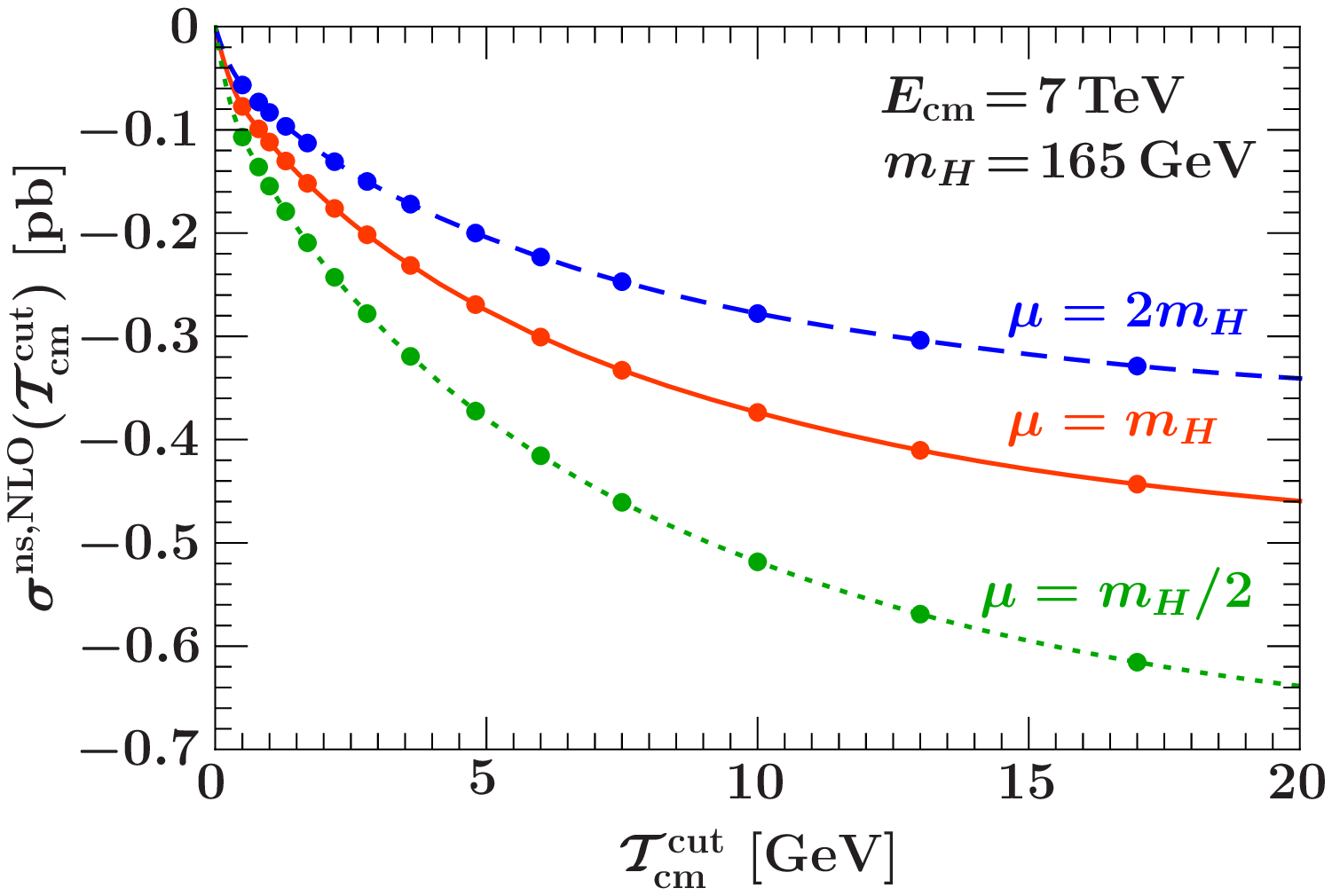}%
\hfill%
\includegraphics[width=0.495\textwidth]{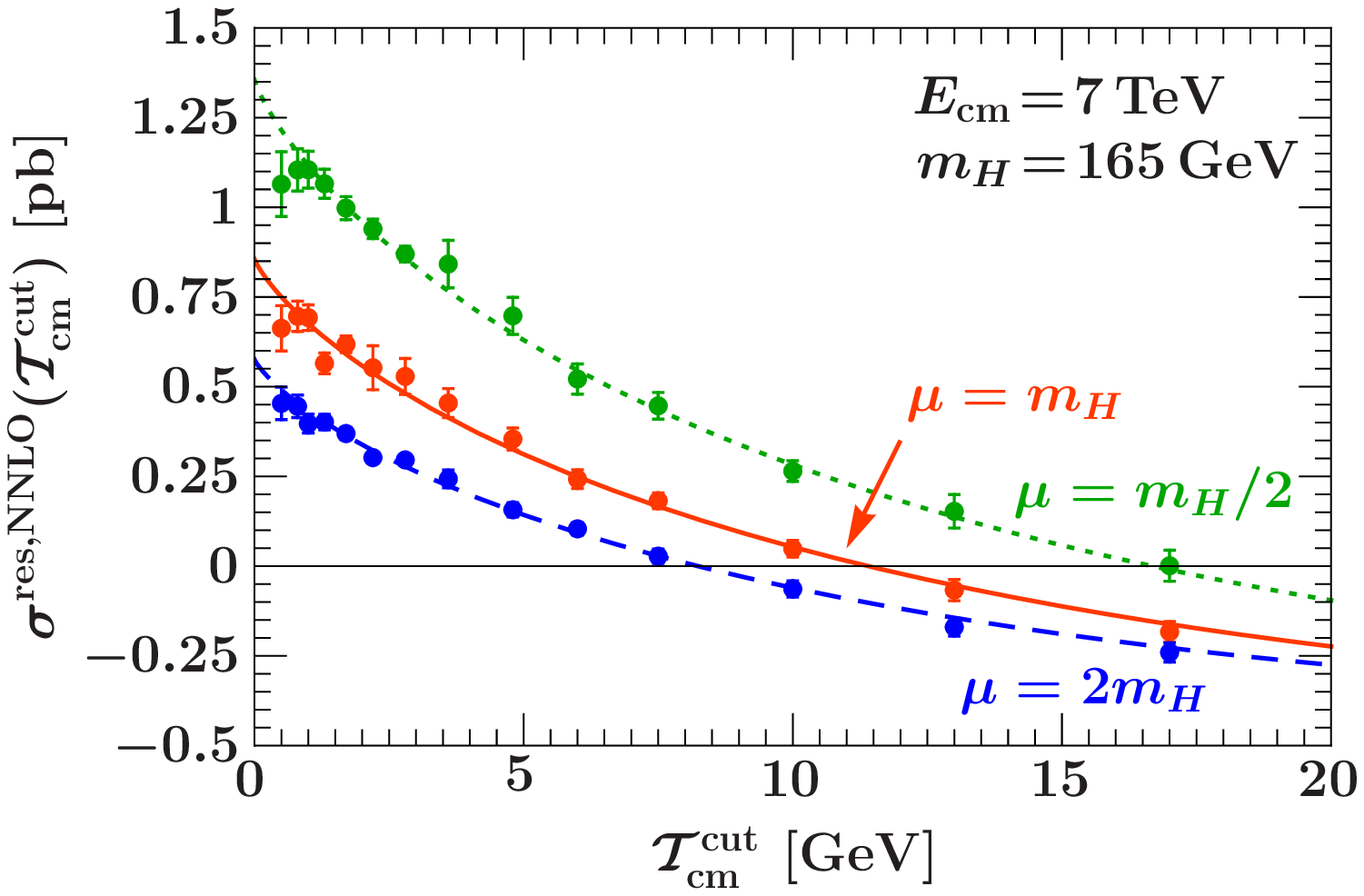}%
\vspace{-0.5ex}
\caption{The left panel shows the nonsingular contribution to the NLO cross section as a function of $\Tau_B^{\cm,\cut}$, for the LHC at $7\TeV$. The residual NNLO cross section shown in the right panel, is the nonsingular NNLO cross section, $\sigma^{\ns,\NNLO}$, plus a constant term $c^\res$, as explained in the text.}
\label{fig:nonsing}
\end{figure}

At tree level, the only nonzero coefficient is $C_{ij}^{-1}$, and the
nonsingular contribution vanishes, $\sigma^{\ns,\LO}(\tau^\cut) = 0$. At
NLO, the $C_{ij}^k$ are nonzero for $k \leq 1$ and are fully contained in the
resummed NNLL result for $\df\sigma^\sing/\df\tau$. Hence, we can obtain them by
expanding our NNLL singular result to fixed next-to-leading order,
\begin{equation} \label{eq:singNLO}
\sigma^{\sing,\NLO}(\tau^\cut)
= \sigma^{\sing,\NNLL}(\tau^\cut) \big\vert_\NLO
\,.\end{equation}
The explicit expressions are given in \app{singular}. Subtracting this from the
full NLO result we get the nonsingular contribution at NLO,
\begin{equation} \label{eq:nsNLO}
\sigma^{\ns,\NLO}(\tau^\cut)
= \sigma^\mathrm{NLO}(\tau^\cut) - \sigma^{\sing,\NNLL}(\tau^\cut) \big\vert_\NLO
\,.\end{equation}
In the left panel of \fig{nonsing} we plot the NLO nonsingular cross section
determined by this procedure for three different choices of $\mu$, namely
$\mu=m_H/2, m_H, 2m_H$.  The scaling of the nonsingular distribution implies
that it involves only integrable functions, therefore the cumulant
$\sigma^{\ns,\NLO}(\tau^\cut)$ vanishes for $\tau^\cut \to 0$.  The $\mu$
dependence of the NLO nonsingular cross section is sizeable since this is the
leading term in this part of the cross section. This $\mu$ dependence is
canceled by the nonsingular terms at NNLO which we turn to next.

At NNLO the singular cross section is determined by a result analogous to
\eq{singNLO}
\begin{align} \label{eq:singNNLO}
\sigma^{\sing,\NNLO}(\tau^\cut)
= \sigma^{\sing,\NNLL}(\tau^\cut) \big\vert_{\NNLO, k\geq 0} + \sigma^{\sing,\NNLO}\big\vert_{k = -1}
\,.\end{align}
The NNLO singular coefficients $C_{ij}^k$ are nonzero for $-1\leq k \leq 3$.
Those for $k \ge 0$ can be obtained by expanding the singular NNLL result to
fixed NNLO.  Their explicit expressions are given in \app{singular}. For $k=-1$
the NNLO contribution to the coefficient $C_{ij}^{-1}$ of the $\delta(\tau)$ is
not fully contained in the NNLL result. Therefore, the $\tau^\cut$-independent
$k=-1$ contribution is not included in the first term on the right-hand side of
\eq{singNNLO}, but in the second term. For this second term we proceed as follows.

First, we write the NNLO contribution to $C_{ij}^{-1}$ as
\begin{equation} \label{eq:Cm1terms}
C_{ij}^{-1} (z_a,z_b, Y,\mu)\big\vert_\NNLO
= \frac{\alpha_s^2(\mu)}{(2\pi)^2} \Bigl[
c_{ij}^\pi(z_a,z_b,Y) + c_{ij}^{\mu}(z_a,z_b,Y, \mu) + c_{ij}^{\res}(z_a,z_b,Y) \Bigr]
\,.\end{equation}
The first term in brackets denotes the $\mu$-independent terms proportional to
$\pi^2$ that are part of the $\pi^2$ summation. The second term
contains all terms proportional to $\ln(\mu/m_H)$, which cancel
the $\mu$ dependence in the NLO result. We can obtain these two contributions
analytically, and they are given in \app{singular}. The remaining
$\mu$-independent terms, $c_{ij}^{\res}$, are currently not known analytically.
They could be obtained when the complete NNLO results for the hard, beam, and soft
functions become available. Using \eq{Cm1terms}, the second term on the right-hand
side of \eq{singNNLO} is given by
\begin{equation}
\sigma^{\sing,\NNLO}\big\vert_{k = -1}
= c^\pi(\mu) + c^\mu(\mu) + c^\res(\mu)
\,,\end{equation}
with ($x=\{\pi, \mu,\res\}$)
\begin{align}
c^x(\mu) &= \sigma_0\,\frac{\alpha_s^4(\mu)}{(2\pi)^2} \Bigl\lvert F^\zero\Bigl(\frac{m_H^2}{4m_t^2}\Bigr)\Bigr\rvert^2
  \int\!\df Y\!\int\! \frac{\df \xi_a}{\xi_a} \frac{\df \xi_b}{\xi_b}\,\sum_{i,j}
  c_{ij}^{x} \Bigl(\frac{x_a}{\xi_a},\frac{x_b}{\xi_b}, Y, \mu \Bigr)\, f_i(\xi_a,\mu)\, f_j(\xi_b,\mu)
\,.\end{align}
The $\mu$ dependence of $c^\mu(\mu)$ cancels that of $\sigma^{\ns,\NLO}(\tau^\cut)$
up to terms of $\ord{\alpha_s^5}$, whereas that of $c^\pi(\mu)$ and $c^\res(\mu)$
only starts at $\ord{\alpha_s^5}$.  Since $x_{a,b} = (m_H/\Ecm)e^{\pm Y}$, the
$c^x(\mu)$ have a nontrivial dependence on $m_H$.

To determine the constant $c^\res$ numerically, we now consider
\begin{equation}
\sigma^\res(\tau^\cut)
\equiv \sigma^\NNLO(\tau^\cut) - \sigma^{\sing,\NNLL}(\tau^\cut) \Big\vert_{\NNLO,k\geq 0} - c^\pi - c^\mu
= c^\res + \sigma^{\ns,\NNLO}(\tau^\cut)
\,.\end{equation}
Since $\sigma^{\ns,\NNLO}(\tau^\cut)$ vanishes as $\tau^\cut \to 0$, the
coefficient $c^\res$ is determined by $\sigma^\res(\tau^\cut)$ as
$\tau^\cut\to 0$, while the nonsingular corrections are given by the remainder
$\sigma^\res(\tau^\cut) - c^\res$. Hence, we can obtain both by fitting our numerical results for
$\sigma^\mathrm{res}(\tau^\cut)$ at different values of $\tau^\cut$ to the
following function:
\begin{equation} \label{eq:fit_ns}
 \sigma^\mathrm{res}(\tau)
= c^\res + a_0\, \tau \ln\tau + a_1\, \tau + a_2\, \tau^2 \ln \tau + a_3\, \tau^2
\,.\end{equation}
The $a_0$ through $a_3$ terms are sufficient to describe the $\tau^\cut$
dependence of $\sigma^{\ns,\NNLO}(\tau^\cut)$ over the whole range of
$\tau^\cut$. The results of the fit for $pp$ collisions at $7\TeV$ and
$m_H=165\GeV$ for $\mu=m_H$, $\mu=m_H/2$, and $\mu=2 m_H$ are shown in
the right panel of \fig{nonsing}. At $\mu=m_H$ this fit gives
\begin{align}
c^\res &= 0.86\pm 0.02\,,
 &a_0 &= 7.6 \pm 0.6 \,,
 &a_1 &= 9.3 \pm 1.5 \,,
 &a_2 &= 3.9 \pm 1.1 \,,
 &a_3 &= -9.9 \pm 1.6 \,.
\end{align}
Similarly, for $p\bar p$ collisions at $1.96\TeV$, $m_H = 165\GeV$, and $\mu = m_H$, we obtain
\begin{align}
c^\res &= 0.028\pm 0.001\,,
 &a_0 &= 0.28 \pm 0.03 \,,
 &a_1 &= 0.44 \pm 0.08 \,,
 &a_2 &= 0.18 \pm 0.06 \,,
\nn\\
 &&a_3 &= -0.42 \pm 0.08 \,.
\end{align}
We have checked that we obtain the same values for $c^\res$ within the
uncertainties when the fit range is restricted to the region of small
$\tau^\cut$. (In this case the fit is not sensitive to $a_2$ and $a_3$, so either
one or both of them must be set to zero.) Note that $\sigma^\NNLO(\tau^\cut)$ and
$\sigma^{\sing,\NNLL}\vert_\NNLO$ both diverge as $\tau^\cut \to 0$. The fact
that the difference in $\sigma^\res(\tau^\cut)$ does not diverge for $\tau^\cut
\to 0$ provides an important cross check between our analytic results for the
NNLO singular terms and the full numerical NNLO result. The numerical
uncertainty from fitting $c^\res(\mu)$ is much smaller than its
$\mu$ dependence, and so can be ignored for our final error analysis.
The $\mu$
dependence of $c^\res(\mu)$ comes from the PDFs and the overall
$\alpha_s^4(\mu)$, where the latter is by far the dominant effect. To see this we
can rescale $c^\res(m_H/2) = 1.35$ and $c^\res(2m_H) = 0.58$ as
$c^\res(m_H/2) \alpha_s^4(m_H)/\alpha_s^4(m_H/2) = 0.91$ and
$c^\res(2m_H) \alpha_s^4(m_H)/\alpha_s^4(2m_H) = 0.83$, giving values
that are very close to the central value $c^\res(m_H) = 0.86$.

\begin{figure}[t]
\includegraphics[width=0.505\textwidth]{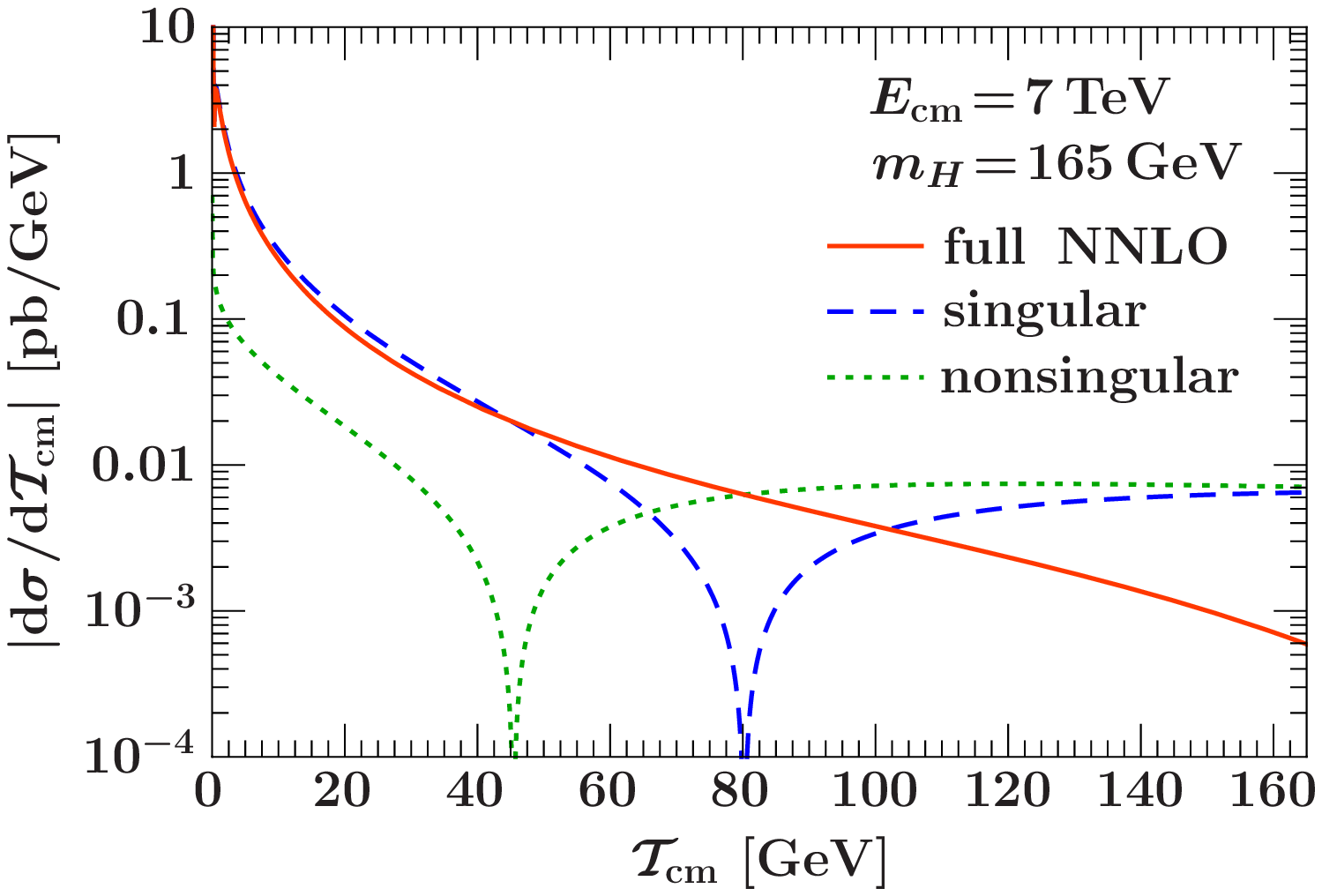}%
\hfill%
\includegraphics[width=0.475\textwidth]{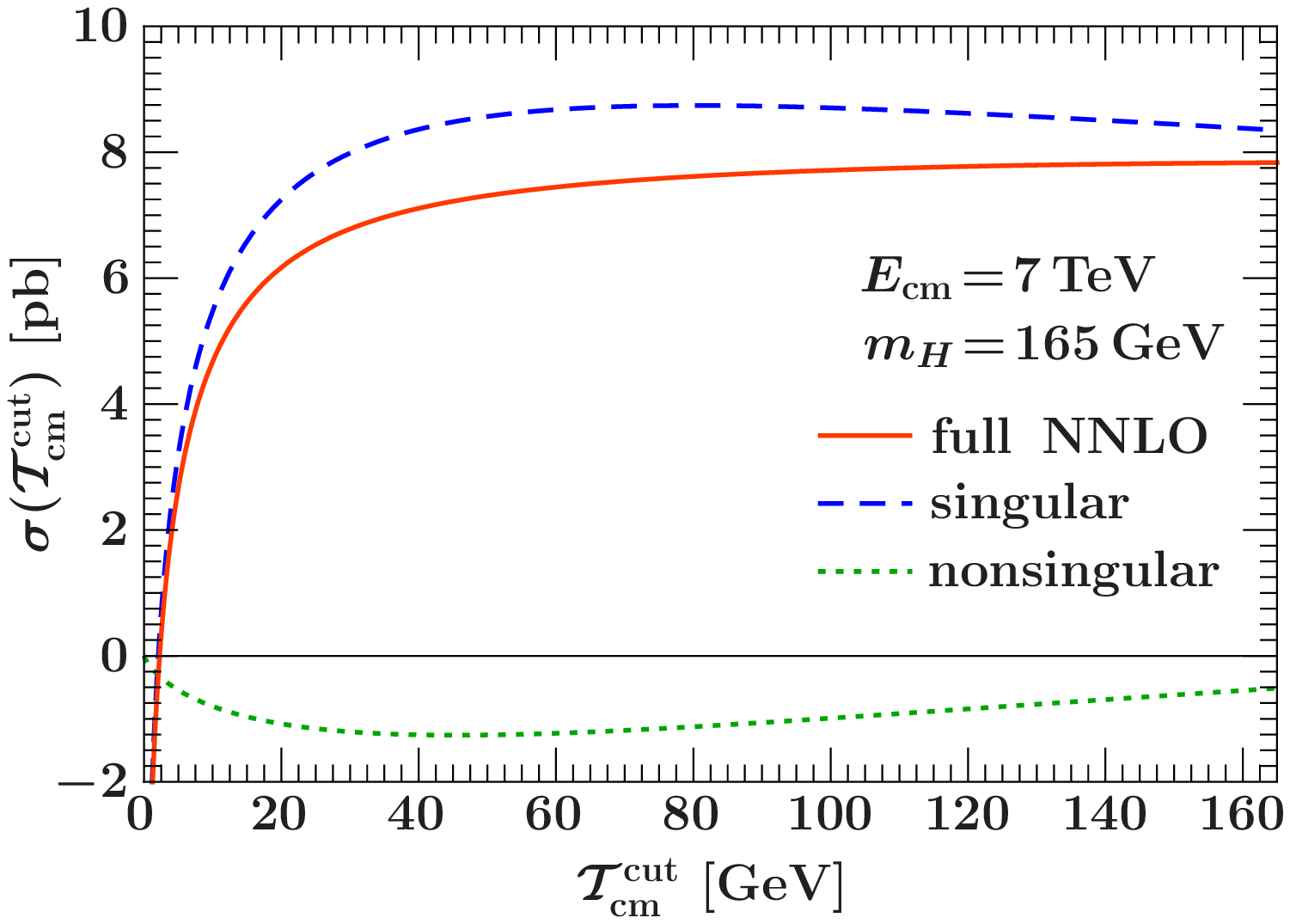}%
\vspace{-0.5ex}
\caption{Comparison of the singular, nonsingular, and full cross sections at NNLO for $\mu = m_H$. The left panel shows the magnitude of the differential cross sections on a logarithmic scale. The right panel shows the corresponding cumulant cross sections.}
\label{fig:singcomp}
\end{figure}

Having determined the nonsingular contributions to the cross section we can
compare their size to the dominant singular terms. In \fig{singcomp} we plot the
singular, nonsingular, and full cross sections at NNLO for $\mu = m_H$.  The
left panel shows the absolute value of these components of the differential
cross sections (obtained by taking the derivative of $\sigma(\tau^\cut)$ with
respect to $\tau^\cut$).  For $\Tcm \ll m_H$ the nonsingular terms are an
order of magnitude smaller than the singular ones. On the other hand for
$\Tcm \gtrsim m_H/2$ the singular and nonsingular terms become equally
important and there is a large cancellation between the two contributions. These
features of the fixed-order cross section will have implications on our choice of
running scales discussed in \subsec{profiles}.

To determine the singular NNLO contributions in \eq{singNNLO} for the above
analysis we only considered the $k\geq 0$ terms contained in
$\sigma^{\sing,\NNLL}$. Of course $\sigma^{\sing,\NNLL}$ also contains some $k =
-1$ terms at NNLO, in particular the $c^\pi(\mu)$ and $c^\mu(\mu)$
contributions, but also parts of the $c^\res(\mu)$ contribution from cross terms
between the NLO matching corrections. Since we know $c^\res(\mu)$ numerically,
we are able to determine the missing $k = -1$ contribution at NNLO numerically,
which corresponds to the sum of the unknown $\mu$-independent NNLO matching
corrections to the hard, beam, and soft functions. It is given by the difference
\begin{equation} \label{eq:cdelta}
c^\delta(\mu)
= \sigma^{\sing,\NNLO} - \sigma^{\sing,\NNLL} \big\vert_\NNLO
= c^\pi(\mu) + c^\mu(\mu) + c^\res(\mu) - \sigma^{\sing,\NNLL} \big\vert_{\NNLO,k = -1}
\,.\end{equation}
Since we include the $\mu$-dependent NNLO matching corrections in $\sigma^{\sing,\NNLL}$, its NNLO
expansion is obtained by setting $\mu_S = \mu_B = \mu_H = \mu$. Thus, we can easily evaluate
\eq{cdelta} numerically. For $m_H=165 \GeV$, we find for the LHC at $7\TeV$,
\begin{equation}
c^\delta(m_H/2) = 0.002
\,,\qquad
c^\delta(m_H) = -0.035
\,,\qquad
c^\delta(2m_H) = -0.028
\,,\end{equation}
and for the Tevatron,
\begin{equation}
c^\delta(m_H/2) = -0.0043
\,,\qquad
c^\delta(m_H) = -0.0026
\,,\qquad
c^\delta(2m_H) = -0.0027
\,.\end{equation}
Comparing this to $c^\res(m_H) = 0.86$ (LHC) and $c^\res(m_H) = 0.028$
(Tevatron), we see that these coefficients are almost fully accounted for by
cross terms between the NLO hard, beam, and soft functions. The remaining NNLO
terms in $c^\delta$ are in fact very small, and our NNLL+NNLO results are
therefore numerically very close to the complete NNLL$^\prime$+NNLO result.

\subsection{Cross Section at NNLL+NNLO}
\label{subsec:NNLLNNLO}

Using the results of
sections~\ref{subsec:hard} to \ref{subsec:nonsingular} our final result
at NNLL+NNLO for the distribution and cumulant is obtained as
\begin{align} \label{eq:NNLL_NNLO}
\frac{\df\sigma^\mathrm{NNLL+NNLO}}{\df \Tcm}
&= \frac{\df\sigma^{\sing,\NNLL}}{\df \Tcm}
+ \frac{\df\sigma^\mathrm{\delta}}{\df \Tcm}
+ \frac{\df\sigma^{\ns,\NNLO+\pi^2}}{\df \Tcm}
\,,\nn\\[1ex]
\sigma^\mathrm{NNLL+NNLO}(\Tcmc)
&= \sigma^{\rm s,NNLL}(\Tcmc) + \sigma^\delta(\Tcmc)
  + \sigma^{\ns,\NNLO+\pi^2}(\Tcmc)
\,.\end{align}
The first term in each equation contains the resummed singular result obtained
from \eq{TauBcm_run} to NNLL order, including the $\mu$-dependent NNLO matching
corrections. The last term contains the NNLO nonsingular corrections determined
in the previous subsection, but including $\pi^2$ summation by using
\begin{align} \label{eq:NNLOnspi2}
  \frac{\df\sigma^{\ns,\NNLO+\pi^2}}{\df \Tcm}
 &= U_H(m_H^2,-\img\mu_\ns,\mu_\ns) \biggl[ \frac{\df\sigma^{\ns,\NNLO}}{\df \Tcm}
  - \frac{\alpha_s(\mu_\ns) C_A}{2\pi}\, \pi^2\, \frac{\df\sigma^{\ns,\NLO}}{\df \Tcm}
  \biggr] \,,
\nn\\
\sigma^{\ns,\NNLO+\pi^2}(\Tcmc)
 &= U_H(m_H^2,-\img\mu_\ns,\mu_\ns) \Bigl[ \sigma^{\ns,\NNLO}(\Tcmc)
  - \frac{\alpha_s(\mu_\ns) C_A}{2\pi}\, \pi^2 \sigma^{\ns,\NLO}(\Tcmc) \Bigr] \,.
\end{align}
Here $U_H(m_H^2,-\img\mu_\ns,\mu_\ns) = \exp[\alpha_s(\mu_\ns) C_A\pi/2 +
\ldots]$ contains the $\pi^2$ summation.  There are two reasons we include the
$\pi^2$ summation for the nonsingular terms.  First, using SCET one can derive
factorization theorems for the nonsingular terms when $\Tcm \ll m_H$, and the
results will involve a combination of leading and subleading hard, jet, and soft
functions. Many of these terms will have the same LL evolution for their hard
functions, and hence they predominantly require the same $\pi^2$ summation. As a
second reason we observe from \fig{singcomp} that there are important
cancellations between the singular and nonsingular cross sections for $\Tcm
\gtrsim m_H/2$. Since the $\pi^2$ summation modifies the cross section for all
$\Tcm$ and $\Tcmc$ values it is important to include it also in the nonsingular
terms to not spoil these cancellations.

The middle terms in \eq{NNLL_NNLO} incorporate the singular NNLO terms that are
not reproduced by our resummed NNLL result. At fixed order, they are given by
\begin{equation}
\frac{\df\sigma^\mathrm{\delta}}{\df \Tcm}\bigg\vert_\NNLO = c^\delta(\mu)\,\delta(\Tcm)
\,,\qquad
\sigma^\delta(\Tcmc) \bigg\vert_\NNLO = c^\delta(\mu)
\,.\end{equation}
As we saw in \subsec{nonsingular}, $c^\delta(\mu)$ turns out to be very small
numerically, which means we might as well neglect the $\sigma^\delta$ term
entirely. We include it for completeness in \eq{NNLL_NNLO}, in order to formally
reproduce the complete NNLO cross section. In fact, at this level other
contributions that we neglect here, such as the bottom-quark contributions or
electroweak corrections, are likely more relevant numerically.

Formally, $c^\delta(\mu)$ is reproduced by the complete two-loop matching
required at NNLL$'$ or N$^3$LL. When it is properly incorporated into the
predictions at that order it is multiplied by a Sudakov exponent that ensures
that the total $\sigma(\Tcmc)\to 0$ as $\Tcmc\to 0$. Hence, we can include it in
the resummed result by multiplying it with the NNLL evolution factors,
\begin{align} \label{eq:de_sm}
\frac{\df\sigma^\delta}{\df \Tcm}
&= c^\delta(\mu_{\ns})\, U_H(m_H^2,\mu_H,\mu)
   \int\! \df t_a \df t_b\, U_B^g(t_a,\mu_B,\mu)\, U_B^g(t_b,\mu_B,\mu)\, U_S\Bigl(\Tcm - \frac{t_a + t_b}{m_H},\mu_S,\mu\Bigr)
\,.\end{align}
The scale where we evaluate $c^\delta(\mu)$ here is an N$^3$LL effect, and so
beyond the order we are working. We choose $\mu = \mu_\ns$, which is the scale
at which we evaluate $\sigma^\ns(\Tcmc)$.

\subsection{Choice of Running Scales}
\label{subsec:profiles}

The factorization theorem in \eq{TauBcm_run} resums the singular cross section
by evaluating the hard, beam, and soft function at their natural scales $\mu_H
\simeq -\img m_H$, $\mu_B \simeq \sqrt{\tau} m_H$, $\mu_S \simeq \tau m_H$ where
they have no large logarithms in fixed-order perturbation theory. Their
renormalization group evolution is then used to connect these functions at a
common scale. This resums logarithms in the ratios of $\mu_H$, $\mu_B$, and
$\mu_S$, which are logarithms of $\tau$. The $\tau$-spectrum has three distinct
kinematic regions where this resummation must be handled differently and we will
do so using $\tau$-dependent scales given by profile functions $\mu_S(\tau)$ and
$\mu_B(\tau)$. Profile functions of this type have been previously used to
analyze the $B\to X_s\gamma$ spectrum~\cite{Ligeti:2008ac} and the thrust event shape in
$e^+e^-\to $ jets~\cite{Abbate:2010xh}.

For $\lqcd/m_H \ll \tau \ll 1$ the scales $\mu_H$, $\mu_B$, $\mu_S$, and $\lqcd$
are all widely separated and the situation is as described above. We define this
region to be $\tau_1< \tau< \tau_2$. In the $\tau < \tau_1$ region the scale
$\mu_S$ drops below $1 \GeV$, we have $\lqcd/m_H \sim \tau$, and nonperturbative
corrections to the soft function become important.  In this case the scales are
$\mu_H \simeq -\img m_H$, $\mu_B \simeq \sqrt{\lqcd m_H}$, and $\mu_S \simeq
\lqcd$.

Finally for $\tau > \tau_2$ we have $\tau \sim 1$, the resummation is not
important, and the nonsingular corrections, which are evaluated at a fixed scale
$\mu_\ns$, become just as important as the singular corrections. In this region
there is only one scale $\img \mu_H = \mu_B = \mu_S = \mu_\ns \simeq m_H$.
Furthermore it is known from $B\to X_s\gamma$ and
thrust~\cite{Ligeti:2008ac,Abbate:2010xh} that there can be important
cancellations between the singular and nonsingular terms in this limit, and that
to ensure these cancellations take place the scales $\mu_B(\tau)$ and
$\mu_S(\tau)$ must converge to $\abs{\mu_H} = \mu_\ns$ in a region, rather than
at a single point. To ensure this we make the approach to $\mu_H$ quadratic for
$\tau_2< \tau< \tau_3$ and set $\mu_B(\tau)=\mu_S(\tau)=\img \mu_H$ for $\tau\ge
\tau_3$ (recall that $\img \mu_H>0$).  As we saw in \subsec{nonsingular}, the
singular and nonsingular contributions in our case become equally important for
$\tau \gtrsim 1/2$. Accordingly, we choose the profile functions such that the
scales converge around this value and stay equal for larger $\tau$.

\begin{figure}[t]
\centering
\includegraphics[width=0.5\textwidth]{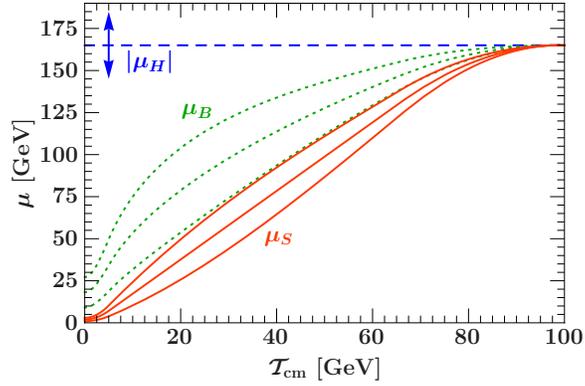}%
\vspace{-0.5ex}
\caption{Profiles for the running scales $\mu_H$, $\mu_B$, and $\mu_S$. The central lines for $\mu_B$ and $\mu_S$ show our central scale choices. The upper and lower curves for $\mu_B$ and $\mu_S$ correspond to their respective variations b) and c) in \eq{scales}.}
\label{fig:scales}
\end{figure}

A transition between these three regions is given by the following running scales
\begin{align}
  \mu_H &= -\img\, \mu
  \,, \nn \\
  \mu_B(\tau) & =
  \Big[1 + e_B\, \theta(\tau_3-\tau) \Big(1 - \frac{\tau}{\tau_3}\Big)^2\,\Big]
  \sqrt{\mu\, \mu_\mathrm{run}(\tau,\mu)}
  \,, \nn \\
  \mu_S(\tau) & = \Big[1 + e_S\, \theta(\tau_3-\tau) \Big(1 - \frac{\tau}{\tau_3}\Big)^2\,\Big] \mu_\mathrm{run}(\tau,\mu)
  \,, \nn \\
  \mu_\ns & = \mu
  \,.
\end{align}
For the profile $\mu_\mathrm{run}(\tau,\mu)$ we use a combination of two
quadratic functions and a linear function as in ref.~\cite{Abbate:2010xh}. For
$\tau> \tau_3$ our choice for $\mu_\mathrm{run}(\tau,\mu)$ ensures that our
cross section formula becomes precisely the fixed-order result.
\begin{align}
& \mu_\mathrm{run}(\tau,\mu) =
\begin{cases}
\mu_0 + a\tau^2/\tau_1 & \tau \leq \tau_1
\,,\\
2a\, \tau + b & \tau_1 \leq \tau \leq \tau_2
\,,\\
\mu - a (\tau-\tau_3)^2/(\tau_3 - \tau_2) & \tau_2 \leq \tau \leq \tau_3
\,,\\
\mu & \tau > \tau_3
\,,\end{cases}
\nn \\
&
a= \frac{\mu_0-\mu}{\tau_1-\tau_2-\tau_3}
\,, \qquad
b = \frac{\mu \tau_1 - \mu_0 (\tau_2 + \tau_3)}{\tau_1-\tau_2-\tau_3}
\,.\end{align}
The expressions for $a$ and $b$ follow from demanding that
$\mu_\mathrm{run}(\tau)$ is continuous and has a continuous derivative. The
value of $\mu_0$ determines the scales at $\tau = 0$, while $\tau_{1,2,3}$
determine the transition between the regions discussed above.  For our central
value we use the following choice of parameters
\begin{equation}
\mu = m_H
\,,\quad
e_B = e_S = 0
\,,\quad
\mu_0 = 2 \GeV
\,,\quad
\tau_1 = \frac{5 \GeV}{m_H}
\,,\quad
\tau_2 = 0.4
\,,\quad
\tau_3 = 0.6
\,.\end{equation}
The corresponding running scales are shown in \fig{scales}.

Since the factorization theorem is not affected by $\ord{1}$ changes of the
renormalization scales, we should vary them to determine the perturbative
uncertainty. For a reasonable variation of the above parameters, the cross
section is most sensitive to $\mu$, $e_B$ and $e_S$. We therefore estimate our
uncertainties from higher order terms in perturbation theory by taking the
envelope of the following three separate variations,
\begin{align} \label{eq:scales}
&\text{a)}&
\mu &= 2^{\pm 1} m_H\,,& e_B &= 0\,,& e_S &= 0
\,,\nn\\
&\text{b)}&
\mu &= m_H\,,&  e_B &= \pm 0.5\,,& e_S &= 0
\,,\nn\\
&\text{c)}&
\mu &= m_H\,,& e_B &= 0\,,& e_S &= \pm 0.5
\,.\end{align}
The effect of variations b) and c) are shown in \fig{scales} by the upper and
lower curves for $\mu_B$ and $\mu_S$, respectively. The effect of variation a)
is to change the overall vertical scale of the $|\mu_H|$, $\mu_B$, and $\mu_S$
curves in \fig{scales} by a factor of $1/2$ or $2$ as indicated by the arrows.
In predictions based on fixed-order perturbation theory only a scale variation
analogous to a) can be considered.

\subsection{PDFs and $\pi^2$ Summation}
\label{subsec:pi2pdf}

In this subsection we briefly discuss the choice of the order of the PDFs for our
resummed results and the effect of the $\pi^2$ summation.

As shown in table~\ref{tab:counting}, by default we use the PDFs that correspond
to the order of the matching corrections, namely LO PDFs at NLL and NLO PDFs at
NNLL. Since the MSTW2008 PDFs \cite{Martin:2009bu} are extracted simultaneously
with the value of $\alpha_s(m_Z)$, by using PDFs at different orders we are
forced to also use different values of $\alpha_s(m_Z)$. Our NNLL+NNLO results
contain two-loop corrections and so at this order our default is to use the
MSTW2008 PDFs~\cite{Martin:2009bu} at NNLO with their $\alpha_s(m_Z) = 0.11707$
and with three-loop, five-flavor running for $\alpha_s(\mu)$. For our NLL$'$+NLO
and NNLL results, which include one-loop matching, we use the
corresponding NLO PDFs with their $\alpha_s(m_Z) = 0.12018$ and two-loop,
five-flavor running for $\alpha_s(\mu)$. At LL and NLL, which only includes
tree-level matching, we use the LO PDFs with $\alpha_s(m_Z) = 0.13939$ and
one-loop, five-flavor running for $\alpha_s(\mu)$.

Note that at NLL (NNLL) there is a slight mismatch in the required running of
$\alpha_s(\mu)$. The resummation at this order requires two-loop (three-loop)
$\alpha_s(\mu)$ running, whereas the used LO (NLO) PDFs employ one-loop
(two-loop) running of $\alpha_s(\mu)$. In this case, we use the following
compromise. We use the appropriate $\alpha_s(m_Z)$ and running consistent with
the PDF set to obtain the numerical value of $\alpha_s$ at some required scale.
At the same time, in the NLL (NNLL) RGE solutions we use the QCD $\beta$
function at the appropriate one higher loop order to be consistent with the RGE.
There is no such mismatch in the $\alpha_s$ running at LL, NLL$'$+NLO, and
NNLL+NNLO, and hence no mismatch for our highest order predictions.

\begin{figure}[t]
\includegraphics[width=0.495\textwidth]{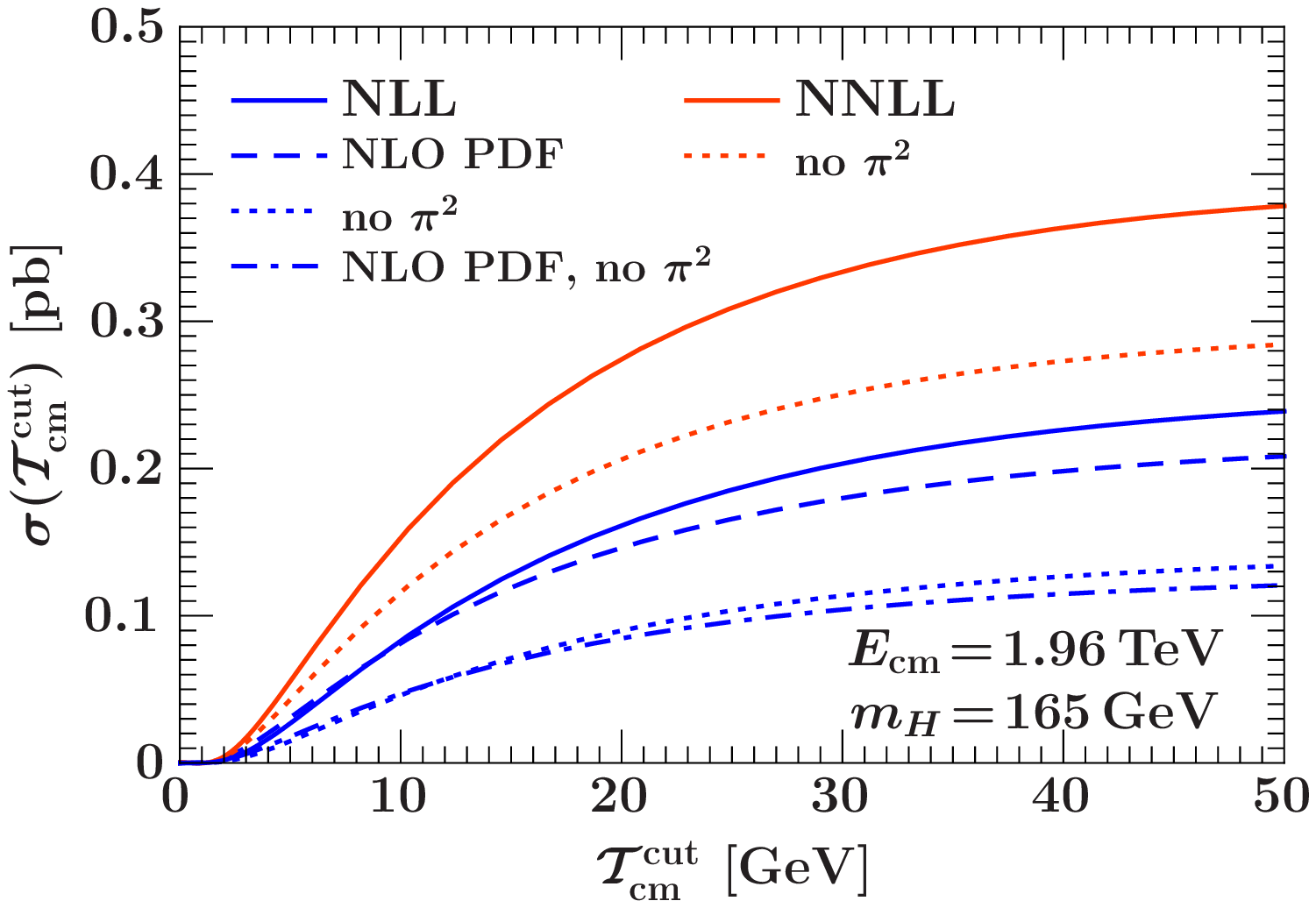}%
\hfill%
\includegraphics[width=0.485\textwidth]{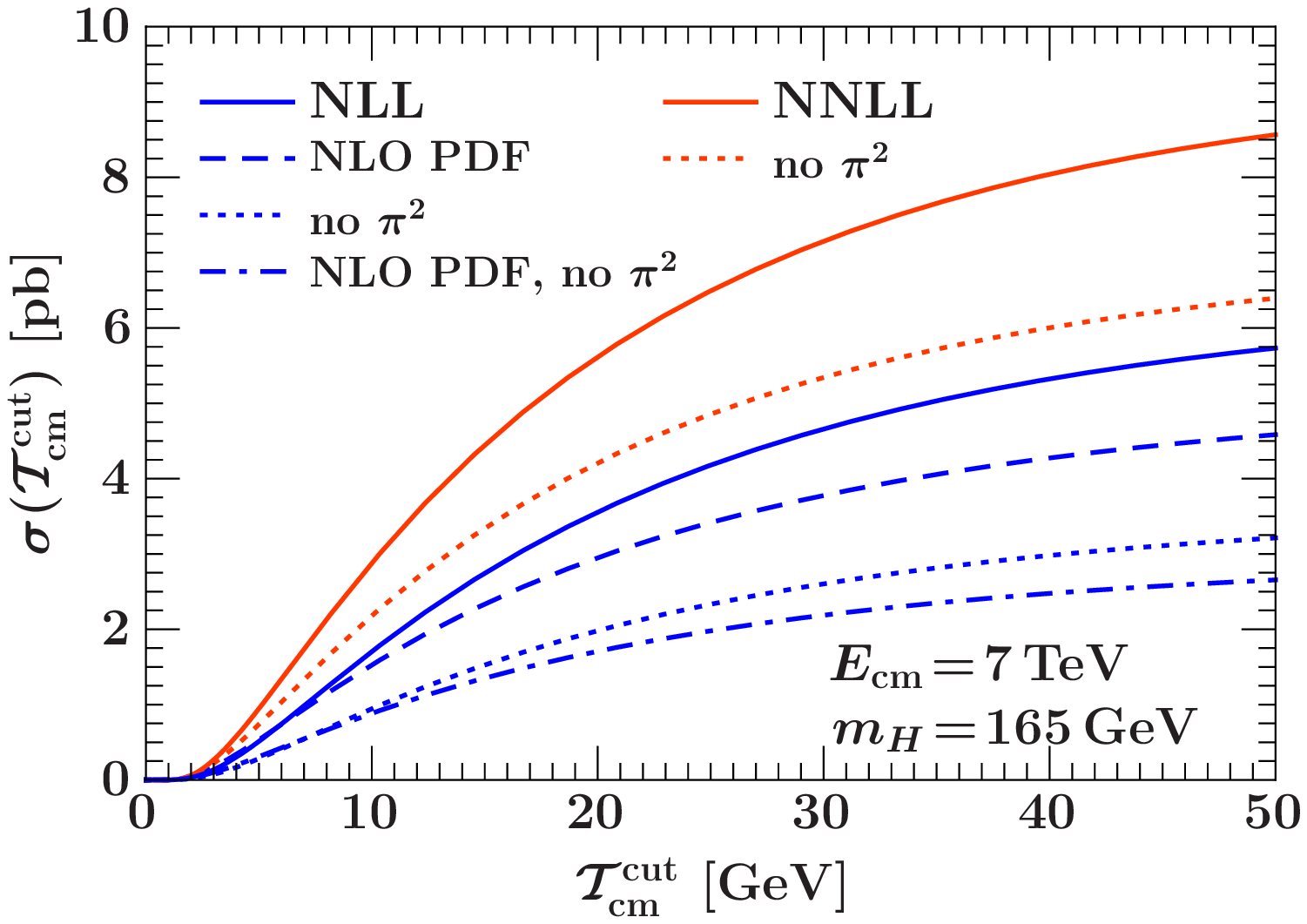}%
\vspace{-0.5ex}
\caption{The effect of $\pi^2$ summation and using different orders for the PDFs
  on the cumulant beam thrust cross section for $m_H=165 \GeV$ at the Tevatron
  (left panel) and the LHC with $7\TeV$ (right panel). Shown are the NLL result
  with/without $\pi^2$ summation and with LO/NLO PDFs, as well as the NNLL
  cross section with/without $\pi^2$ summation and NLO PDFs. See the text for
  further explanations.}
\label{fig:dsdTauB_nopi2_LHC}
\end{figure}

Various results which test the effect of $\pi^2$ resummation and the treatment
of PDFs are shown for the cumulant cross section, $\sigma(\Tcmc)$, for $m_H=165
\GeV$ in \fig{dsdTauB_nopi2_LHC}. The left panel shows the Tevatron case and
right panel shows the LHC with $\Ecm = 7 \TeV$.  The lower four blue curves show
the NLL and the upper two orange curves the NNLL results. (The nonsingular
corrections do not affect this discussion much, so for simplicity we do not
include them in this figure.) The solid lines correspond to our default results,
while the dashed, dot-dashed, and dotted are variations with other choices for
the PDFs or $\pi^2$ summation.

As discussed in \subsec{hard}, the hard function contains large $\alpha_s^n
\pi^{2m}$ terms, with $m \leq n$, which in our default results are summed by
evaluating the hard function at $\mu_H = -\img m_H$. The $\pi^2$ summation is
switched off by taking $\mu_H = m_H$ instead, which is shown by the dotted lines
in \fig{dsdTauB_nopi2_LHC}. The effect of $\pi^2$ summation is very large. It
almost doubles the NLL cross section and increases the NNLL cross section by
about $30\%$. From the fact that the NLL results with $\pi^2$ summation (blue
solid line) is very close to the NNLL result without $\pi^2$ summation (orange
dotted line) we can conclude that the large corrections from NLL to NNLL are
caused by the large $\pi^2$ terms in the virtual hard-matching contributions.
This result for the cross section with a cut on $\Tcm$ agrees with the
observations made in ref.~\cite{Ahrens:2008qu} for the total cross section.
Similarly, we have checked that the $\pi^2$ summation in the NNLL result brings
it much closer to the total NNLO result. As a result, the convergence of the
perturbative series is significantly improved by including the $\pi^2$
summation, and we will make use of this for our main results.

We can also explore the effect of using the NLO PDFs already at NLL, which
amounts to including some higher-order terms in the NLL result but allows us to
use the same value for $\alpha_s(m_Z)$ at NLL and NNLL. The corresponding
results are shown by the blue dashed (with $\pi^2$ summation) and dot-dashed
(without $\pi^2$ summation) in \fig{dsdTauB_nopi2_LHC}. For our default (solid
blue curve) we use the LO PDFs in the NLL cross section, which increases it by
about $10\%$ compared to using NLO PDFs at NLL. Thus it moves the NLL result in
the direction of the NNLL result. The main reason for the upward shift is the
higher value of $\alpha_s$ associated with the lower-order PDFs, since the Higgs
cross section has an overall $\alpha_s^2$.

\subsection{Nonperturbative Corrections}
\label{subsec:Nonperturbative}

%
\begin{figure}[t]
\includegraphics[width=0.495\textwidth]{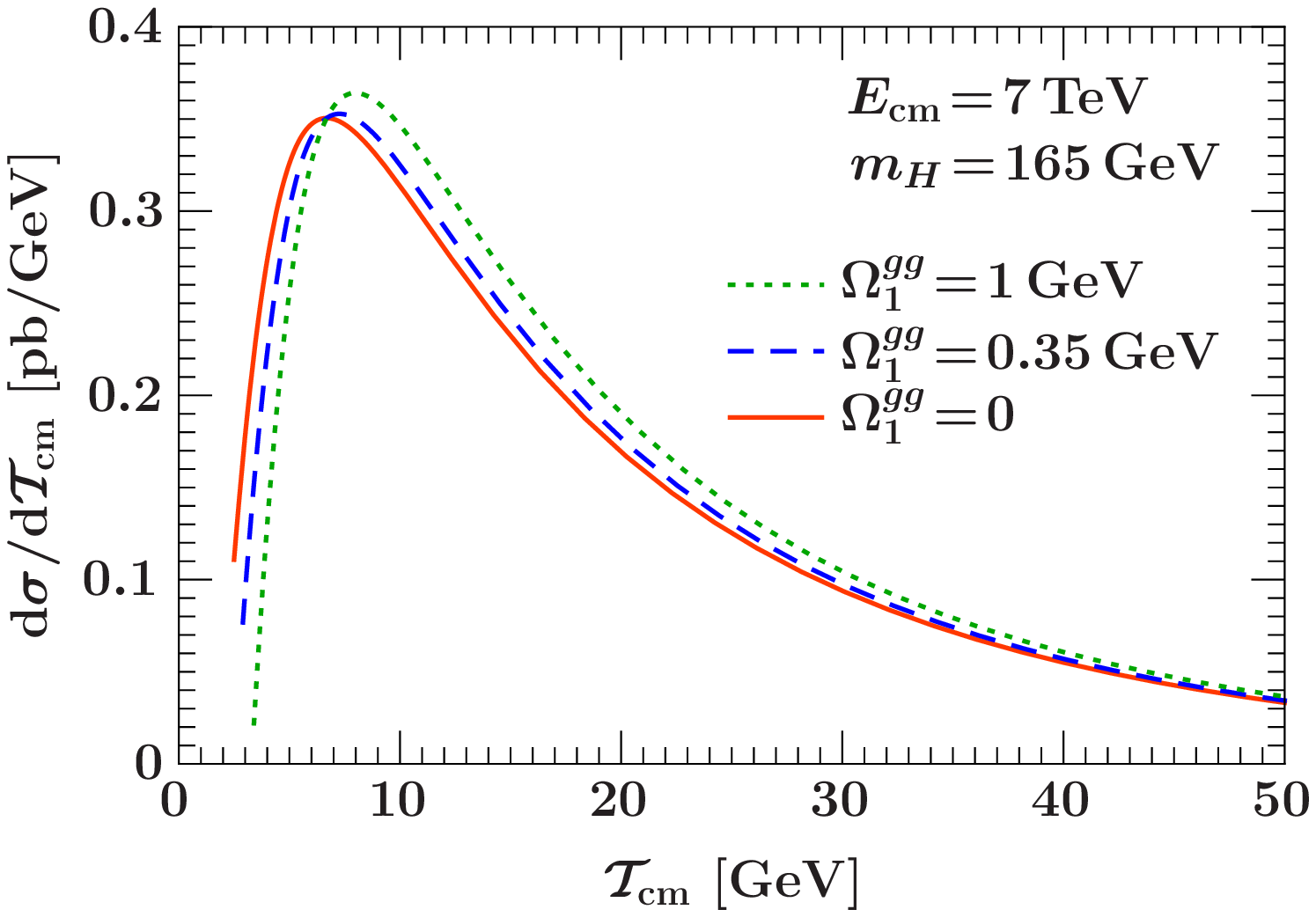}%
\hfill\includegraphics[width=0.485\textwidth]{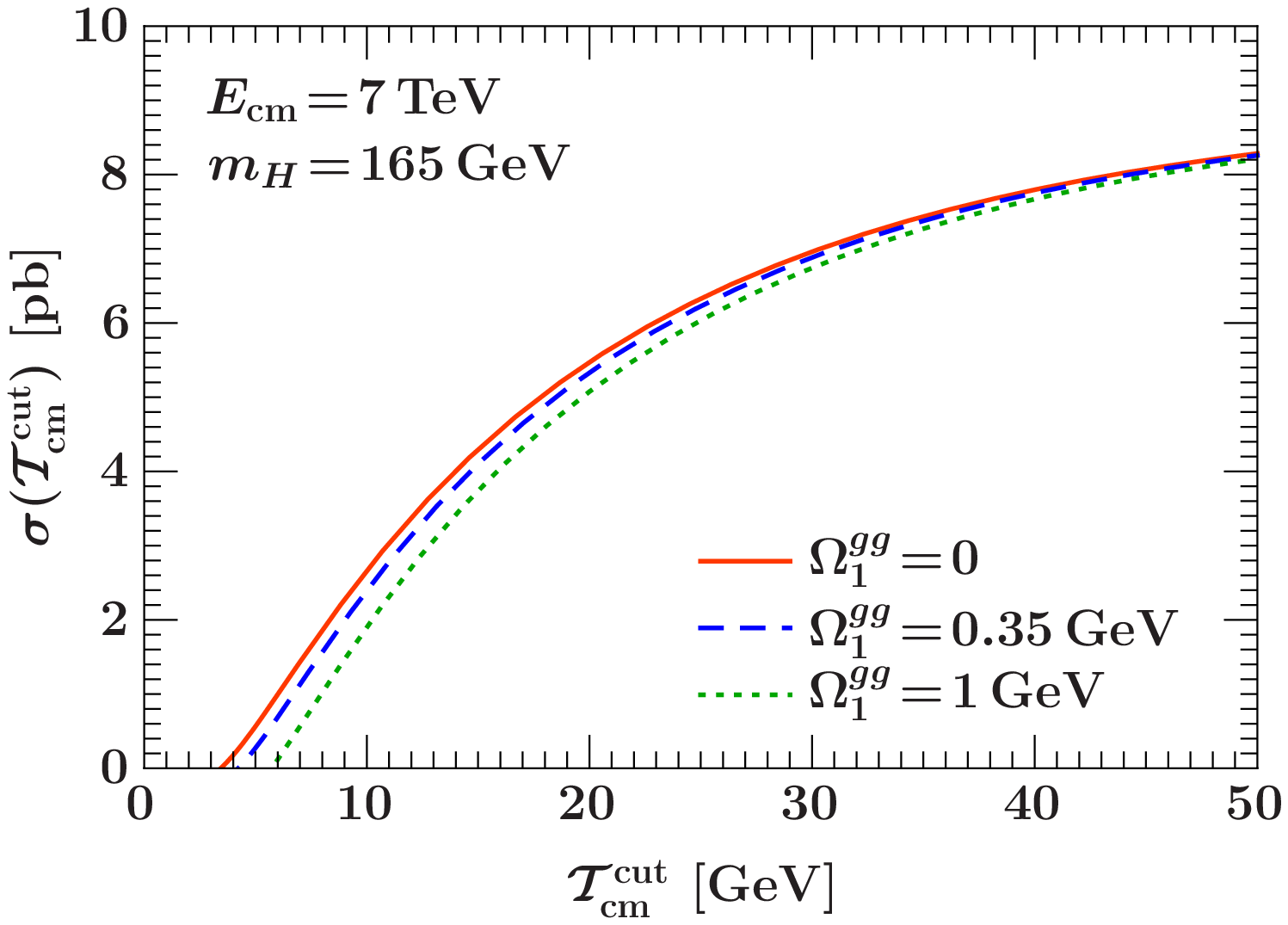}%
\vspace{-0.5ex}
\caption{Shift to the NNLL+NNLO perturbative cross section, shown by solid curves with
  $\Omega_1^{gg}=0$, caused by the leading nonperturbative hadronization corrections,
  shown by the dashed and dotted curves for $\Omega_1^{gg}=0.35\GeV$
  and $\Omega_1^{gg}=1.0\GeV$, respectively.  }
\label{fig:sigNP}
\end{figure}

As discussed in \subsec{soft}, for $\Tcm\ll m_H$ the leading nonperturbative
hadronization corrections to the beam thrust spectrum are given by a
nonperturbative soft function $F^{gg}$.  For $\Tcm \gg \lqcd$ the dominant
nonperturbative effect can be described by an OPE that yields a single
nonperturbative parameter $\Omega_1^{gg} \sim \lqcd$, leading to a shift in the
beam thrust spectrum by $\Tcm \to \Tcm - 2 \Omega_1^{gg}$, and in the cumulant
by $\Tcmc\to \Tcmc-2\Omega_1^{gg}$.  To illustrate the size of this
nonperturbative effect we consider two values for $\Omega_1^{gg}$. First,
$\Omega_1^{gg}=0.35\GeV$ is motivated by the fit result for an analogous
parameter for dijet quark production in $e^+e^-$
collisions~\cite{Abbate:2010xh}. Second, $\Omega_1^{gg}=1.0\GeV$ is motivated by
a potential enhancement by $C_A/C_F=9/4$ from Casimir scaling for adjoint Wilson
lines. This choice also reproduces roughly the size of hadronization effects for
Higgs production in \Pythia. Using \eqs{OPEdsdT}{OPEsigc}, results from the OPE
for the LHC with $\Ecm=7\TeV$ are shown in \fig{sigNP}.  Comparing the OPE
results for the distribution, shown in the left panel, to the full convolution
with a model soft function using \eq{SF}, we find that the OPE works well for
the entire displayed spectrum when $\Omega_1^{gg}=0.35\GeV$ and for $\Tcm >
10\GeV$ when $\Omega_1^{gg}=1.0\GeV$. (Thus, for $\Omega_1^{gg}=0.35\GeV$ the
peak is perturbative.) Examining the right panel of \fig{sigNP}, we see that at
$\Tcmc=20\GeV$ a power correction of $\Omega_1^{gg}=0.35\GeV$ reduces
$\sigma(\Tcmc)$ by $3\%$, while for $\Omega_1^{gg}=1.0\GeV$ the reduction is by
$7\%$.  (The results for the Tevatron are very similar, giving reductions by
$2\%$ and $6\%$, respectively, for $\Tcmc=20\GeV$.) The sign of this
nonperturbative shift is predicted by the factorization theorem, while its
magnitude is determined by $\Omega_1^{gg}$.  Examining the $\Tcm$ spectra from
\Pythia before and after hadronization, we find that the hadronization
correction in \Pythia is consistent with the nonperturbative shift discussed
here with a value $\Omega_1^{gg}=1.0\GeV$ for both the Tevatron and LHC.

\section{Numerical Results}
\label{sec:results}

In this section we present our numerical results for the Higgs production cross
section for both the differential beam thrust spectrum, $\df\sigma/\df\Tcm$, and
the cumulant, $\sigma(\Tcmc)$, which gives the integrated cross section with a
cut on beam thrust, $\Tcm \leq \Tcmc$. We are mostly interested in the region of
small $\Tcm$ or $\Tcmc$, which corresponds to the $0$-jet region.  We will show
resummed results up to NNLL+NNLO order and also compare with the results
obtained in fixed-order perturbation theory at NNLO using
\Fehip~\cite{Anastasiou:2004xq, Anastasiou:2005qj}.  An explanation of the
various orders is given at the beginning of \sec{calc} and in
table~\ref{tab:counting}.  Since our focus in this section is on the
perturbative results and their uncertainties, we will not include the
nonperturbative hadronic correction discussed in \subsec{Nonperturbative}
(i.e. we take $\Omega_1^{gg}=0$).

The perturbative uncertainties in the resummed predictions are estimated as
explained in detail in \subsec{profiles}. For the fixed-order results we use
$\mu = m_H/2$ as the default choice for the central value, which tends to give a
better convergence for the total cross section, mimicking the effect of the
$\pi^2$ summation. The perturbative scale uncertainties at fixed order are then
evaluated using $\mu = m_H$ and $\mu = m_H/4$. (We follow
ref.~\cite{Anastasiou:2009bt} and do not vary the renormalization and factorization
scales independently.) Since our focus here is on the perturbative
uncertainties, we do not add PDF and $\alpha_s(m_Z)$ uncertainties in our plots.
We have checked that they are essentially independent of the cut on beam thrust
and the same as for the total inclusive cross section.

We show results for both the Tevatron and the LHC. For the LHC we always use
$\Ecm = 7\TeV$. The results for higher $\Ecm$ are qualitatively similar, except for
the overall increased cross section. For most of our plots we use $m_H = 165\GeV$,
which is near the $WW$ threshold and where the current Tevatron limits are most sensitive.
We also show some plots that illustrate the dependence of our results on $m_H$.

\subsection{Convergence of Resummed Predictions}
\label{subsec:convergence}

\begin{figure}[p]
\includegraphics[width=0.5\textwidth]{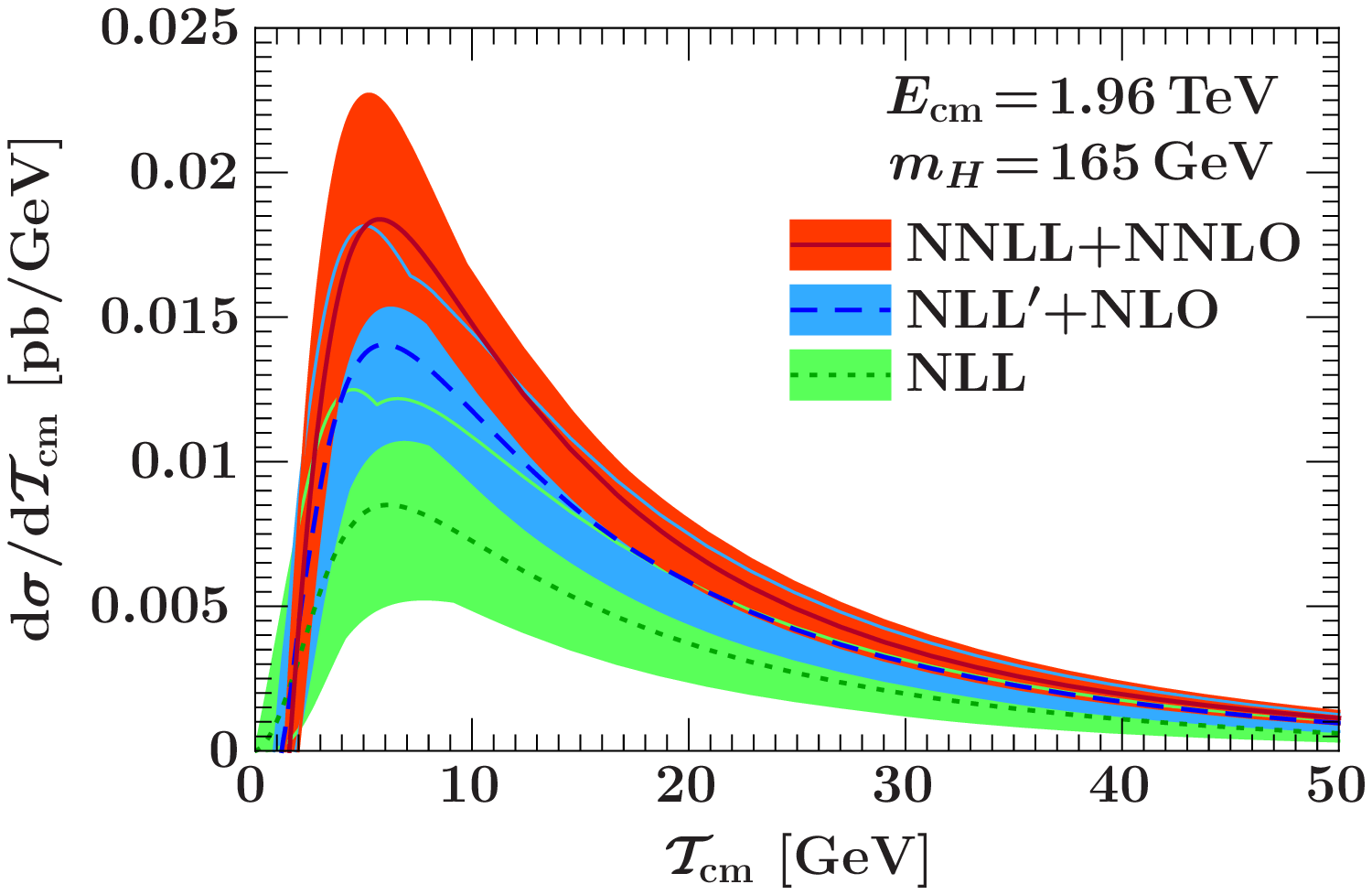}%
\hfill%
\includegraphics[width=0.48\textwidth]{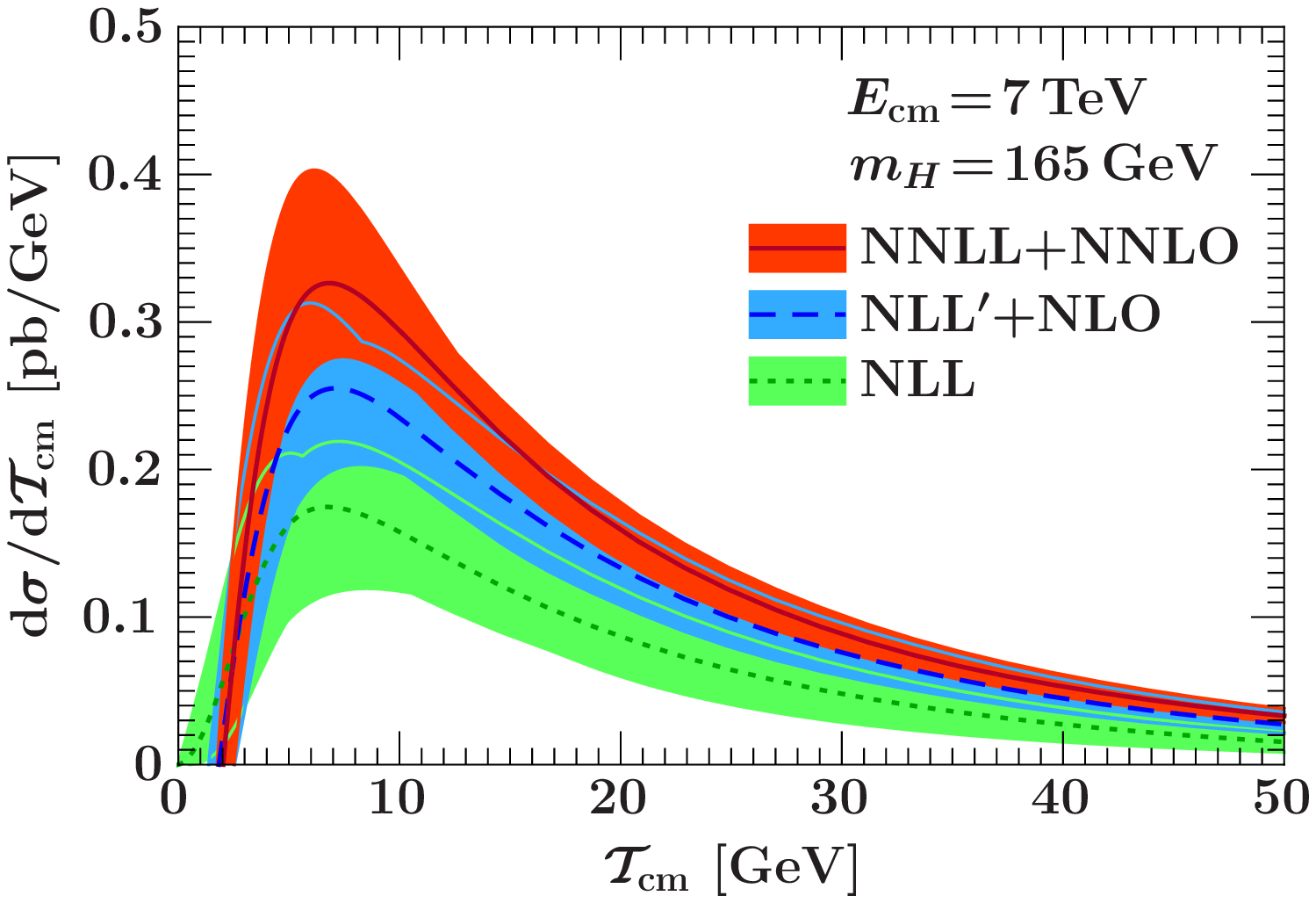}%
\vspace{-0.75ex}
\caption{The beam thrust spectrum for Higgs production for $m_H = 165\GeV$ at the
  Tevatron (left) and the LHC for $\Ecm = 7\TeV$ (right). The bands show the
  perturbative scale uncertainties as explained in \subsec{profiles}.}
\label{fig:dsdTauBcm}
\end{figure}
\begin{figure}[p]
\includegraphics[width=0.495\textwidth]{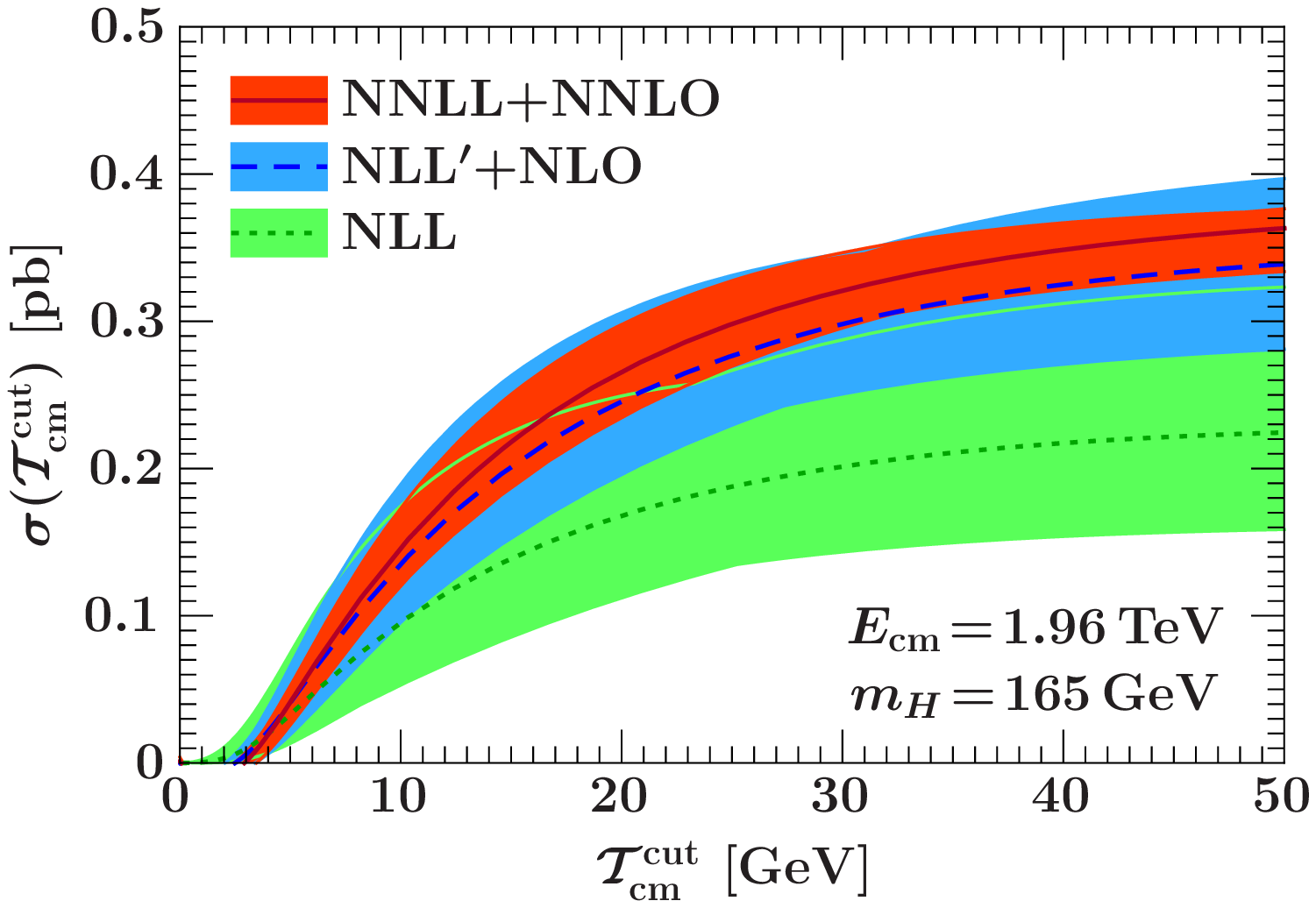}%
\hfill%
\includegraphics[width=0.485\textwidth]{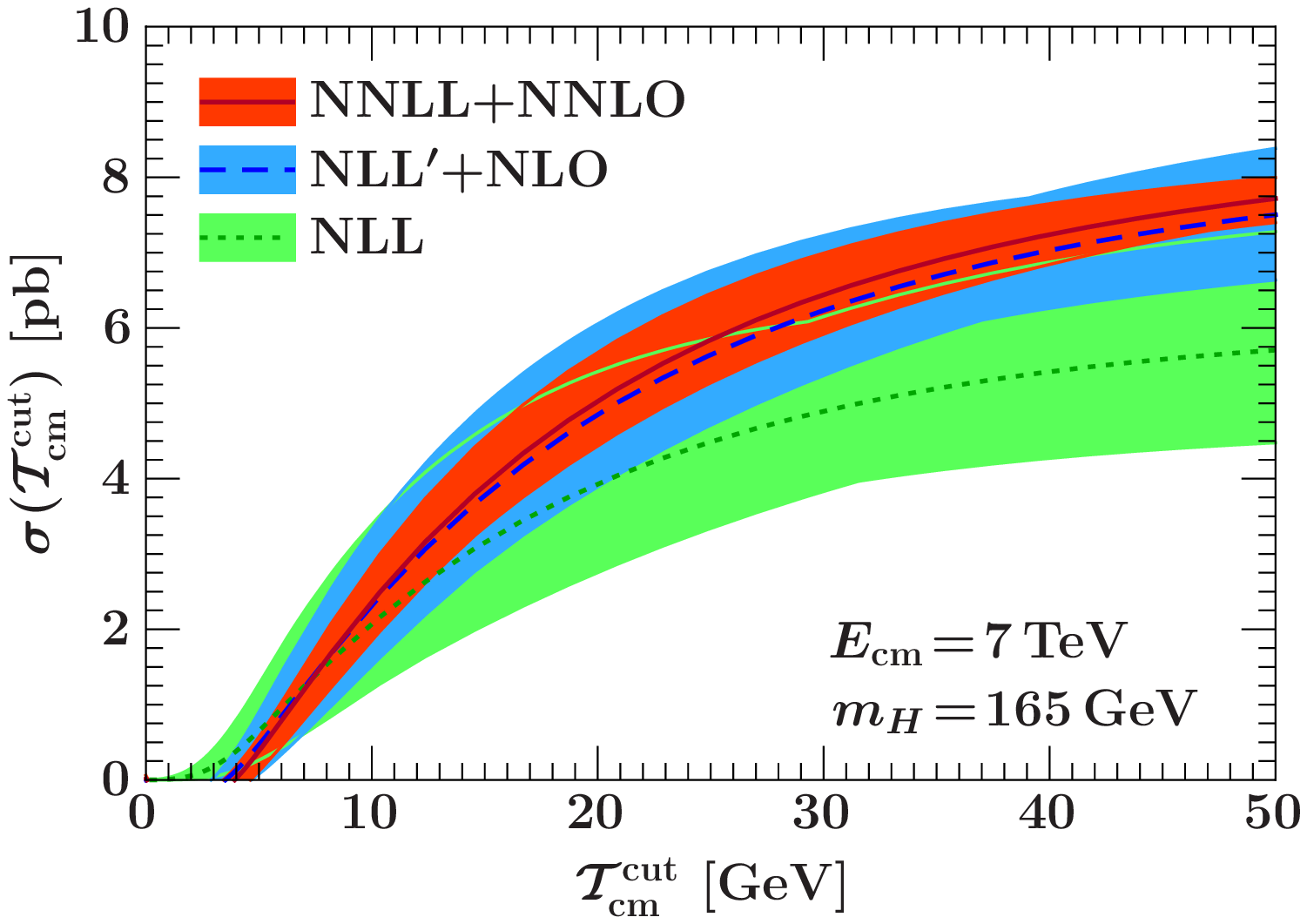}%
\vspace{-0.75ex}
\caption{Higgs production cross section as a function of $\Tcmc$ for $m_H = 165\GeV$ at the
  Tevatron (left) and the LHC with $\Ecm = 7\TeV$ (right). The bands show the
  perturbative scale uncertainties as explained in \subsec{profiles}.}
\label{fig:sigTauBcmcut}
\end{figure}
\begin{figure}[p]
\includegraphics[width=0.495\textwidth]{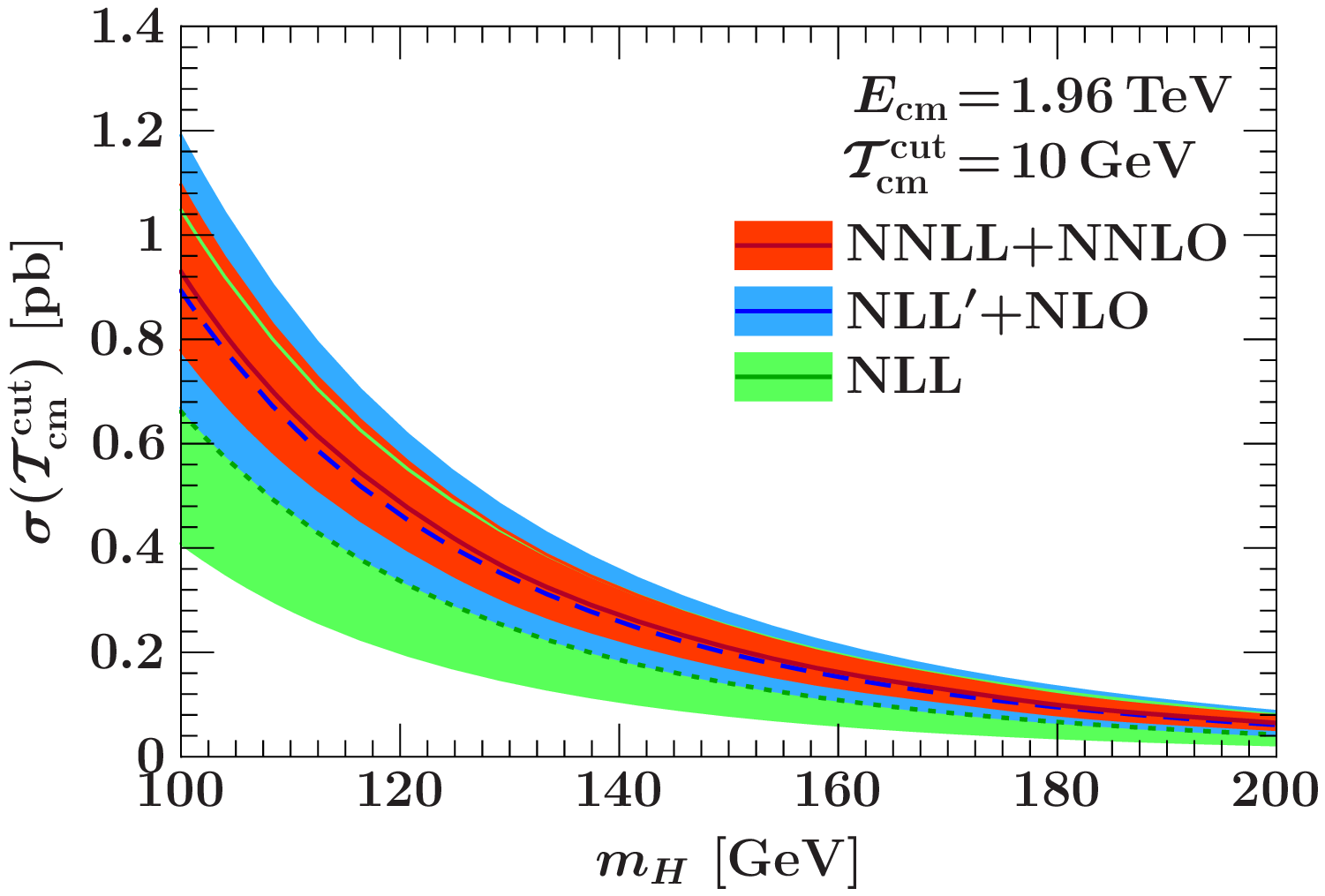}%
\hfill%
\includegraphics[width=0.485\textwidth]{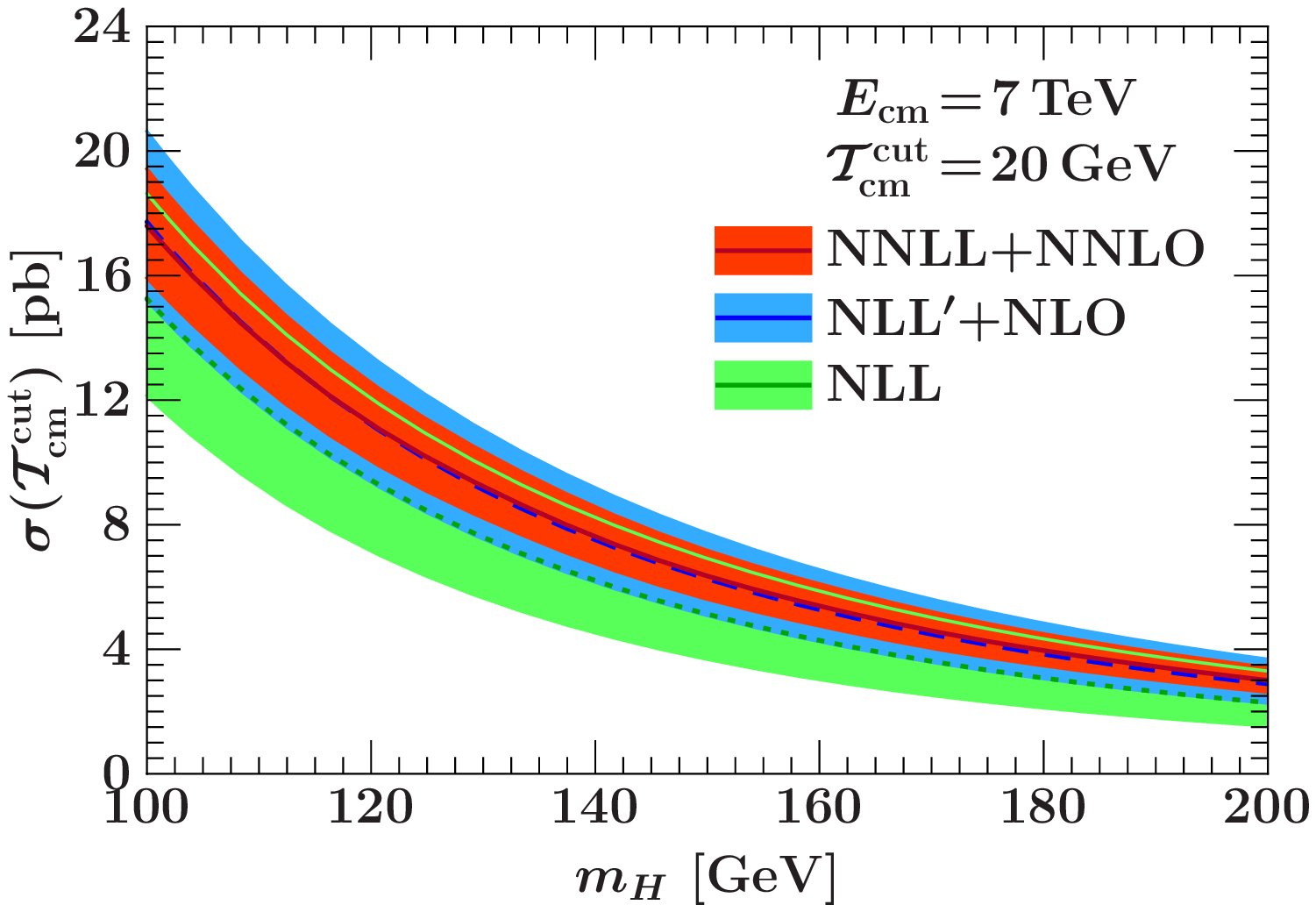}%
\vspace{-0.75ex}
\caption{Higgs production cross section with a cut on beam thrust as function of $m_H$ at the
Tevatron for $\Tcmc = 10\GeV$ (left) and the LHC with $\Ecm = 7\TeV$ and $\Tcmc=20\GeV$ (right).
The bands show the perturbative scale uncertainties as explained in \subsec{profiles}.}
\label{fig:sigmHcm}
\end{figure}
\afterpage{\clearpage}

To study the convergence of the resummed results, we consider results at three different orders:
NLL, NLL$'$+NLO, and NNLL+NNLO, which contain the matching and nonsingular corrections
at LO, NLO, and NNLO, respectively. We choose NLL instead of LL as our lowest order to compare to, since
NLL is the lowest order where we get a useful approximation with
appropriately large scale uncertainties. (The LL results are lower than the NLL ones and also have
a smaller scale uncertainty, which means that they do not contain enough information
yet to provide a reasonable lowest-order approximation.)

In \figs{dsdTauBcm}{sigTauBcmcut} we show the convergence for the differential
spectrum and cumulant, respectively for the Tevatron (left panels) and the LHC
(right panels). In \fig{sigmHcm} we show the cumulant for fixed $\Tcmc$ as a
function of the Higgs mass. We see that the perturbative corrections are rather
large, as is typical for Higgs production. The convergence within our
perturbative uncertainty bands is reasonable for the differential spectrum and
quite good for the cumulant, both for different $\Tcmc$ and different $m_H$.
The large step from NLL to NLL$'$+NLO is mostly due to the NLO matching
corrections.  As we saw in \fig{singcomp}, for $\Tcmc \ll m_H$ the nonsingular
terms are much smaller than the singular corrections that we have computed
analytically. One can also see this by comparing the size of the NLO nonsingular
terms in the left panel of \fig{nonsing} with the full cross section in the
right panel of \fig{sigTauBcmcut}.

The beam thrust spectrum in \fig{dsdTauBcm} is peaked in the $0$-jet region at
small beam thrust $\Tcm \simeq 5 \GeV$ with a large tail towards higher values.
The peak in the spectrum is a perturbative prediction; hadronization effects
only have a mild effect on the peak structure as shown above in
\subsec{Nonperturbative}. For the beam thrust spectrum of Drell-Yan, which is
Fig.~3 of ref.~\cite{Stewart:2010pd}, the peak occurs at smaller values, around
$\Tcm \sim 2 \GeV$, and the tail of the spectrum falls off much faster. The
reason for the shifted peak and higher tail for Higgs production compared to
Drell-Yan is that the incoming gluons emit much more initial-state radiation
than quarks.  This is also the main reason why the perturbative uncertainties
are still rather large even at the highest order, NNLL+NNLO. One can also see
that at the LHC the tail is somewhat higher and the peak less pronounced than at
the Tevatron.  Correspondingly, the cumulant at the Tevatron starts to level out
earlier than at the LHC. The reason is that due to the higher center-of-mass
energy at the LHC, more phase space is available for initial-state radiation.

\begin{figure}[t]
\includegraphics[width=0.495\textwidth]{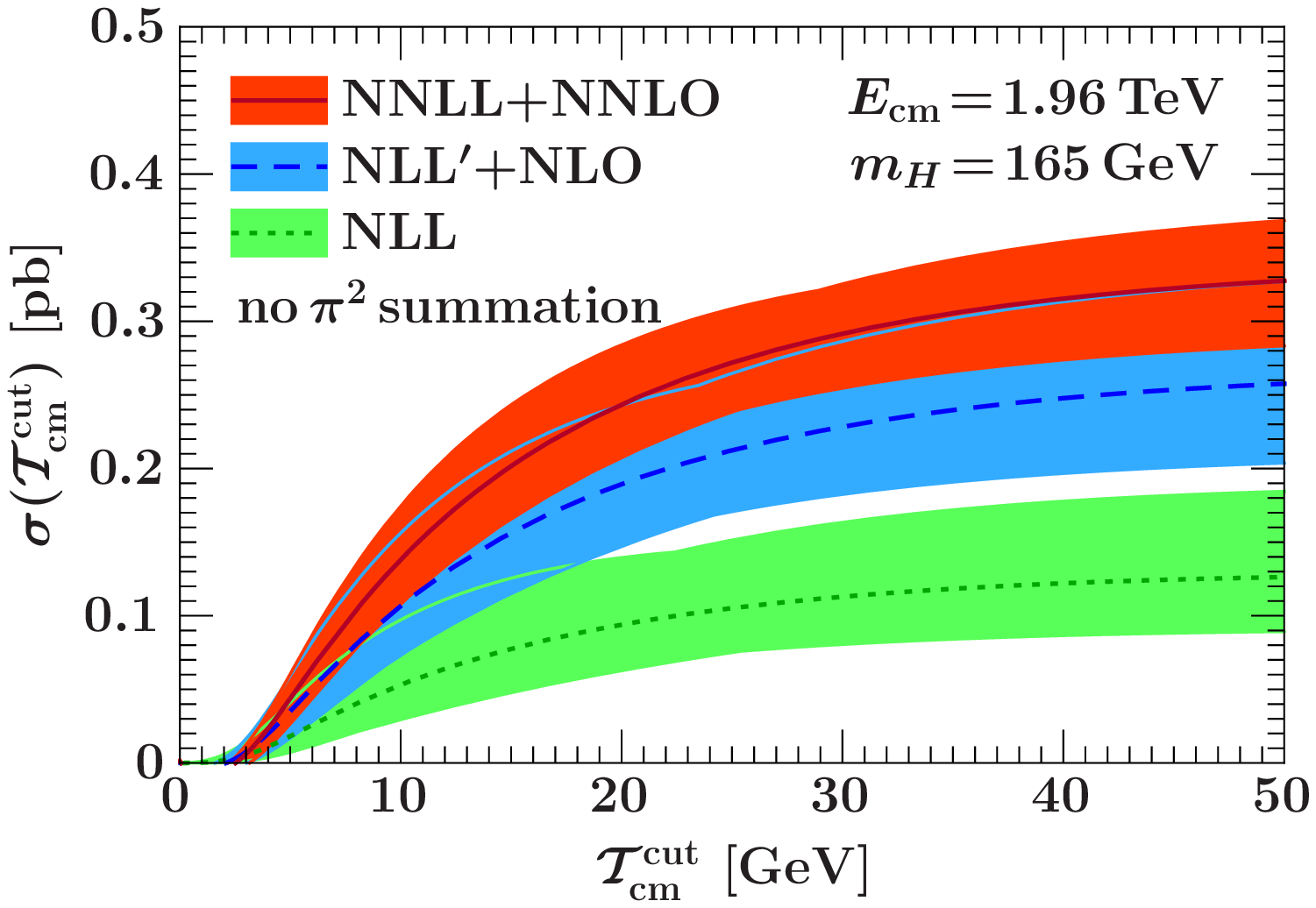}%
\hfill%
\includegraphics[width=0.485\textwidth]{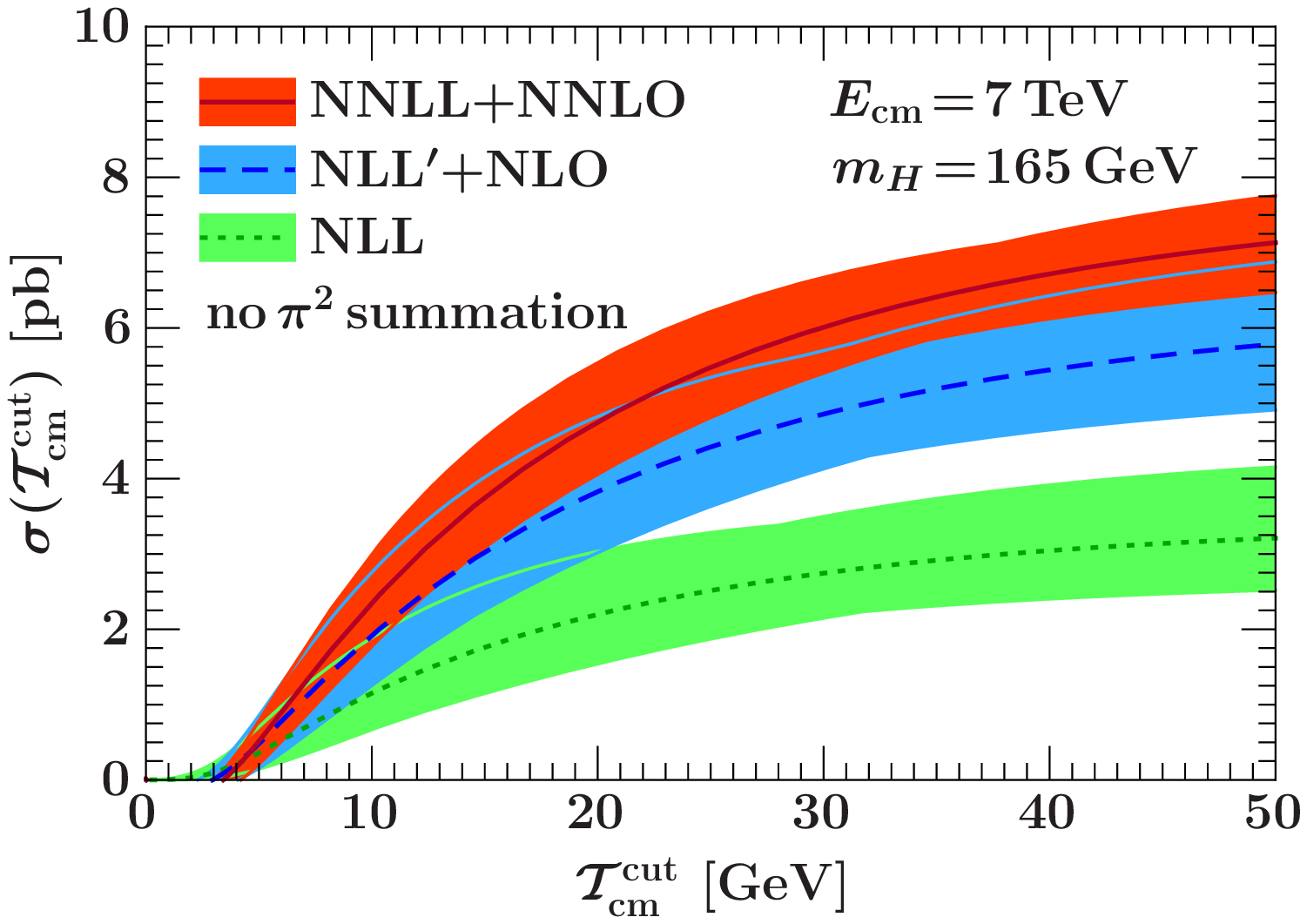}%
\vspace{-0.5ex}
\caption{Same as \fig{sigTauBcmcut} but without the $\pi^2$ summation.}
\label{fig:sigTauBcmcut_nopi2}
\end{figure}

In \fig{sigTauBcmcut_nopi2} we illustrate what happens if we turn off the
$\pi^2$ summation. Comparing with \fig{sigTauBcmcut}, we see again that the
$\pi^2$ summation significantly improves the convergence, and by reducing the
overall size of the fixed-order corrections it also reduces the size of the
perturbative uncertainties.

\subsection{Comparison of Resummed and Fixed-Order Predictions}
\label{subsec:fixedorder}

\begin{figure}[t]
\includegraphics[width=0.5\textwidth]{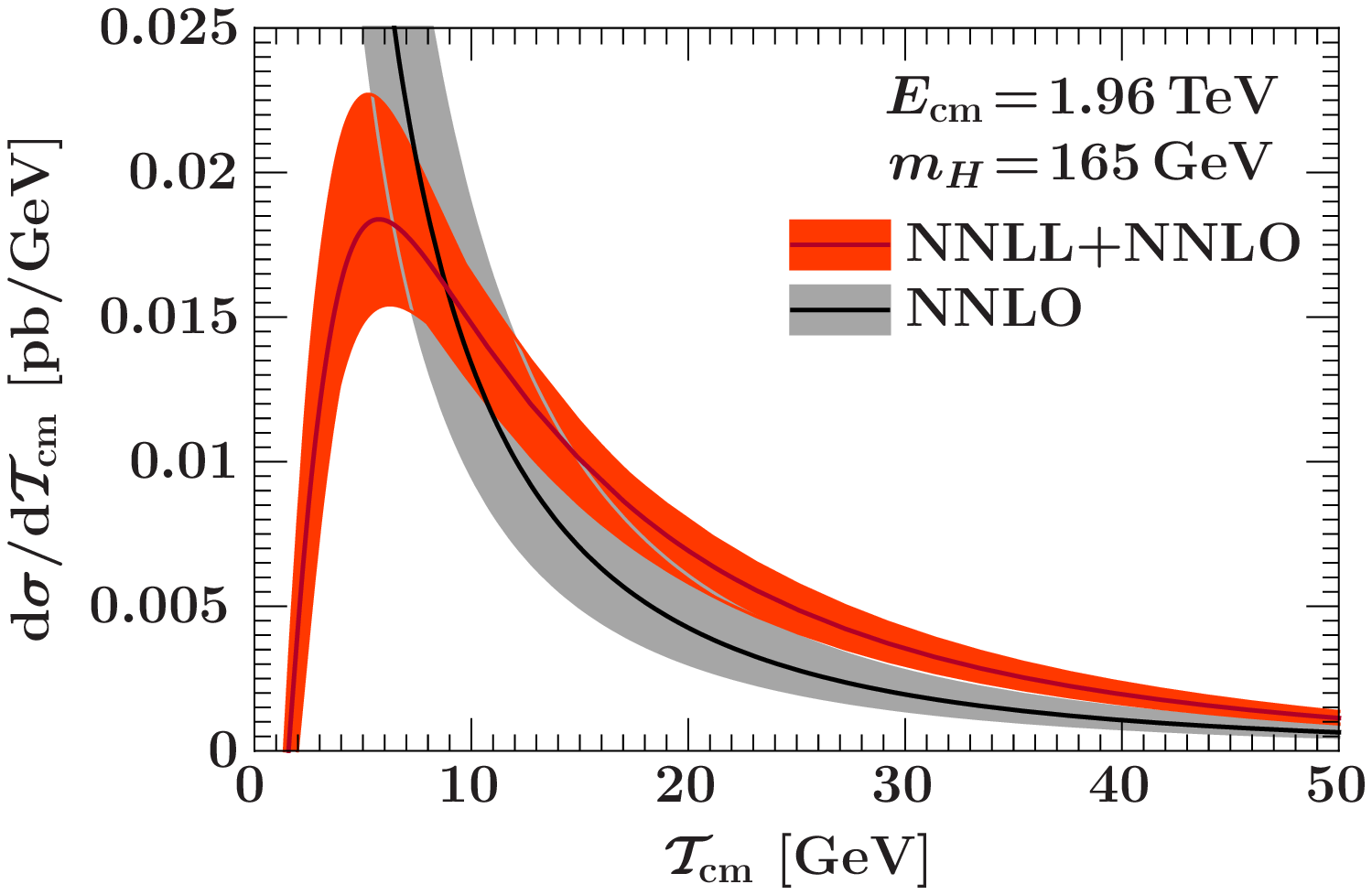}%
\hfill%
\includegraphics[width=0.48\textwidth]{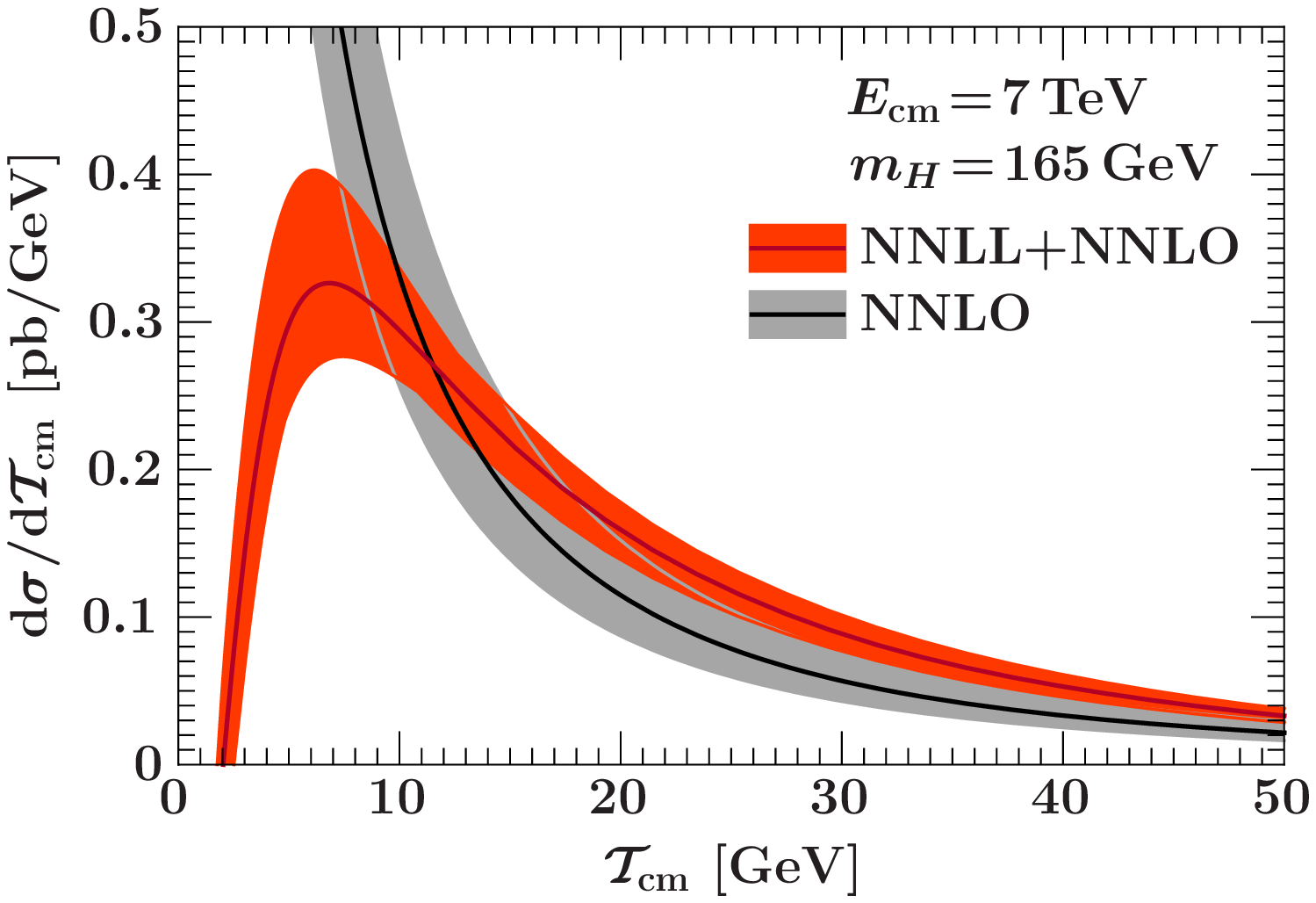}%
\vspace{-0.5ex}
\caption{Comparison of the beam thrust spectrum at NNLL+NNLO to the fixed NNLO result at the Tevatron (left) and the LHC with $\Ecm = 7\TeV$ (right). The bands show the perturbative scale uncertainties.}
\label{fig:dsdTauBcm_compNNLO}
\end{figure}

In this subsection, we compare our best resummed result at NNLL+NNLO to the NNLO fixed-order prediction without any resummation. In \fig{dsdTauBcm_compNNLO} we compare both predictions for the differential beam-thrust spectrum for the Tevatron (left panel) and the LHC (right panel). For small $\Tcm$ the large logarithms of $\Tcm/m_H$ dominate the cross section. The NNLO cross section contains terms up to $\alpha_s^2 \ln^3(\Tcm/m_H)/\Tcm$ and diverges as $\Tcm \to 0$, so we do not expect it to provide a good description of the spectrum at small $\Tcm$. In the NNLL+NNLO calculation, the series of logarithms is summed to all orders, which regulates the divergences and yields a reliable prediction for the cross section. The resummation also enhances the radiative tail in the spectrum, because it essentially sums up the effects of multiple emissions from ISR.

\begin{figure}[t]
\includegraphics[width=0.495\textwidth]{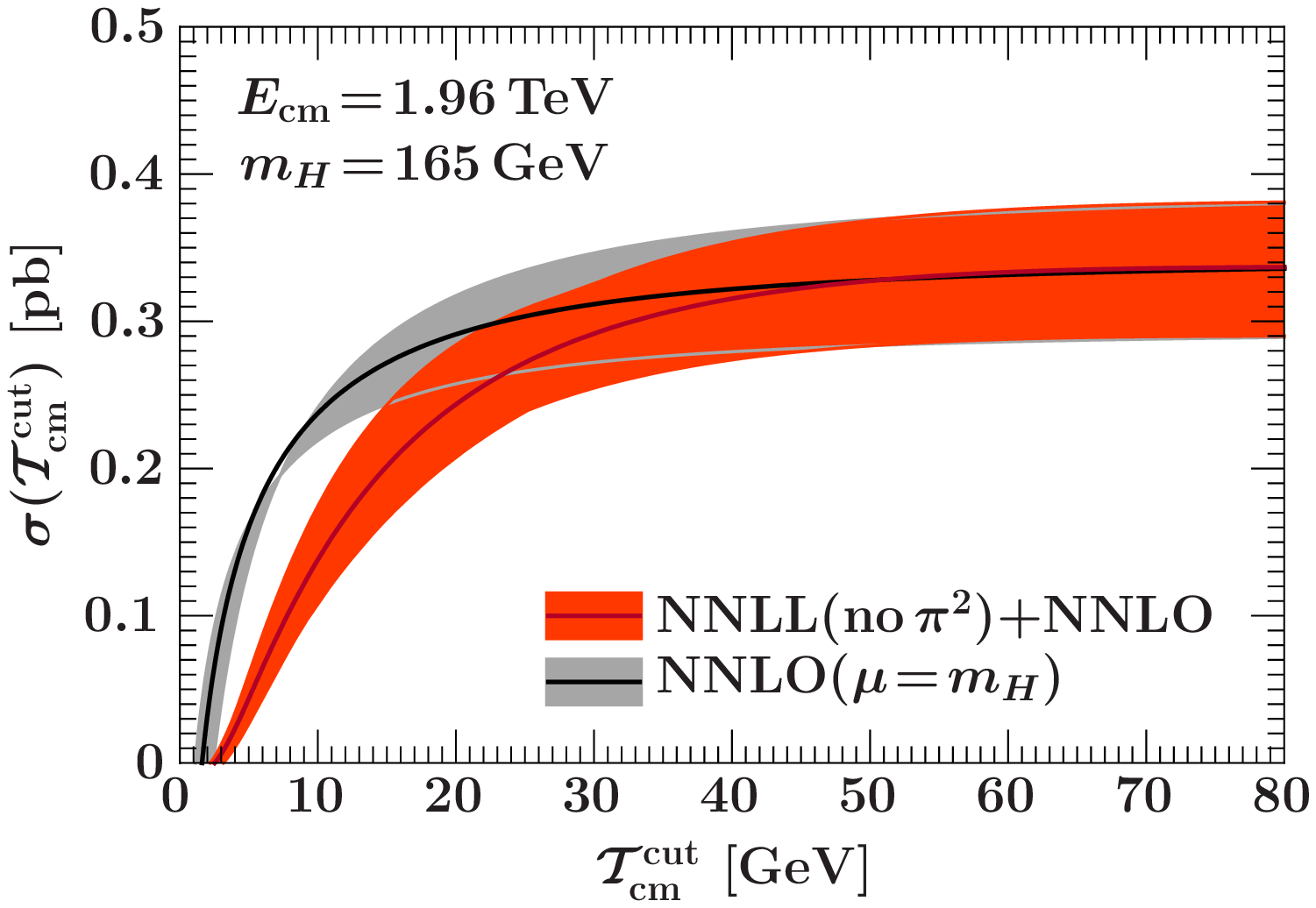}%
\hfill%
\includegraphics[width=0.485\textwidth]{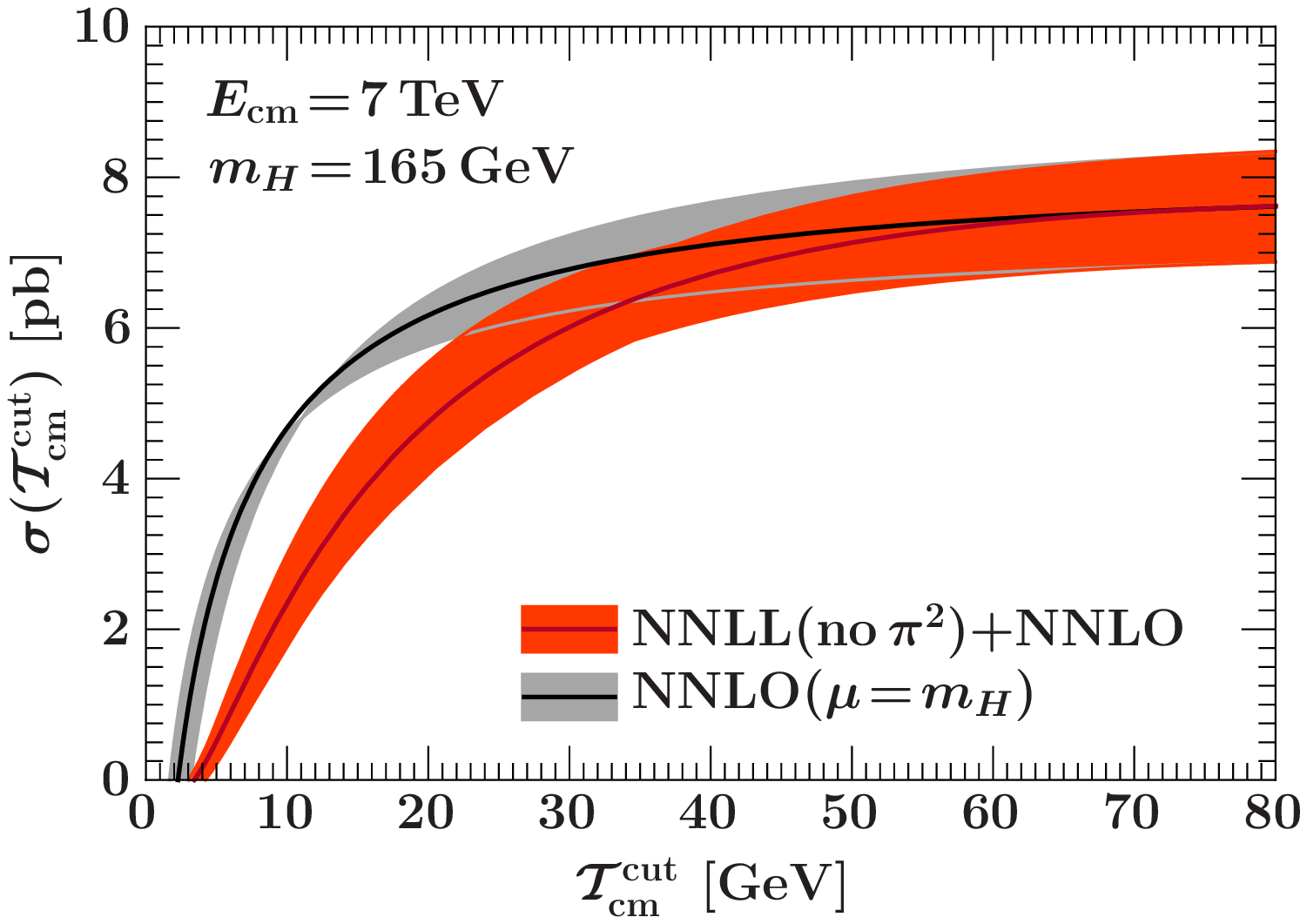}%
\vspace{-0.5ex}
\caption{Illustration that the NNLL+NNLO resummed result reproduces the fixed NNLO result at large beam thrust for the Tevatron (left) and the LHC (right).
The resummed result has the $\pi^2$ summation switched off and $\mu = m_H$ is used for the central value of the fixed-order result. The bands show the perturbative uncertainties. See text for further explanations.
}
\label{fig:matchNNLO}
\end{figure}

In \fig{matchNNLO} we illustrate that the NNLL+NNLO result correctly reproduces the NNLO result for large $\Tcmc$. To see this we need to switch off the $\pi^2$ summation, because the NNLL+NNLO result would otherwise contain higher order $\pi^2$ terms at large $\Tcmc$ that are absent at fixed NNLO. Furthermore, we use $\mu = m_H$ for the NNLO central value, and $\mu = 2m_H$ and $\mu = m_H/2$ for the NNLO scale uncertainties. In this way, the NNLL+NNLO and the NNLO are evaluated at the same scales for large $\Tcm \geq 0.6\,m_H$, where the logarithmic resummation is switched off and our running scales satisfy [see \subsec{profiles}] $\mu_S(\Tcmc) = \mu_B(\Tcmc) = \mu_\ns = \mu_H = m_H$. In \fig{matchNNLO} we see that with these choices the NNLL+NNLO indeed smoothly merges into the NNLO result, including the scale uncertainties, at large $\Tcmc$, as it should. Examining \fig{matchNNLO} for smaller $\Tcmc$ values we see that the resummed
result starts to deviate from the fixed-order one for $\Tcmc \lesssim 40\GeV$ at
the Tevatron and $\Tcmc \lesssim 50\GeV$ at the LHC.

\begin{figure}[p]
\includegraphics[width=0.49\textwidth]{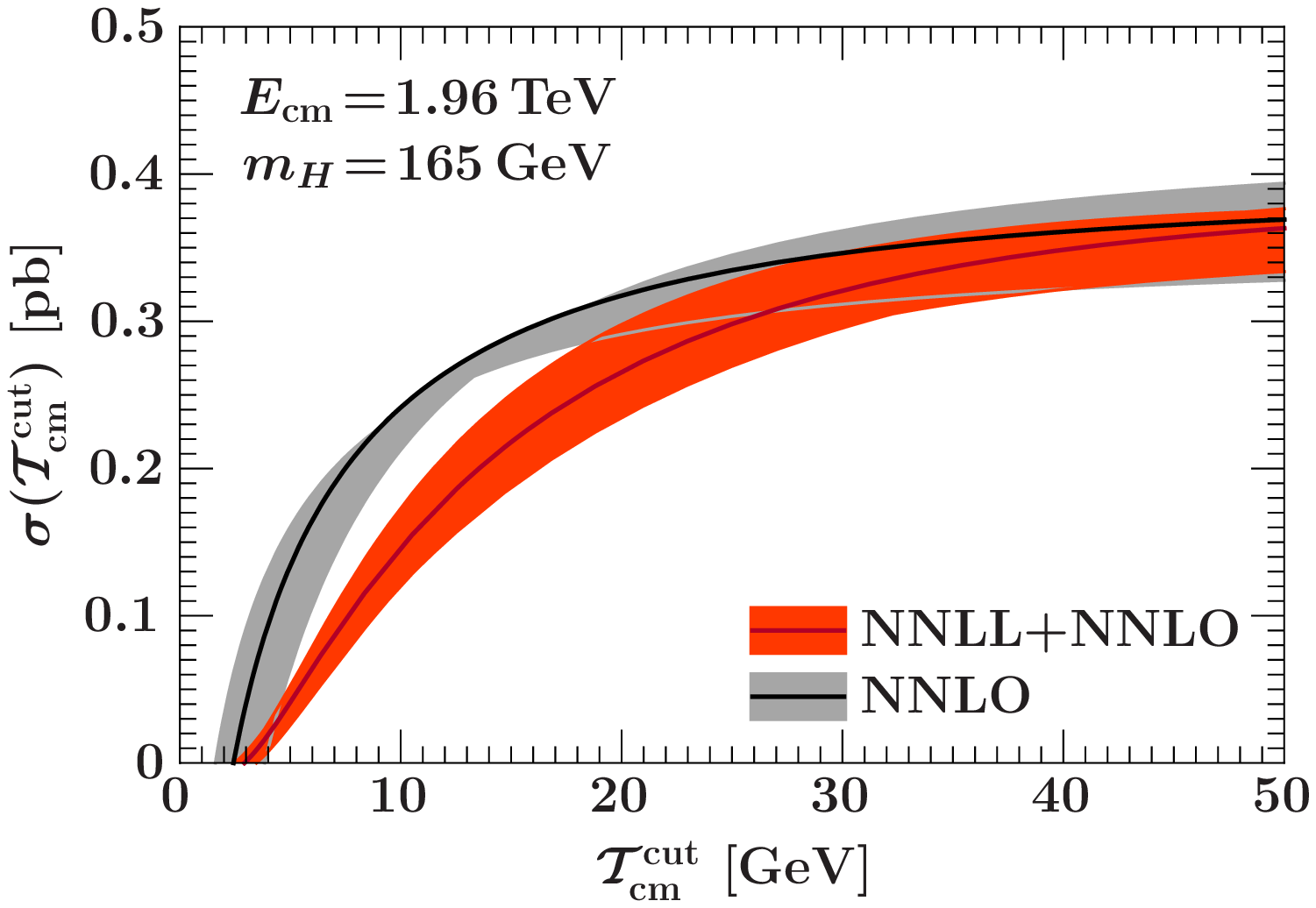}%
\hfill%
\includegraphics[width=0.49\textwidth]{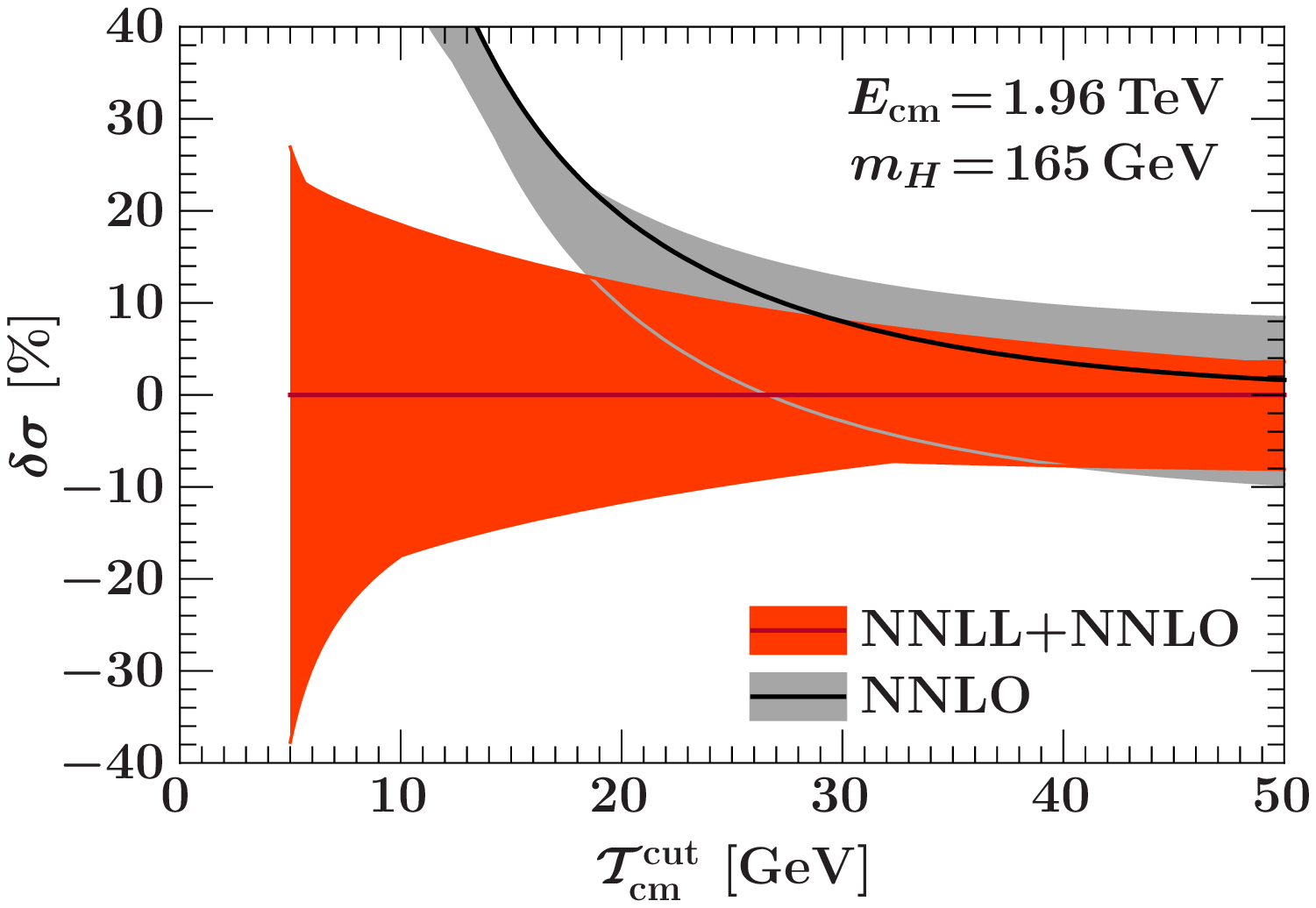}%
\vspace{-0.5ex}
\caption{Comparison of the NNLL+NNLO result for the Higgs production cross section as a function of $\Tcmc$ to the fixed NNLO result for the Tevatron. The bands show the perturbative scale uncertainties. The left plot shows the cumulant cross section. The right plot shows the same information as percent difference relative to the NNLL+NNLO central value.}
\label{fig:compNNLO_Tev}
\end{figure}

\begin{figure}[p]
\includegraphics[width=0.485\textwidth]{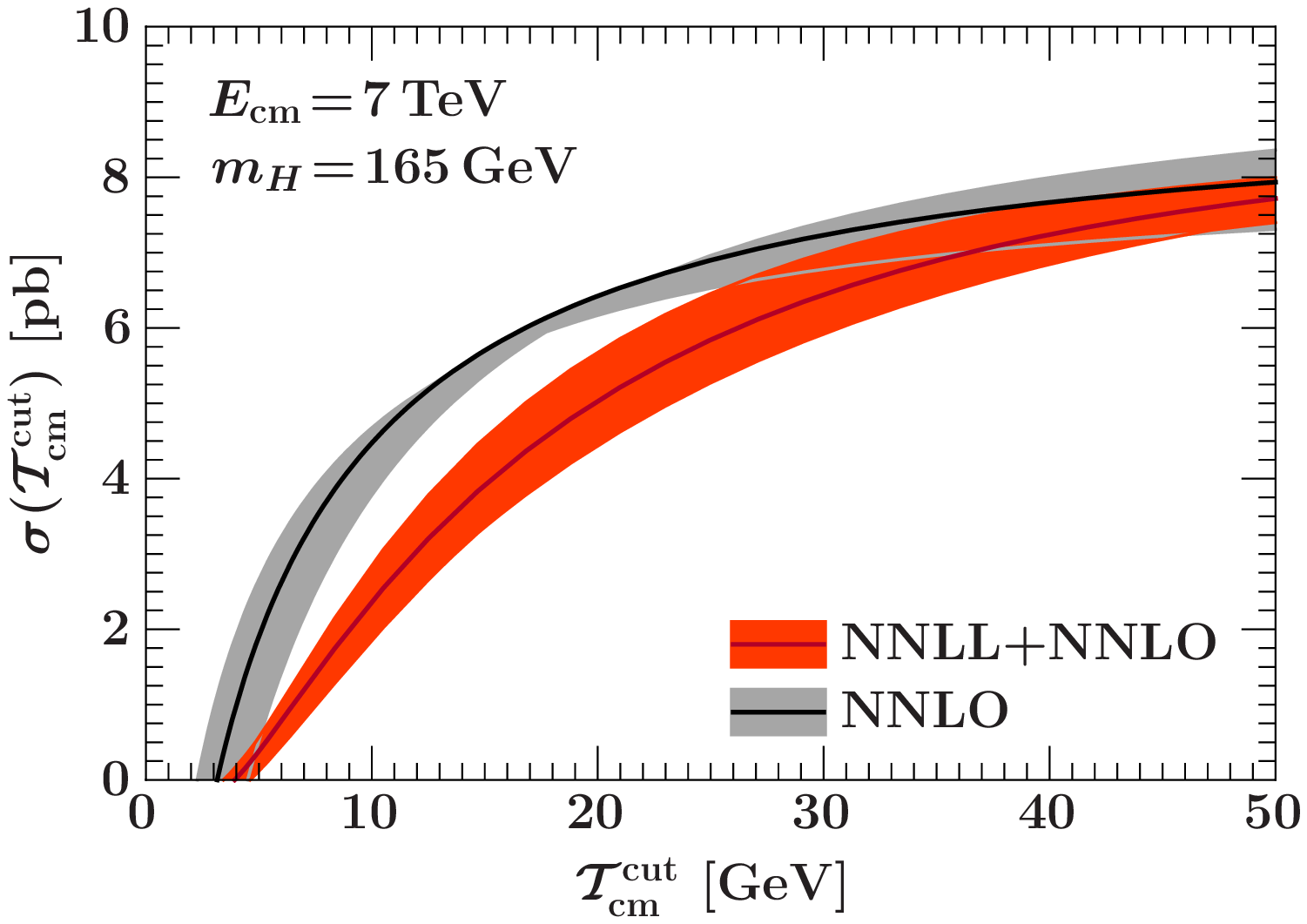}%
\hfill%
\includegraphics[width=0.495\textwidth]{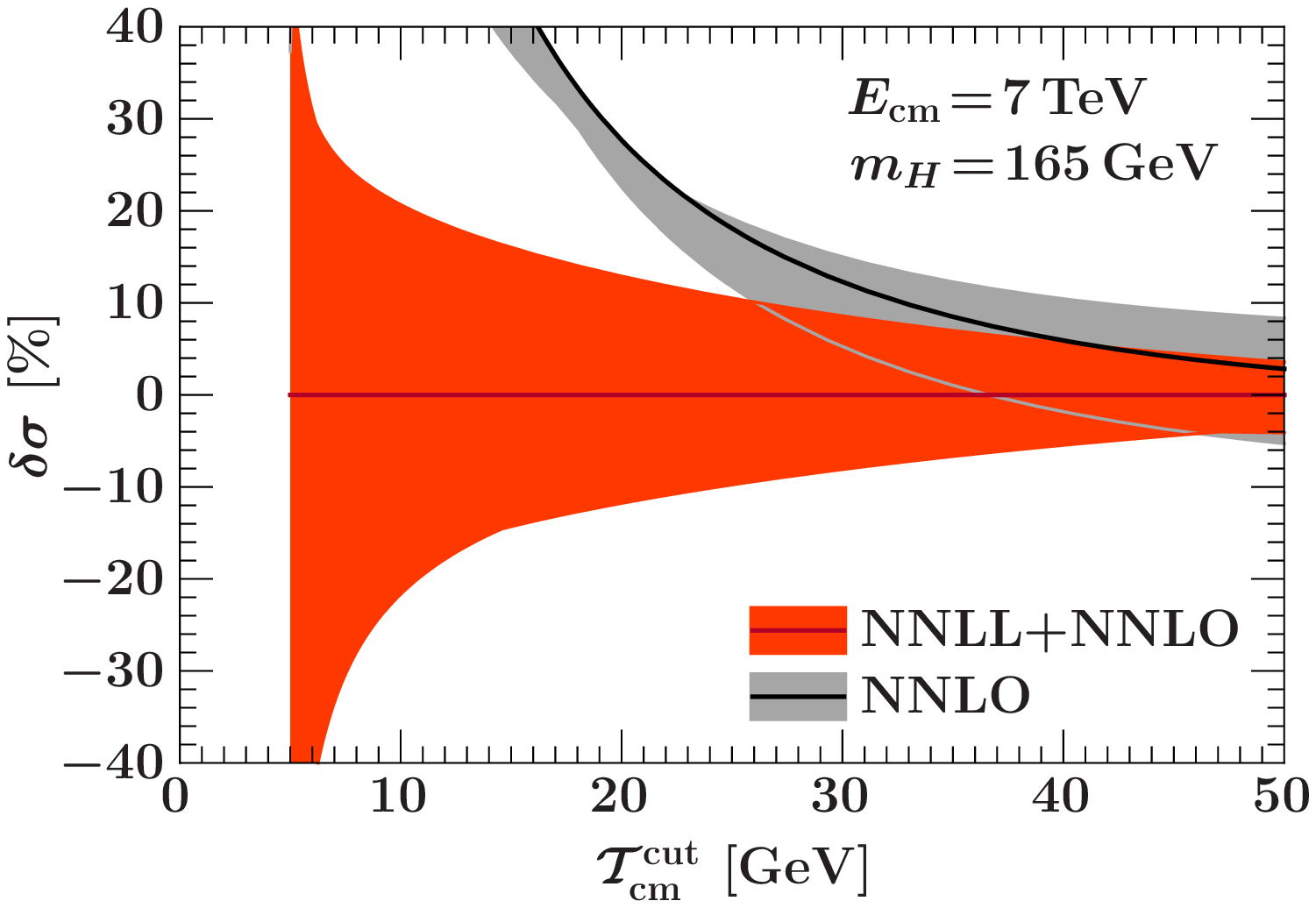}%
\vspace{-0.5ex}
\caption{Same as \fig{compNNLO_Tev} but for the LHC with $\Ecm = 7\TeV$.}
\label{fig:compNNLO_LHC}
\end{figure}

\begin{figure}[p]
\includegraphics[width=0.485\textwidth]{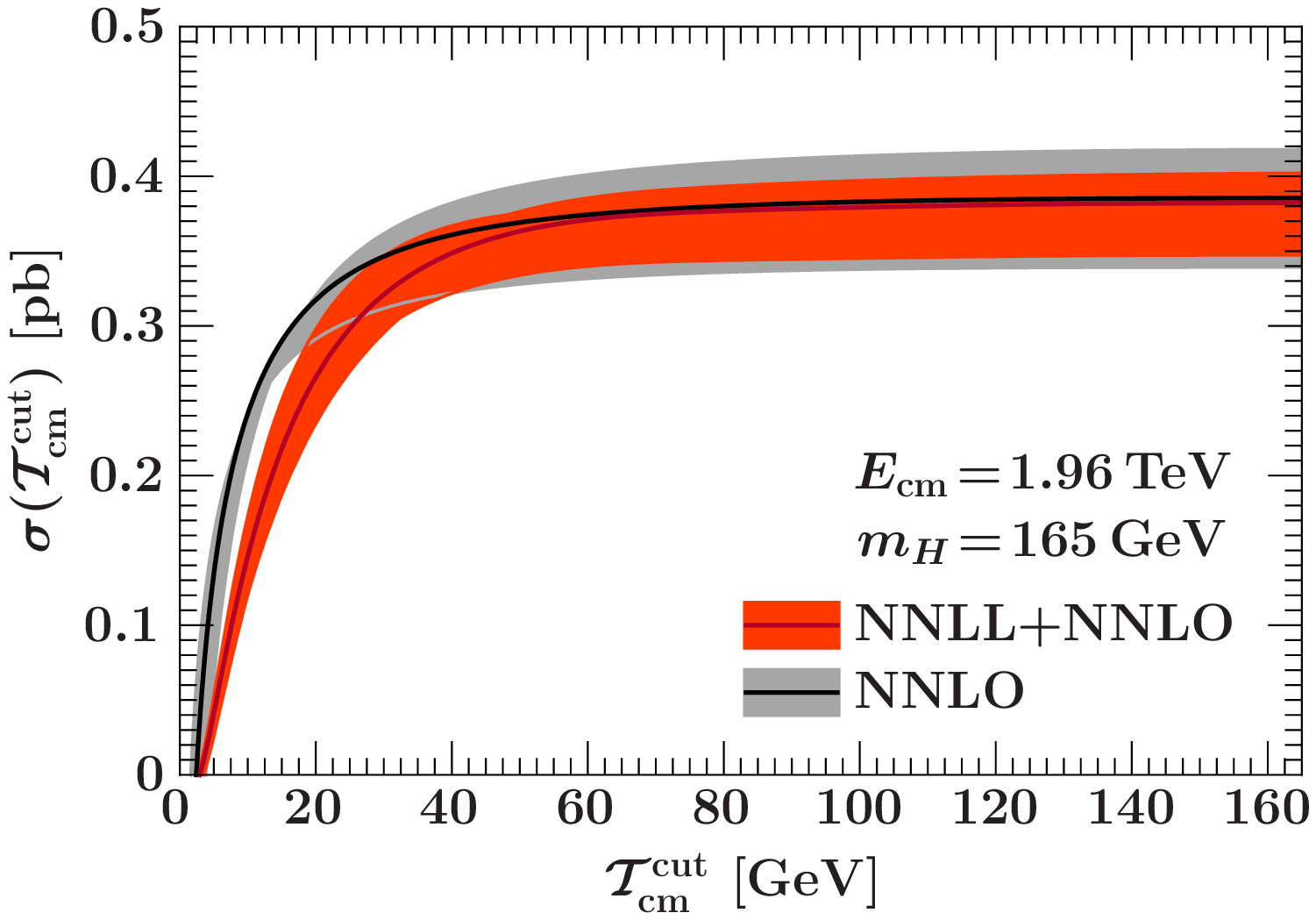}%
\hfill%
\includegraphics[width=0.495\textwidth]{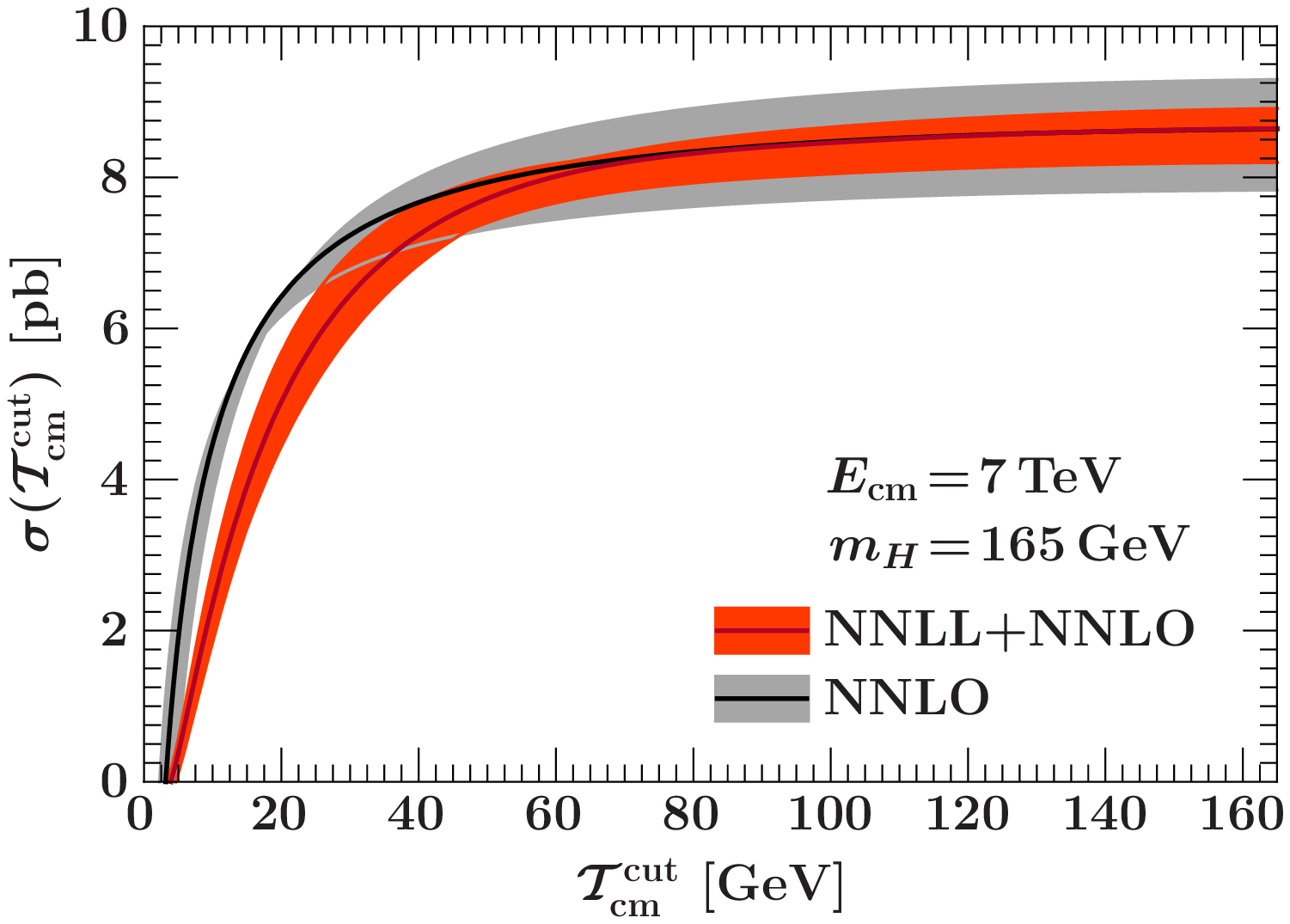}%
\vspace{-0.5ex}
\caption{Same as the left panels of \figs{compNNLO_Tev}{compNNLO_LHC} but plotted up to $\Tcmc = m_H$.}
\label{fig:compNNLO_full}
\end{figure}
\afterpage{\clearpage}

In \figs{compNNLO_Tev}{compNNLO_LHC} we compare the full NNLL+NNLO including
$\pi^2$ summation to the NNLO (using again the default $\mu = m_H/2$ as the
central value) for the Tevatron and LHC, respectively. The left panels show the
cumulant cross section, and the right panels show the same results as the
relative difference in percent to the central NNLL+NNLO curve, which makes it
easy to read off uncertainties. The relative plots are cut off below $\Tcmc =
5\GeV$ because the resummed cross section goes to zero there.  The central value
of the NNLL+NNLO leaves the fixed-order uncertainty band at $\Tcmc \simeq
25\GeV$ at the Tevatron and at $\Tcmc \simeq 35\GeV$ at the LHC.  Hence, for any
lower values the resummation should be taken into account.  At $\Tcmc = 20\GeV$
the central values of the NNLL+NNLO and the NNLO already differ by $20\%$ at the
Tevatron and over $25\%$ at the LHC, which both quickly grows beyond $50\%$
towards $\Tcmc = 10\GeV$.%
\footnote{One might expect that a better agreement could be achieved by using a
  dynamical $\Tcmc$-dependent scale in the fixed-order prediction. We have
  checked that using the intermediate scale $\mu = \mu_B$ in the fixed-order
  result however does not improve its behavior relative to the resummed result,
  but in fact makes it a bit worse.}  This clearly shows that it is important to
resum the higher-order logarithms that are missing in the fixed-order prediction
in order to obtain reliable predictions in the $0$-jet region. This also means
that one cannot expect the scale variation in the fixed-order result to give a
realistic estimate of the size of the missing higher-order terms, and hence it
should not be used to estimate the perturbative scale uncertainty either. In
contrast, since the resummation takes into account the presence of large
logarithms for small $\Tcmc$, we are able to obtain reliable estimates of the
perturbative uncertainties. The perturbative uncertainties at small $\Tcmc$ are
larger than those in the fixed-order result, namely $15-20\%$ at NNLL+NNLO for
$\Tcmc = 15-20\GeV$.  Implications of this for the Higgs search are taken up in
\sec{conclusions}.

\begin{figure}[t]
\includegraphics[width=0.495\textwidth]{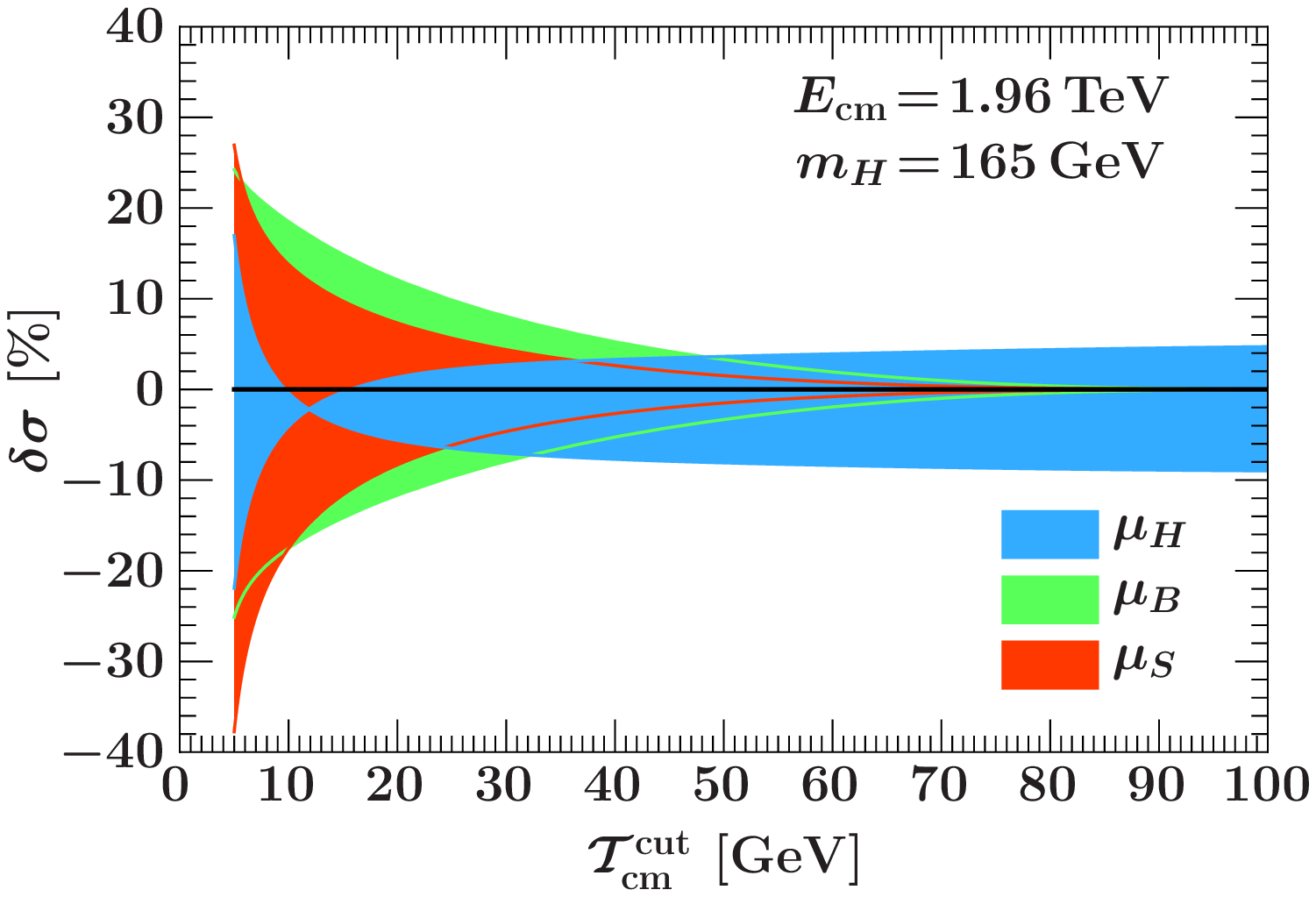}%
\hfill%
\includegraphics[width=0.495\textwidth]{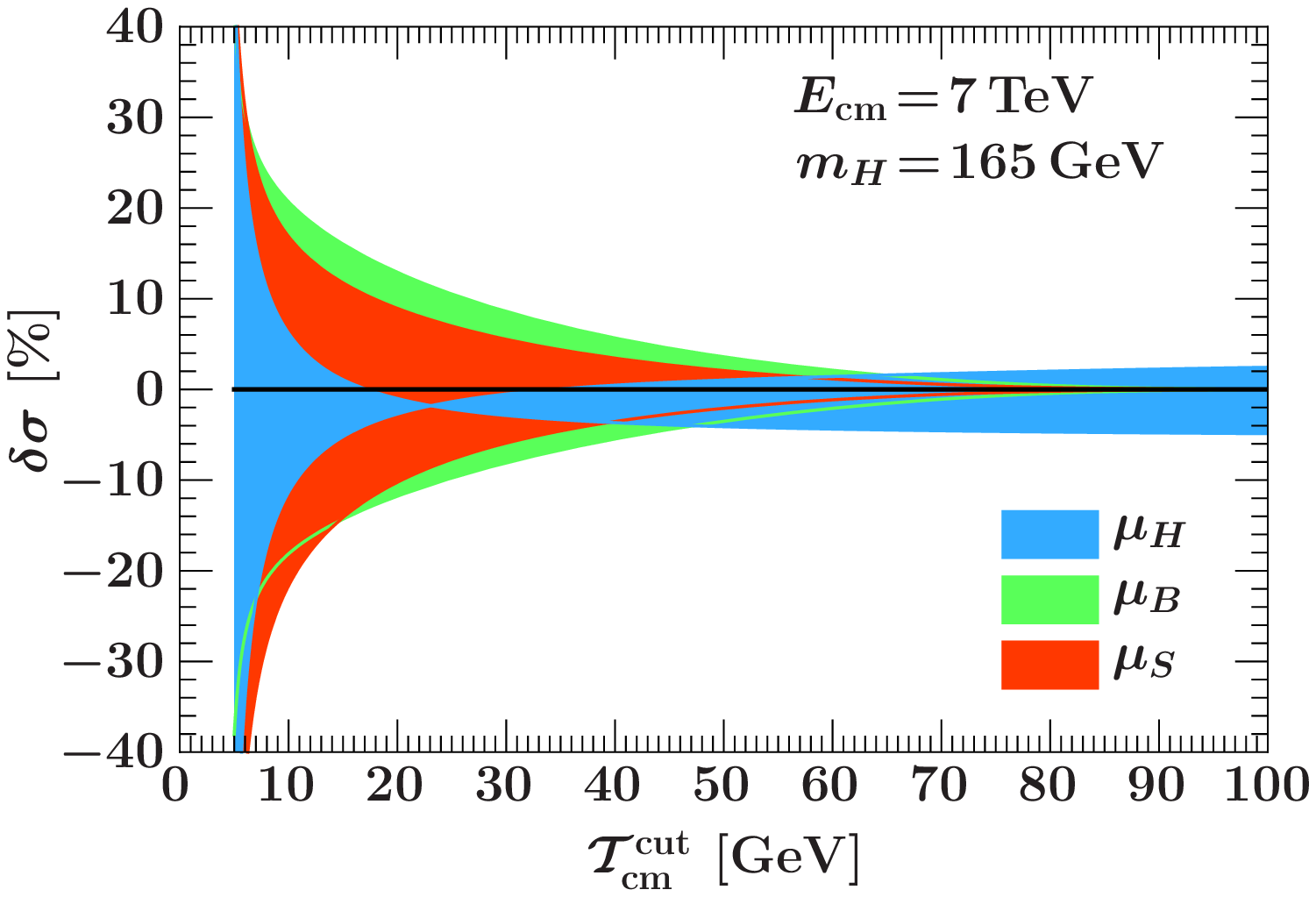}%
\vspace{-0.5ex}
\caption{Contribution to the relative uncertainties in the NNLL+NNLO results shown in \figs{compNNLO_Tev}{compNNLO_LHC} from the individual scale variations. Here, the
$\mu_H$, $\mu_B$, and $\mu_S$ variations correspond to cases a), b), and c) in \eq{scales}
and are shown in \fig{scales}.}
\label{fig:compscales}
\end{figure}

\begin{figure}[t]
\includegraphics[width=0.495\textwidth]{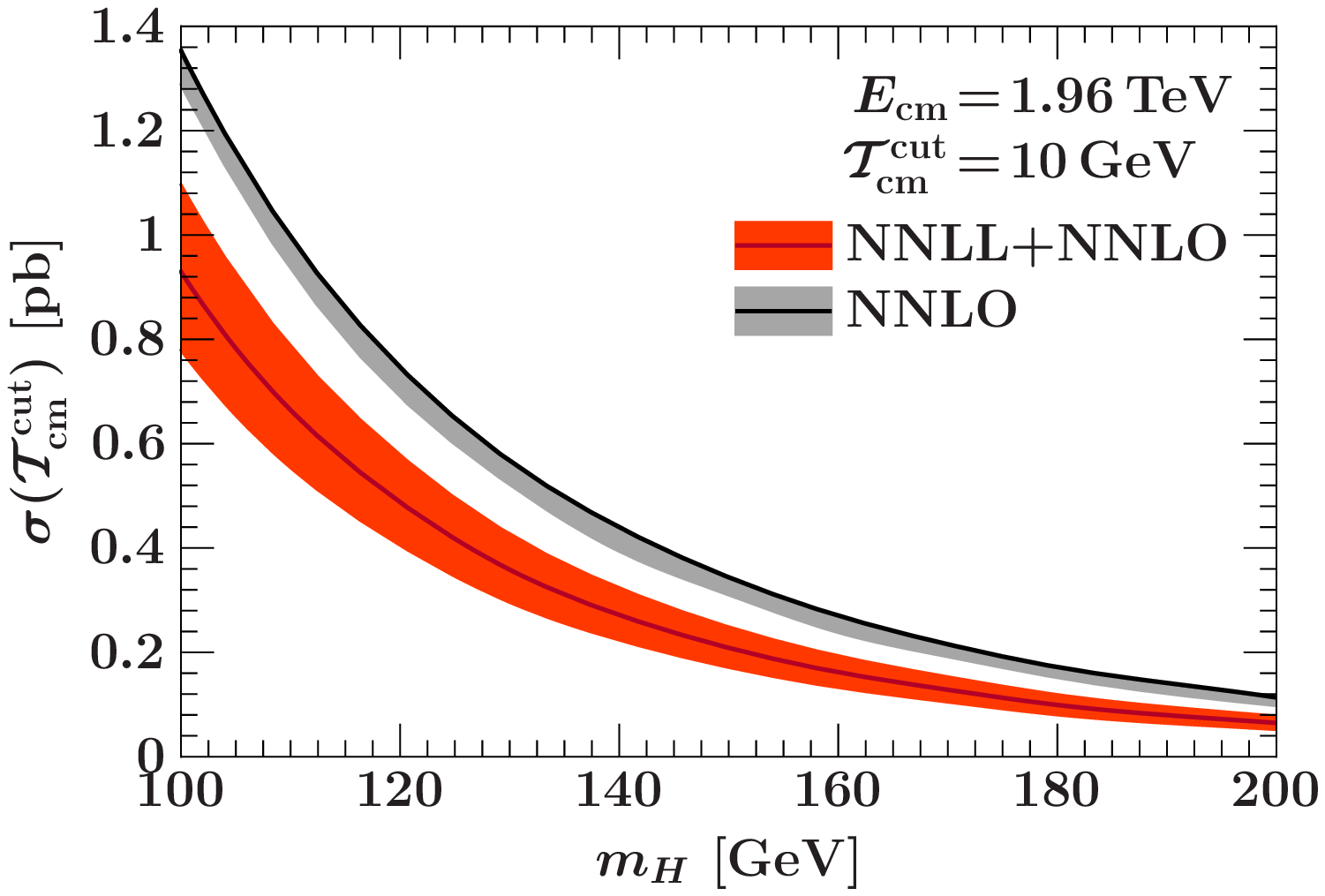}%
\hfill%
\includegraphics[width=0.485\textwidth]{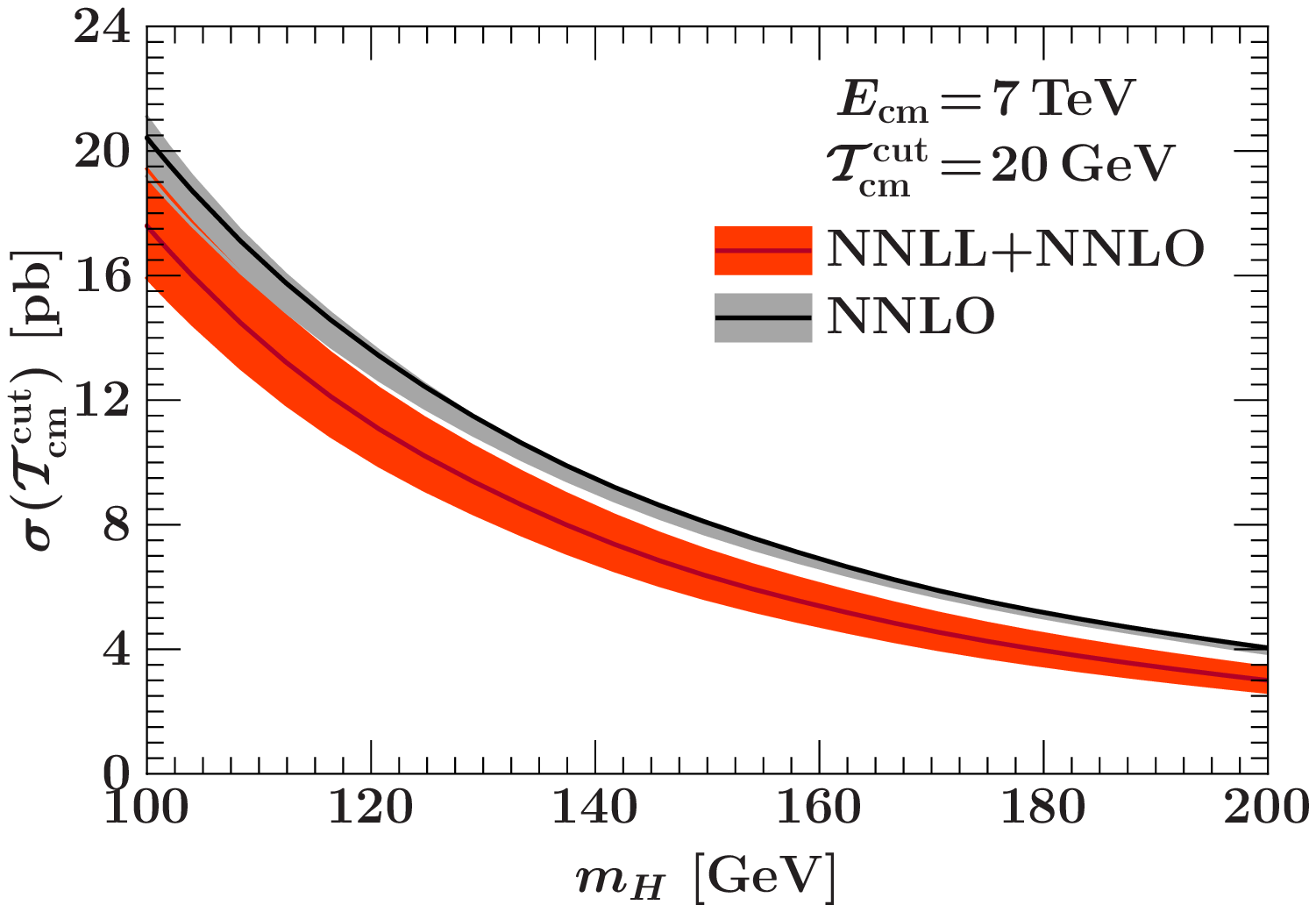}%
\vspace{-0.5ex}
\caption{Comparison of the NNLL+NNLO result to the fixed NNLO result for the Higgs production cross section with a cut on beam thrust as function of $m_H$ at the Tevatron for $\Tcmc=10\GeV$ (left) and the LHC with $\Ecm = 7\TeV$ and $\Tcmc=20\GeV$ (right). The bands show the perturbative scale uncertainties.}
\label{fig:sigmHcm_compNNLO}
\end{figure}

In \fig{compNNLO_full} we show the same comparison of NNLL+NNLO and NNLO
cumulants, but plotted up to $\Tcmc = m_H$.  We can see that the central value
of the resummed result including the $\pi^2$ resummation almost exactly
reproduces the NNLO result which uses $\mu = m_H/2$. However, the $\pi^2$
summation included in the NNLL+NNLO results leads to a reduction of the scale
uncertainties in the inclusive cross section compared to those at NNLO.  For the
uncertainties at the LHC we find $+3\%$ and $-5\%$ and for the Tevatron $+5\%$
and $-9\%$. In \fig{compscales} we show the relative uncertainties at NNLL+NNLO
from the individual scale variations in \eq{scales}. We can see that at small
$\Tcmc$ the uncertainties are dominated by $\mu_B$ and $\mu_S$, i.e.\
variations b) and c) in \eq{scales}. By construction, those variations go to
zero at large $\Tcmc$, where the uncertainties are now completely determined by
variation a) in \eq{scales} and denoted $\mu_H$ in the figure, which is equivalent
to the fixed-order scale variation.

Finally, \fig{sigmHcm_compNNLO} shows a comparison of NNLL+NNLO to NNLO as a
function of the Higgs mass for a fixed value of $\Tcmc$. For smaller Higgs
masses, the logarithms get smaller and the cross section increases, which
reduces the relative differences. This does not change however our overall
conclusions for the importance of the resummation for both the central value
and determining the perturbative uncertainties.

\subsection{Discussion of $K$-Factors}
\label{subsec:Kfactor}

Using our results we can also address the origin of the large NLO and NNLO
$K$-factors that are typically observed for Higgs production.
It is sometimes argued that the origin of
these large $K$-factors for the inclusive cross section are large perturbative
corrections due to hard emissions. This is based on the observation that by
vetoing hard jets, the fixed-order $K$-factors are reduced. As a result the
fixed-order perturbative series in the presence of a jet veto actually appears
to be better behaved than for the inclusive cross section.

Our results show that once the large jet-veto logarithms are properly summed,
the $K$-factor is mostly independent of the jet veto. Hence, it is not caused by
hard real emissions, but rather mostly by hard virtual corrections and to a
lesser extent by collinear and soft virtual and real radiation. In our analysis
this can be examined directly by comparing the convergence of perturbation
theory for the hard, jet, and soft functions. As we have seen in
\subsec{pi2pdf}, and was already observed in ref.~\cite{Ahrens:2008qu}, by
summing the large $\pi^2$ terms in the hard virtual corrections, the $K$-factors
are substantially reduced.

There is a simple reason why the large $K$-factors in fixed-order perturbation
theory are reduced: At NLO, the jet veto reduces the cross section just because
it cuts out available phase space, and since at LO the cross section is not yet
affected by the jet veto, the NLO $K$-factor is reduced accordingly. Essentially,
the large negative phase-space logarithms resulting from the jet veto happen to
cancel the large positive corrections from hard virtual corrections. A similar
effect appears at NNLO where the jet veto reduces the available phase space for
the second jet. Since the hard virtual corrections are independent of the jet
veto one can always choose a particular value for the jet-veto cut such that
they are exactly canceled by the large phase-space logarithms at a given order
in perturbation theory. However, to conclude that the jet veto in general
renders the fixed-order perturbative series better behaved one would have to
know that the same level of cancellation will happen at each order in
perturbation theory.  Since these two types of corrections are a priori not
related there is no reason to believe this will be the case. Instead, to obtain
reliable predictions both types of large corrections should be rendered as
convergent as possible. For the hard virtual corrections the $\pi^2$ summation
improves the convergence, and for the large phase-space logarithms this is
achieved by the resummation carried out here. With the resulting cross section
we can then obtain more realistic estimates for higher-order theoretical
uncertainties as discussed in \subsec{fixedorder}.

\section{Conclusions}
\label{sec:conclusions}

A major effort at the LHC and Tevatron is devoted to the search for the Higgs
boson. In the current Tevatron exclusion limits and the early LHC searches the
$H \to WW \to \ell^+ \nu \ell^- \bar \nu$ channel plays an important role, since
it is the dominant decay channel for Higgs masses above $\sim 130\GeV$. A large
background for this channel are $t\bar t \to W W b \bar b$ events, which must be
eliminated by a veto on central jets, so the resulting measurement is $pp\to H +
0$ jets. Such a jet veto causes large double logarithms in the cross section,
which need to be summed to all orders in perturbation theory in order to obtain
reliable theoretical predictions and uncertainties for the $H + 0$ jet
production cross section.

In this paper we have studied Higgs production from gluon fusion, using the
inclusive event shape beam thrust, $\Tau_\cm$, to implement a central jet veto.
As beam thrust characterizes the event as a whole and does not require a jet
algorithm, it is well suited to analytic calculations. This allows us to resum
the jet-veto logarithms to NNLL order, based on the factorization theorem for
the beam thrust cross section derived in ref.~\cite{Stewart:2009yx}. In our
analysis we also include the full set of NNLO corrections, such that our final
result at NNLL+NNLO provides the best possible theoretical prediction at any
value of beam thrust.

Our main results are presented in figs.~\ref{fig:compNNLO_Tev} and
\ref{fig:compNNLO_LHC}. We find that in the $0$-jet region at small beam thrust,
the resummation of jet-veto logarithms is crucial, and our central value at
NNLL+NNLO for the cross section with a cut on beam thrust, $\Tcm \leq \Tcmc$,
differs significantly from the fixed-order prediction at NNLO.  We also find
substantially larger perturbative scale uncertainties arising from the jet veto
compared to those at NNLO. Since fixed-order perturbation theory is not reliable
in the presence of large jet-veto logarithms, one also cannot expect its scale
variation to yield a reliable estimate of perturbative uncertainties due to
neglecting higher-order corrections.

At present, the jet-veto logarithms are taken into account in the experimental
analyses using the leading-logarithmic resummation provided by parton-shower
Monte Carlo programs, usually supplemented with some reweighting procedure to
reproduce the total NNLO cross section. While this might yield a reasonable
central value that takes into account the dominant effect of the logarithmic
resummation, the uncertainties in the so predicted $0$-jet cross section cannot
be taken as those of the inclusive NNLO cross section. They could at best be
equal to those in our NNLL+NNLO results. In fact, they are probably larger than
that, since we include the resummation at two orders higher than the LL
resummation of parton showers. As we have seen in \subsec{convergence}, already
one order lower, at NLL$'$+NLO, the perturbative uncertainties are much larger
than those in our NNLL+NNLO results.

The conventional method to veto central jets is to use a jet algorithm and
require $p_T^\jet < p_T^\cut$ for all jets in the event. As we saw in \sec{Intro},
$p_T^\cut$ can be related to $\Tcmc$ by associating $\Tcmc/m_H \simeq
(p_T^\cut/m_H)^{\sqrt{2}}$ or $\Tcmc = p_T^\cut$, where the former works well at
NNLO while the latter is favored by \Pythia 8. Hence, we can use the perturbative uncertainties in
our results based on $\Tcmc$ as a benchmark for the perturbative uncertainties
from large logarithms relevant for $p_T^\cut$, on which the current experimental analyses are based.
For example, the perturbative uncertainties for typical values $p_T^\cut \simeq 20-30\GeV$
at the LHC can be as large as those for $\Tcmc\simeq 10-15\GeV$,
which are $15-20\%$ at NNLL+NNLO (for $m_H = 165\GeV$ and $\Ecm = 7\TeV$).
In the current Tevatron analyses, the
perturbative scale uncertainty in the $0$-jet cross section for $p_T^\cut =
15\GeV$ is taken as $7\%$ from the fixed-order analysis in
ref.~\cite{Anastasiou:2009bt}. In contrast, the perturbative uncertainties at
NNLL+NNLO for values $\Tcmc \leq 15 \GeV$ are significantly larger, e.g. about $20\%$
for $\Tcmc = 10\GeV$ and $m_H = 165\GeV$. In light of this, the current Tevatron
exclusion limits should be reconsidered with an increased theory uncertainty for
the $0$-jet Higgs cross section. We note that our conclusions about theoretical
uncertainties are based on a systematic study of the jet veto, and are therefore
independent of the arguments in ref.~\cite{Baglio:2010um} proposing the use of a
larger range for the fixed-order scale variation.

To implement our NNLL+NNLO results in the experimental searches, one could
reweight the partonic beam-thrust spectrum obtained from Monte Carlo to our
results.  This would incorporate both the higher-order resummation of large
logarithms in the $0$-jet region and at the same time the full NNLO result for
the total cross section. This also allows a consistent treatment of the
perturbative uncertainties in both regions. To illustrate this, consider
dividing the total inclusive cross section, $\sigma_\mathrm{total}$, into a
$0$-jet bin, $\sigma_0 \equiv \sigma(\Tcmc)$, and a remaining $(\geq 1)$-jet
bin, $\sigma_{\geq 1} \equiv \sigma_\mathrm{total} - \sigma(\Tcmc)$. This
separation into exclusive jet bins causes large logarithms of $\Tcmc$ in both
$\sigma_0$ and $\sigma_{\geq 1}$ which cancel in their sum. This implies that
the uncertainties due to $\Tcmc$ in both bins are anti-correlated.  In
particular, if one wants to consider the theory uncertainties in $\sigma_0$ and
$\sigma_{\geq 1}$ simultaneously in a Gaussian fashion, one has to consider the
full theory covariance matrix
\begin{equation} \label{eq:C}
C = \begin{pmatrix}
           \Delta_0^2 & \Delta_0\, \Delta_{\geq 1}\,\rho_{0,\geq 1} \\
           \Delta_0\, \Delta_{\geq 1}\,\rho_{0,\geq 1} & \Delta_{\geq1}^2
         \end{pmatrix}
\approx \begin{pmatrix}
          \Delta_0^2 & -\Delta_0^2 \\
          -\Delta_0^2 & \quad\Delta_0^2 + \Delta_\mathrm{total}^2
        \end{pmatrix}
\,,\end{equation}
where $\Delta_0$ and $\Delta_{\geq 1}$ are the theory uncertainties of
$\sigma_0$ and $\sigma_{\geq 1}$, and $\rho_{0,\geq1}$ is their correlation
coefficient.  In the second step above we used the approximation that the
uncertainties in $\sigma_0$ and $\sigma_\mathrm{total}$ are uncorrelated and are
added in quadrature in $\sigma_{\geq 1}$. From our results in \fig{compscales}
we can see that this is reasonable since the uncertainties in $\sigma_0$ are
dominated by the lower scales $\mu_B$ and $\mu_S$, while those in
$\sigma_\mathrm{total}$ are determined by $\mu_H$ (which in this case is
combined in quadrature with the others rather then by taking the envelope). The
uncertainty squared for the total cross section, $\sigma_0 + \sigma_{\ge 1}$, is
given by the sum of all entries in the matrix $C$ in \eq{C}. Due to the
anti-correlation the uncertainties for the individual jet bins can be larger than
that for the total cross section.  Our numerical results for the theory
uncertainties for the 0-jet bin directly give $\Delta_0$.  The full correlation
can be taken into account by reweighting the Monte Carlo to both the central
value curve as well as the results obtained from the individual scale
variations.

It would also be useful to have a benchmark theoretical uncertainty that is
desired for the experimental searches, since with further effort our NNLL+NNLO
results can be extended to NNLL$^\prime$+NNLO or N$^3$LL+NNLO, which has the
potential to reduce the uncertainty in the resummed perturbation theory.  Given
that the theoretical predictions and their uncertainties are very sensitive to
the jet veto, it would also be useful to implement the jet veto in the
experimental analyses directly in terms of beam thrust, for which resummed
theory predictions are available. In addition, a benchmark experimental study of
the beam-thrust spectrum can be made with Drell-Yan pairs, as advocated in
ref.~\cite{Stewart:2010pd}.

In this paper, we have restricted ourselves to studying the case of $gg\to H +
0$ jets. The same methods can be used to calculate the dominant irreducible
background from direct $WW$ production, i.e., the process $pp\to WW + 0$ jets
using beam thrust for the jet veto. The generalization of beam thrust to
processes with $N$ signal jets is provided by the event shape $N$-jettiness
introduced in ref.~\cite{Stewart:2010tn}. It can be used in an analogous fashion
to study the exclusive $H+1$ jet and $H+2$ jet cross sections, the latter being
relevant for Higgs production from vector-boson fusion.

\begin{acknowledgments}
  We thank the ATLAS group at the University of Pennsylvania and Elliot Lipeles
  for stimulating discussions.  We also thank Lance Dixon, Steve Ellis, Zoltan
  Ligeti, Aneesh Manohar, Kerstin Tackmann, Jesse Thaler, Joey Huston,
  Fabian St\"ockli and especially Kirill
  Melnikov and Frank Petriello for useful discussions and comments on the
  manuscript.  For hospitality during part of this work we thank the Max-Planck
  Institute for Physics (Werner-Heisenberg Institute), the CERN theory group,
  and the Center for the Fundamental Laws of Nature at Harvard. This work was
  supported in part by the Office of Nuclear Physics of the U.S.\ Department of
  Energy under the Contract DE-FG02-94ER40818, by the Department of Energy under
  the grant DE-SC003916, and by a Friedrich Wilhelm Bessel award from the
  Alexander von Humboldt foundation.
\end{acknowledgments}

\appendix

\section{NLO Calculation of the Gluon Beam Function}
\label{app:calculation}

In this section we compute the NLO gluon beam function. We first recall its definition, renormalization, and operator product expansion in terms of standard PDFs and Wilson coefficients $\cI_{ij}$, as discussed in detail in ref.~\cite{Stewart:2010qs}. For a discussion of the basic SCET ingredients relevant to our context we refer to refs.~\cite{Stewart:2009yx, Stewart:2010qs}. We then give the details of the one-loop matching calculation, which provides a check on the one-loop anomalous dimension of the gluon beam function and yields the $\cI_{ij}$ at NLO.

\subsection{Definition and General Results}
\label{app:beamdef}

As usual, we use lightlike vectors $n^\mu = (1,\vec{n})$ and $\bn^\mu=(1,-\vec{n})$, satisfying $n^2 = \bn^2 = 0$, $n \cdot \bn=2$, to write a vector $p^\mu$ in light-cone coordinates
\begin{equation}
  p^\mu = p^+ \frac{\bn^\mu}{2} + p^- \frac{n^\mu}{2} + p_\perp^\mu
  \qquad\text{with}\qquad
  p^+ = n \cdot p
  \,, \qquad
  p^- = \bn \cdot p
  \,.
\end{equation}

The bare gluon beam function operator is defined as
\begin{align} \label{eq:op_def}
\op_g^\bare(\w b^+, \w)
&= -\theta(\w) \int \! \frac{\df y^-}{4\pi}\, e^{\img b^+ y^-/2}\, e^{-\img\hp^+ y^-/2}\,
 \cB_{n\perp\mu}^c \Bigl(y^- \frac{n}{2}\Bigr) \bigl[\delta(\w - \bnP_n) \cB_{n\perp}^{\mu c}(0) \bigr]
\nn\\
&= -\w\,\theta(\w) \, \cB_{n\perp\mu}^c(0)\, \delta(\w b^+ - \w\hp^+) \bigl[\delta(\w - \bnP_n) \cB_{n\perp}^{\mu c}(0) \bigr]
\,.\end{align}
The SCET gluon field $\cB_{n\perp}^\mu$ describes $n$-collinear gluons and includes zero-bin subtractions~\cite{Manohar:2006nz}. It contains collinear Wilson lines $W_n$ to render it invariant under collinear gauge transformations. It is the field after the BPS field redefinition~\cite{Bauer:2001yt} to decouple soft gluons and is thus invariant under soft gauge transformations. Hence, $\op_g^\bare$ is fully gauge invariant. The gluon beam function is defined by the proton matrix element of the corresponding renormalized operator $\op_g(t, \w, \mu)$,
\begin{equation} \label{eq:B_def}
B_g(t, x = \w/P^-,\mu) = \Mae{p_n(P^-)}{\op_g(t,\w,\mu)}{p_n(P^-)}
\,.\end{equation}
The proton states $\ket{p_n(P^-)}$ have lightlike momentum $P^\mu = P^- n^\mu/2$, i.e.\ $\vec{n}$ is chosen in the direction of the proton, and the matrix elements are always implicitly averaged over the proton spin.

In \eq{op_def}, $\hp^+$ is the momentum operator of the residual
plus momentum and acts on everything to its right. The additional
phase in the position-space operator is included to allow us to
write the $b^+$ dependence in terms of $\delta(\w b^+ - \w\hp^+)$,
as in the second line of \eq{op_def}. Hence, $b^+$ measures the
total plus momentum of real initial-state radiation, i.e.\ of any
intermediate state inserted between the fields. The label operator
$\delta(\w - \bnP_n)$ in \eq{op_def} only acts inside the square
brackets and forces the total sum of the minus labels of all
fields in $\cB_{n\perp}$ to be equal to $\w$. In \eq{B_def} it
sets the fraction of the proton's light-cone momentum carried into
the hard collision to $x = \w/P^->0$. At the time of the
collision, the annihilated collinear gluon propagates in a jet of
initial-state radiation, and by momentum conservation the small
plus momentum of the gluon is $-b^+ < 0$. Hence, it is spacelike and $-t = \w
(-b^+) < 0$ measures its transverse virtuality. The operator
$\op_g(t, \w)$ is RPI-III invariant, because the transformation of
the overall $\w$ compensates that of $\delta(\w - \bnP_n)$.
Therefore, the gluon beam function only depends on the RPI-III
invariant variables $x$ and $t$.

The RGE of the beam function follows from the renormalization of $\op_g(t, \w, \mu)$. To all orders in perturbation theory, it takes the form
\begin{equation} \label{eq:Bg_RGE}
\mu\,\frac{\df}{\df\mu} B_g(t,x, \mu) = \int\!\df t'\,
  \gamma^g_B(t - t', \mu)\, B_g(t', x, \mu)
\,,\end{equation}
with the anomalous dimension
\begin{equation} \label{eq:gaB_gen}
\gamma_B^g(t, \mu)
= -2 \Gamma^g_{\cusp}[\alpha_s(\mu)] \,\frac{1}{\mu^2}\cL_0\Bigl(\frac{t}{\mu^2}\Bigr) + \gamma_B^g[\alpha_s(\mu)]\,\delta(t)
\,.\end{equation}
Here, $\cL_0(x) = [\theta(x)/x]_+$ is defined in \eq{plusdef}, $\Gamma_\cusp^g(\alpha_s)$ is the gluon cusp anomalous dimension, and $\gamma_B^g(\alpha_s)$ denotes the non-cusp part, whose coefficients up to three loops are given in \app{rge}. The RGE of the beam function is identical to that of the jet function. It sums double logarithms of $t/\mu^2$ and unlike the PDF evolution does not change the $x$ dependence and does not mix different parton types. The solution of \eq{Bg_RGE} was already given in \eq{Bgrun}.

The standard PDFs are defined in SCET by the matrix elements~\cite{Bauer:2002nz}
\begin{equation} \label{eq:f_def_SCET}
f_i(\xi = \w'/P^-,\mu) = \Mae{p_n(P^-)}{\oq_i(\w',\mu)}{p_n(P^-)}
\,,\end{equation}
where $\oq_i(\w', \mu)$ are the \MSbar renormalized operators corresponding to the bare operators
\begin{align} \label{eq:oq_def}
\oq^\bare_g(\w')
& = -\w'\theta(\w')\, \cB_{n\perp\mu}^c(0) \bigl[\delta(\w' - \bnP_n) \cB_{n\perp}^{\mu c}(0) \bigr]
\,, \nn \\
\oq^\bare_q(\w')
&= \theta(\w')\, \bar{\chi}_n(0) \frac{\bnslash}{2} \bigl[\delta(\w' - \bnP_n) \chi_n(0)\bigr]
\,, \nn \\
\oq^\bare_{\bar q}(\w')
&= \theta(\w')\, \tr \Bigl\{\frac{\bnslash}{2} \chi_n(0) \bigl[\delta(\w' - \bnP_n) \bar\chi_n(0)\bigr] \Bigr\}
\,.\end{align}
The $\theta(\w')$ here separates the quark and antiquark PDFs and is included to
keep analogous definitions for the PDFs and beam functions.

Compared to \eq{oq_def}, the fields in the beam-function operator in \eq{op_def} are separated along the $n$ light-cone with
large separation $y^-$ corresponding to the small momentum $b^+ \ll \w$.
Hence, by performing an operator product expansion about the limit $y^-\!\to 0$ we can expand the beam-function operator $\op_g(t, \w, \mu)$ in terms of a sum over $\oq_i(\w', \mu)$,
\begin{equation} \label{eq:op_OPE}
\op_g(t, \w, \mu) = \sum_j \int\! \frac{\df \w'}{\w'}\,
\cI_{gj}\Bigl(t,\frac{\w}{\w'},\mu \Bigr) \oq_j(\w',\mu) + \ORd{\frac{y^-}{\w}}
\,,\end{equation}
where the functional form of the matching coefficients $\cI_{ij}(t, z, \mu)$ is again determined by RPI-III invariance.
The proton matrix element of \eq{op_OPE} yields the OPE for the gluon beam function quoted in \eq{Bg_OPE},
\begin{equation} \label{eq:B_OPE}
B_g(t,x,\mu)
= \sum_j \int \! \frac{\df \xi}{\xi}\, \cI_{gj}\Bigl(t,\frac{x}{\xi},\mu \Bigr) f_j(\xi,\mu)
  \biggl[1 + \ORd{\frac{\lqcd^2}{t}}\biggr]
\,.\end{equation}
The power corrections scale like $(\lqcd^2/\w)/b^+$, where $t = \w b^+$ and $\lqcd^2/\w$ is the typical plus momentum of the partons in the proton, and involve higher-twist proton structure functions. For $t\sim\lqcd^2$ the OPE would require an infinite set of higher-twist proton structure functions, which means the beam functions essentially become nonperturbative $b^+$-dependent PDFs. On the other hand, for $t \gg \lqcd^2$, we can compute the matching coefficients $\cI_{ij}$ in \eq{B_OPE} in perturbation theory at the beam scale $\mu_B^2 \simeq t$.

The matching calculation is carried out as usual by taking partonic quark and gluon matrix elements of the operators on both sides of \eq{op_OPE}, which we denote as
\begin{align}
B_{g/j}(t, z = \w/p^-, \mu)
&= \Mae{j_n(p^-)}{\op_g(t, \w, \mu)}{j_n(p^-)}
\,, \nn \\
f_{i/j}(z = \w/p^-, \mu)
&= \Mae{j_n(p^-)}{\oq_i(\w, \mu)}{j_n(p^-)}
\,,
\end{align}
and analogously for the bare quantities. Here, $\ket{j_n(p^-)}$ is an $n$-collinear gluon or quark state, $\ket{g_n(p^-)}$ or $\ket{q_n(p^-)}$, with momentum $p^\mu = p^- n^\mu/2$ and $p^- > 0$, and we average the matrix elements over both color and spin. We denote the partonic momentum fractions by $z = \w/p^-$ to distinguish them from the hadronic $x$ and $\xi$. The operators are normalized such that at tree level we simply have
\begin{equation} \label{eq:tree}
B^\zero_{g/j}(t, z, \mu) = \delta_{gj}\, \delta(t)\, \delta(1 - z)
\,,\qquad
f^\zero_{i/j}(z, \mu) = \delta_{ij}\, \delta(1 - z)
\,,\end{equation}
which yields
\begin{equation}
\cI^\zero_{gj}(t, z, \mu) = \delta_{gj}\, \delta(t)\, \delta(1 - z)
\,,\qquad
B^\zero_g(t, x, \mu) = \delta(t)\, f_g(x, \mu)
\,.\end{equation}
Beyond tree level $x$ and $\xi$ will be different, because the momentum fraction $\xi$ provided by the PDFs is modified by the collinear radiation encoded in the $\cI_{gj}(t, z, \mu)$.

Expanding \eq{op_OPE} to NLO and using the above tree-level results we find
\begin{align} \label{eq:B_fact_1loop}
  B_{g/g}^\one(t,z,\mu)
  &= \cI_{gg}^\one(t,z,\mu) + \delta(t) f_{g/g}^\one(z,\mu)
\,, \nn \\
  B_{g/q}^\one(t,z,\mu)\,
  &= \cI_{gq}^\one(t,z,\mu) + \delta(t) f_{g/q}^\one(z,\mu)
\,.\end{align}
Hence, to compute the one-loop coefficients $\cI_{gj}^\one(t, z, \mu)$
we have to compute the one-loop gluon and quark matrix
elements of the gluon operators $\op_g(t, \w, \mu)$ and
$\oq_g(\w, \mu)$. (The quark and antiquark coefficients are identical, $\cI_{gq} = \cI_{g\bar q}$,
so we do not need to consider separate matrix elements with external antiquarks.) The corresponding one-loop
calculation for the quark operators to determine the matching
coefficients $\cI_{qj}$ for the quark beam function is given in
detail in ref.~\cite{Stewart:2010qs}. There, the calculation is
performed in two different ways, first using different regulators
for IR and UV, and second using dimensional regularization for
both IR and UV. In the first case, the UV and IR divergences can
be separated allowing one to obtain the UV renormalization of the
operators and at the same time check that the IR divergences in
the beam function match those in the PDFs, so the matching
coefficients are IR finite. In the second case, the calculation is
much quicker, because most diagrams are scaleless and vanish, but
it does not allow one to distinguish the IR and UV divergences.
This means one can only check that the IR divergences cancel in
the matching if the renormalization of both beam function and PDF
are already known. Alternatively, one can assume that the IR
divergences cancel and use the known renormalization of the PDFs
to obtain the renormalization of the beam function. In our case,
the renormalization of the gluon beam function is already known
from the general analysis of ref.~\cite{Stewart:2010qs}, so we
only perform the second, simpler, calculation, regulating both UV
and IR in dimensional regularization using the \MSbar scheme.

\subsection{The Gluon PDF at One Loop}
\label{app:gluon_PDF}

As explained in ref.~\cite{Stewart:2010qs}, the definition in \eq{f_def_SCET} is equivalent to the standard definition of the PDFs in QCD.
In particular, the collinear fields in \eq{oq_def} do not require zero-bin subtractions, because the soft region does not contribute to the PDFs. Therefore, we can use the standard \MSbar renormalization of the PDFs,
\begin{equation} \label{eq:f_ren}
f_i^\bare(\xi) = \sum_j \int\! \frac{\df \xi'}{\xi'}\, Z^f_{ij}\Bigl(\frac{\xi}{\xi'},\mu\Bigr) f_j(\xi',\mu)
\,,\end{equation}
where $j = \{q, \bar{q}, g\}$, and the entries in the matrix $Z^f_{ij}(z, \mu)$ are a series of $1/\eps$ poles with coefficients in terms of the renormalized \MSbar coupling $\alpha_s(\mu)$. At one loop with $d = 4 - 2\eps$, we have for the standard gluon PDF~\cite{Altarelli:1977zs},
\begin{align} \label{eq:f_Z_1loop}
  Z^f_{gg}(z,\mu) &= \delta(1-z) + \frac{1}{\eps} \frac{\alpha_s(\mu)}{2\pi}\,
  \theta(z)\, \Bigl[C_A P_{gg}(z) + \frac{1}{2} \beta_0\, \delta(1-z)\Bigr]
  \,, \nn \\
  Z^f_{gq}(z,\mu) &= \frac{1}{\eps} \frac{\alpha_s(\mu) C_F}{2\pi}\,
  \theta(z)\, P_{gq}(z)
  \,,
\end{align}
where $\beta_0 = (11 C_A - 4 n_f T_f)/3$, is the lowest order coefficient of the QCD $\beta$ function, and the $g\to gg$ and $q\to gq$ splitting functions are
\begin{align} \label{eq:split}
P_{gg}(z)
&= 2 \cL_0(1-z)z + 2\theta(1-z)\Bigl[\frac{1-z}{z} +  z(1-z)\Bigr]
\nn\\
&= 2\theta(1-z)\Bigl[\frac{z}{(1-z)_+} + \frac{1-z}{z} +  z(1-z)\Bigr]
\,,\nn\\
P_{gq}(z) &= \theta(1-z)\, \frac{1+(1-z)^2}{z}
\,,\end{align}
with $\cL_0(1-z) = [\theta(1-z)/(1-z)]_+$ defined in \eq{plusdef}.

Expanding \eq{f_ren} to one loop, we get
\begin{align} \label{eq:f_ren_1loop}
f_{g/j}^{\bare\, \one}(z) = Z^{f\one}_{gj}(z,\mu) + f_{g/j}^\one(z,\mu)
\,.\end{align}
In pure dimensional regularization, the only Lorentz invariant quantity $f_{g/j}^\bare(z)$ can depend on is $z$. Since $z$  is dimensionless, all loop diagrams contributing to the PDF calculation vanish, $f_{g/j}^{\bare\,\one}(z) = 0$, meaning that the IR and UV divergences in $f_{g/j}^{\bare\,\one}(z)$ are equal with opposite signs. Thus, from \eqs{f_ren_1loop}{f_Z_1loop} we get
\begin{align}\label{eq:f_g}
f^\one_{g/g}(z,\mu) &= - \frac{1}{\eps} \frac{\alpha_s(\mu)}{2\pi}\, \theta(z)
\Bigl[C_A P_{gg}(z) + \frac{1}{2} \beta_0\, \delta(1-z)\Bigr]
\,, \nn \\
f^\one_{g/q}(z,\mu) &= - \frac{1}{\eps} \frac{\alpha_s(\mu) C_F}{2\pi}\, \theta(z) P_{qg}(z)
\,,\end{align}
where the $\eps$ poles  are IR divergences.

\subsection{The Gluon Beam Function at One Loop}
\label{app:calc}

\begin{figure}[t!]
\subfigure[]{\includegraphics[scale=0.75]{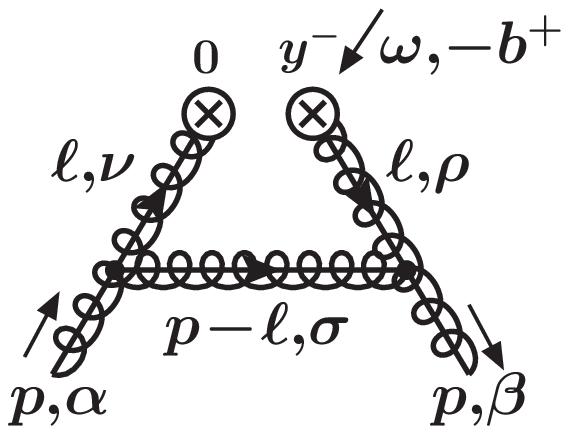}\label{fig:Bone_a}}%
\hfill%
\subfigure[]{\includegraphics[scale=0.75]{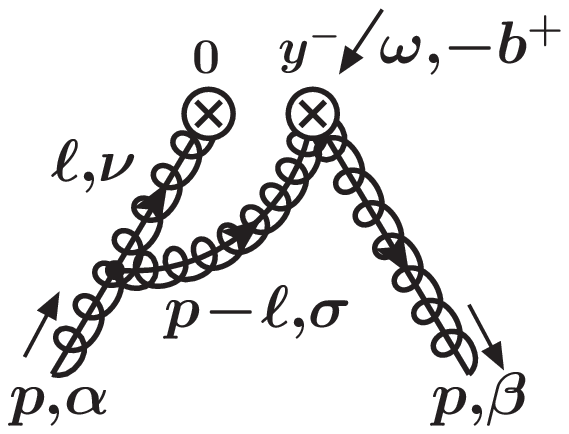}\label{fig:Bone_b}}%
\hfill%
\subfigure[]{\includegraphics[scale=0.75]{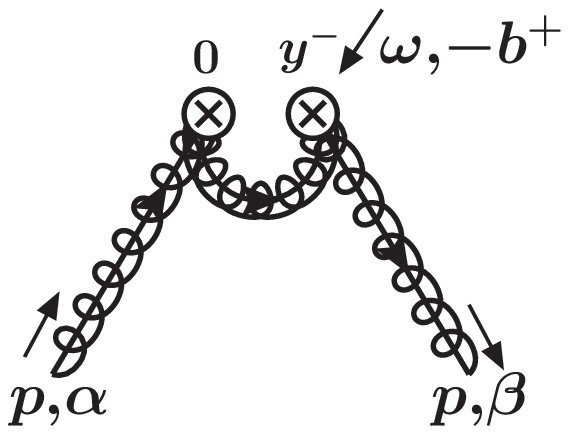}\label{fig:Bone_c}}%
\\
\subfigure[]{\includegraphics[scale=0.75]{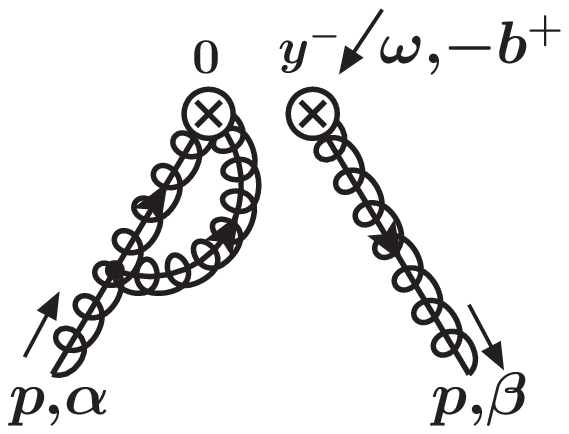}\label{fig:Bone_d}}%
\hfill%
\subfigure[]{\includegraphics[scale=0.75]{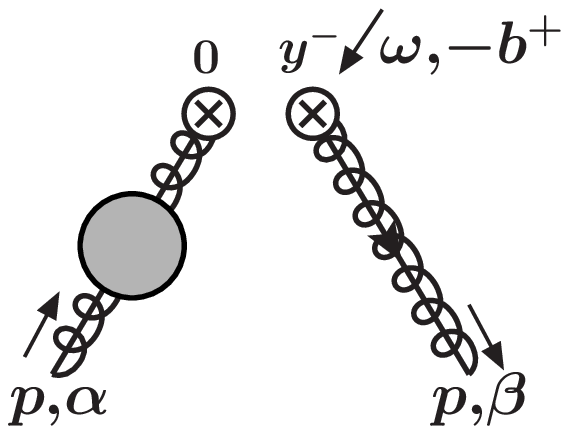}\label{fig:Bone_e}}%
\hfill%
\subfigure[]{\includegraphics[scale=0.75]{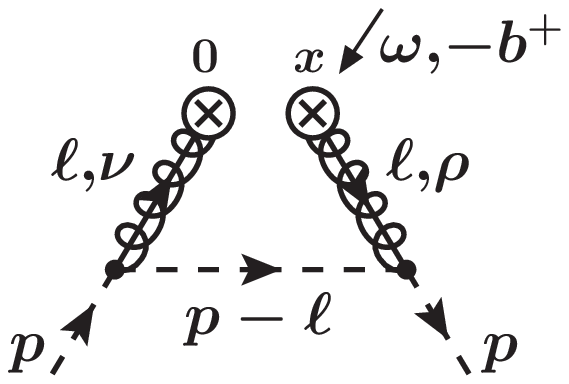}\label{fig:Bone_f}}%
\vspace{-0.5ex}
\caption{One-loop Feynman diagrams for the gluon beam function. The minus momentum $\w$ is incoming at the vertex and the $b^+$ momentum is outgoing. Graphs (b), (d), and (e) have symmetric counterparts, which are equal to the ones shown. The corresponding diagrams with the external lines crossed or involving a four-gluon vertex vanish and are not shown.}
\label{fig:Bone}
\end{figure}

We now turn to the one-loop calculation of the gluon beam
function. The relevant one-loop diagrams are shown in \fig{Bone},
with the Feynman rules given below in \eq{Brules}. Regulating both the
UV and IR in dimensional regularization, the virtual diagram in
\fig{Bone_d} and its symmetric counterpart vanish because there is
only the $p^-$ momentum of the external gluon flowing into the
loop, which is insufficient to give a nonzero Lorentz-invariant
result for the loop integral. The diagram in \fig{Bone_e},
corresponding to the contribution from wave-function
renormalization, vanishes for the same reason. Thus, in the
on-shell scheme, both the wave-function renormalization constant
$Z_\xi$ as well as the residue $R_\xi$ entering the LSZ formula
are equal to one. (A different scheme would give contributions to
both $Z_\xi$ and $R_\xi$ that cancel each other in the final
result.) The diagram in \fig{Bone_c} vanishes, because the
external gluons have perpendicular polarization, so contracting
the two $\cB_{n\perp}^\one$ in \eq{Brules} leads to $\bn \cdot \bn = 0$ in the numerator.

Hence, the only diagrams we have to compute are those in
\figs{Bone_a}{Bone_b} (plus its symmetric counterpart), which
determine $B_{g/g}^{\bare\,\one}(t, z)$, and \fig{Bone_f}, which
determines $B_{g/q}^{\bare\,\one}(t, z)$. Since the diagrams have real
radiation in the intermediate state, we can compute them as the
discontinuity of the matrix element of the corresponding
time-ordered product of fields~\cite{Stewart:2010qs}
\begin{align}
B_{g/j}^\bare(t ,\w/P^-)
&= -\theta(\w) \Disc_{t>0}\,
   \int\! \frac{\df y^-}{4\pi}\,e^{\img t y^-/(2\w)}
\nn\\ & \quad\times
   \MAe{j_n(p^-)}{T\Bigl\{\cB_{n\perp\mu}^c\Bigl(y^- \frac{n}{2}\Bigr)
   \bigl[\delta(\w-\bnP_n) \cB_{n\perp}^{\mu c}(0)\bigr]\Bigr\}}{j_n(p^-)}
\,.
\end{align}

\begin{figure}[t!]
\hfill%
\subfigure[]{\includegraphics[scale=0.75]{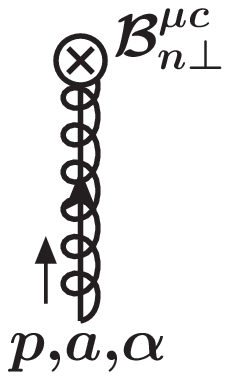}\label{fig:Brules0}}%
\hfill%
\subfigure[]{\includegraphics[scale=0.75]{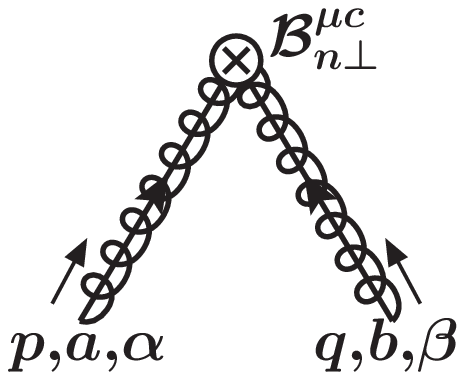}\label{fig:Brules1}}%
\hspace*{\fill}%
\vspace{-0.5ex}
\caption{The SCET Feynman rules for the gluon field strength at $\ord{g^0}$ (a) and
$\ord{g}$ (b).}
\label{fig:Brules}
\end{figure}

For our calculation we need the Feynman rules for the SCET gluon field strength
\begin{equation}
\cB_{n\perp}^\mu = \frac{1}{g} \Bigl[ W_n^\dagger\, \img D_{n\perp}^\mu W_n \Bigr]
\end{equation}
with one and two external gluons. They are illustrated in \fig{Brules} and are given by
\begin{align} \label{eq:Brules}
\text{\fig{Brules0}} &= \delta^{ca} \Bigl(g_\perp^{\mu \al} -
\frac{p_\perp^\mu \bn^\al}{\bn \sdt p}  \Bigr) \equiv \delta^{ca} \cB^{(0)\,\mu \al}_{n\perp}(p)
\,,\\\nn
\text{\fig{Brules1}}  &=
 \img g f^{cab} \biggl[
   \frac{g_\perp^{\mu \bt} \bn^\al}{\bn \sdt p}
   -\frac{g_\perp^{\mu \al} \bn^\bt}{\bn \sdt q}
   +\Bigl(\frac{p_\perp^\mu}{\bn \sdt q} -\frac{q_\perp^\mu}{\bn \sdt p}\Bigr)
   \frac{\bn^\al \bn^\bt}{\bn \sdt (p+q)}
      \biggr]
 \equiv ig f^{cab} \cB^{(1)\,\mu \al \bt}_{n\perp}(p, q)
\,,\end{align}
where $p$ and $q$ are both taken as incoming. We abbreviate the triple gluon vertex as
\begin{equation}
g f^{abc} [g^{\mu \nu} (p_1 - p_2)^\rho + g^{\nu \rho} (p_2 - p_3)^\mu + g^{\rho \mu} (p_3 - p_1)^\nu]
\equiv g f^{abc} V_3^{\mu \nu \rho} (p_1, p_2, p_3)
\,,\end{equation}
where all momenta are incoming and momentum conservation holds, $p_1 + p_2 + p_3 = 0$. Finally, \fig{Bone_e} requires the collinear quark-gluon vertex, which we write as
\begin{equation} \label{eq:quarkglue1}
\img g\, T^a \Bigl(
n^\mu + \frac{\pslash_\perp \gamma_\perp^\mu}{\bn \sdt p}
+ \frac{\gamma_\perp^\mu \qslash_\perp }{\bn \sdt q}
- \frac{\pslash_\perp \qslash_\perp}{\bn \sdt p \, \bn \sdt q}\, \bn^\mu
\Bigr) \frac{\bnslash}{2}
\equiv \img g\, T^a V^\mu_n (p,q)\, \frac{\bnslash}{2}
\,,\end{equation}
where $p$ and $q$ are the momenta of the outgoing and incoming quark lines, respectively, so the gluon carries incoming momentum $p - q$.

Let us start with the diagram in \fig{Bone_a}. The average over the polarizations of the external gluons with momentum $p^\mu=p^- n^\mu/2$ gives
\begin{equation}
\frac{1}{d-2} \sum_\mathrm{pol}\, \ve^\al {\ve^*}^\beta = -\frac{g_\perp^{\alpha\beta}}{d-2}
\,, \end{equation}
where $d = 4 - 2\eps$. We also average over the color of the external gluons. This choice for the polarizations and colors of the external states is just a matter of calculational convenience. Using states with fixed polarization and color gives identical results.
To make our expressions more palatable, we first perform the color algebra, which yields
\begin{equation}
\frac{\delta^{ab}}{N_c^2-1} f^{aec} f^{bde}\, \delta^{fc}\, \delta^{fd} = -C_A
\,.\end{equation}
The diagram is then given by
\begin{align}
&\text{\fig{Bone_a}}
\nn\\ & \qquad
= -\img \Bigl(\frac{e^{\gamma_E} \mu^2}{4\pi}\Bigr)^{\eps} g^2 C_A\, \theta(\w)\,
  \Disc_{t>0}\, \int\! \frac{\df^d \ell}{(2\pi)^d}\,
  \frac{\delta(\ell^-\! - \w) \delta(\ell^+\! + b^+)}{(\ell^2 + \img 0)^2[(\ell-p)^2 + \img 0]}
\\ & \qquad\quad\times
  \frac{g_{\perp\al}^\bt}{d-2}\,
  V_3^{\al \si \nu}(p,\ell-p,-\ell) V_{3\,\bt \rho \si}(-p,\ell,p-\ell)\,
   \cB^\zero_{n\perp \mu \nu}(\ell)\, \cB^{\zero\,\mu\rho}_{n\perp}(-\ell)
\nn \\ & \qquad
= -\img \Bigl(\frac{e^{\gamma_E} \mu^2}{4\pi}\Bigr)^{\eps}
  \frac{g^2 C_A}{(2\pi)^2}\, \theta(z)\,
  \Disc_{t>0}\, \int_0^1\! \df \alpha\, (1-\alpha) \int\! \frac{\df^{d-2} \vec \ell_\perp}{(2\pi)^{d-2}}\,
  \frac{(5 + 4/z^2 ) \, \vec\ell_\perp^2 + (1 + 1/z ) t}{[\vec \ell_\perp^2 + t(1-\alpha/z)]^3}
\nn \\ & \qquad
= \frac{\alpha_s(\mu) C_A}{2\pi}\, \theta(z)\theta(1-z)
   2\Bigl[\frac{1-z}{z} + z(1-z) + \frac{z}{2}\Bigr]
  \Gamma(1+\eps) (e^{\gamma_E} \mu^2)^\eps
  \frac{\sin \pi\eps}{\pi \eps} \frac{\theta(t)}{t^{1+\eps}} \Bigl(\frac{z}{1-z}\Bigr)^\eps
\,.\nn\end{align}
In the second line we integrated $\ell^+$ and $\ell^-$ and introduced a Feynman parameter $\alpha$. The last line follows from the standard integrals and the discontinuities listed in \app{loops}. Expanding in $\eps$, using \eq{plus_exp}, we arrive at
\begin{equation}
  \text{\fig{Bone_a}} =  \frac{\alpha_s(\mu)C_A}{2\pi}\, \theta(z) \theta(1-z)
   2\Bigl[\frac{1-z}{z} + z(1-z) + \frac{z}{2}\Bigr]
  \biggl[\delta(t)\Bigl(-\frac{1}{\eps} + \ln\frac{1-z}{z} \Bigr)
    + \frac{1}{\mu^2} \cL_0\Bigl(\frac{t}{\mu^2}\Bigr) \biggr]
\,.\end{equation}
The color algebra for the diagram in \fig{Bone_b} yields
\begin{equation}
\frac{\delta^{ab}}{N_c^2-1} f^{aec}\, \de^{fc} f^{feb} = -C_A
\,.\end{equation}
Including a factor of two for its mirror graph, we get
\begin{align}
  \text{\fig{Bone_b}} & = -2\img \Bigl(\frac{e^{\gamma_E} \mu^2}{4\pi}\Bigr)^{\eps}
  g^2 C_A\, \theta(\w)\, \Disc_{t>0}\, \int\! \frac{\df^d \ell}{(2\pi)^d}\,
  \frac{\delta(\ell^-\! - \w) \delta(\ell^+\! + b^+)}{(\ell^2 + \img 0) \,[(\ell-p)^2 + \img 0]}
  \nn \\ & \quad \times
  \frac{g_{\perp \al}^\bt}{d-2}\,
   V_3^{\al \si \nu}(p,\ell-p,-\ell) \,
   \cB^{\zero}_{n\perp\, \mu \nu}(\ell)\,
   {\cB^{\one \mu}_{n\perp}}_{ \si \bt}(p-\ell,-p)
  \nn \\
  & = -\img \Bigl(\frac{e^{\gamma_E} \mu^2}{4\pi}\Bigr)^{\eps} \frac{g^2 C_A}{(2\pi)^2}\,
   \theta(z)\, \frac{1+z}{1-z}\, \Disc_{t>0}\, \int_0^1\! \df \alpha
  \int\! \frac{\df^{d-2} \vec\ell_\perp}{(2\pi)^{d-2}}\, \frac{1}{[\vec \ell_\perp^2 + t(1-\alpha/z)]^2}
 \nn \\
  & = \frac{\alpha_s(\mu)C_A}{2\pi}\, \theta(z) (1+z)z^{1+\eps} \frac{\theta(1-z)}{(1-z)^{1+\eps}}\,
  \Gamma(1+\eps)(\eps^{\gamma_E} \mu^2)^\eps \frac{\sin \pi \eps}{\pi \eps}\,
  \frac{\theta(t)}{t^{1+\eps}}
\,,\end{align}
using again the relations listed in \app{loops}. Expanding in $\eps$ yields
\begin{align}
  \text{\fig{Bone_b}} & =  \frac{\alpha_s(\mu) C_A}{2\pi}\, \theta(z)
  \biggl\{ \biggl[-\frac{1}{\eps} \delta(t) + \frac{1}{\mu^2} \cL_0\Bigl(\frac{t}{\mu^2}\Bigr)\biggr]
  \Bigl[-\frac{2}{\eps} \delta(1 - z) + \cL_0(1 - z) z(1+z)\Bigr]
  \nn\\ & \quad +
  \frac{2}{\mu^2} \cL_1\Bigl(\frac{t}{\mu^2}\Bigr) \delta(1 - z) +
  \delta(t) z(1+z) \Bigl[\cL_1(1 - z)  - \cL_0(1 - z)\ln z - \frac{\pi^2}{12} \delta(1 - z) \Bigr] \biggr\}
\,.\end{align}

Note that crossing the external lines of these diagrams corresponds to $p\to-p$ (the polarization and color are symmetric and hence unaffected). By changing $\ell \to -\ell$ this yields the same as the original diagram with the replacement $\delta(\ell^-\! - \w) \delta(\ell^+\! + b^+) \to \delta(\ell^-\! + \w) \delta(\ell^+\! - b^+)$ which leads to
\begin{equation} \label{eq:cross}
\Delta \equiv \Bigl(1-\frac{\al}{z}\Bigr) t \to \Bigl(1+\frac{\al}{z}\Bigr) t
\end{equation}
in the denominator of the loop integrals. Since $\alpha, z>0$ these graphs have no discontinuity for $t>0$ and vanish.
Other one-loop diagrams that could in principle connect the two gluon fields, e.g.\ involving a four gluon vertex, also vanish, because they require to cut one of the external lines entering the gluon operator, which sets $\Delta = t$ in the loop integral and leads to a vanishing discontinuity for $t>0$.

Adding up the two nonvanishing diagrams in \figs{Bone_a}{Bone_b}, we find
\begin{align}\label{eq:B_ggbare}
  B_{g/g}^{\bare\one}(t,z) &
  = \frac{\alpha_s(\mu) C_A}{2\pi}\, \theta(z)
  \biggl\{ \biggl[ \frac{2}{\eps^2}\, \delta(t)
  - \frac{2}{\eps}\, \frac{1}{\mu^2} \cL_0\Bigl(\frac{t}{\mu^2}\Bigr) \biggr] \delta(1-z)
  - \frac{1}{\eps}\, \delta(t) P_{gg}(z)
  \nn \\ & \quad
  + \frac{2}{\mu^2} \cL_1\Bigl(\frac{t}{\mu^2}\Bigr) \delta(1-z)
  + \frac{1}{\mu^2} \cL_0\Bigl(\frac{t}{\mu^2}\Bigr) P_{gg}(z)
  \nn \\ & \quad
  + \delta(t)\Bigl[\cL_1(1-z) \frac{2(1-z+z^2)^2}{z} - P_{gg}(z) \ln z - \frac{\pi^2}{6} \delta(1-z) \Bigr] \biggr\}
\,,\end{align}
where $P_{gg}(z)$ is the $g\to gg$ splitting function given in \eq{split}.

The mixing contribution of the quark PDF into the gluon beam function
originates from the diagram in \fig{Bone_f}. The spin and color averages are
\begin{equation}
\frac{1}{2} \sum_\mathrm{spins}\, u_n(p)\bar{u}_n(p) = \frac{1}{2} \pslash
\,,\qquad
\frac{1}{N_c}\,\tr[T^a T^b]\, \delta^{ab} = C_F
\,.\end{equation}
The diagram is thus given by
\begin{align} \label{eq:Bone_f}
&\text{\fig{Bone_f}}
\nn\\ & \qquad
= -\img \Bigl(\frac{e^{\gamma_E} \mu^2}{4\pi}\Bigr)^{\eps}
  g^2 C_F\, \theta(\w)\,\Disc_{t>0}\, \int\! \frac{\df^d \ell}{(2\pi)^d} \,
  \frac{(p^-\! - \ell^-)\, \delta(\ell^-\! - \w) \delta(\ell^+\! + b^+)}{(\ell^2 + \img 0)^2 [(p-\ell)^2 + \img 0]}\,
\\ & \qquad\quad \times
  \bar u_n(p) V^\rho_n (p, p-\ell) V_{n\, \nu}(p-\ell, p) \frac{\bnslash}{2}\, u_n(p)\,
 \cB^{\zero \mu \nu}_{n\perp}(\ell)\,  \cB^{\zero}_{n\perp \mu \rho}(-\ell)
  \nn \\ & \qquad
= \img \Bigl(\frac{e^{\gamma_E} \mu^2}{4\pi}\Bigr)^{\eps} \frac{g^2 C_F}{(2\pi)^2}\,
  \theta(z) \Bigl(\frac{d-2}{1-z} + \frac{4}{z^2} \Bigr)
  \Disc_{t>0}\, \int_0^1\! \df \alpha\, (1-\alpha)
  \!\int\! \frac{\df^{d-2} \vec\ell_\perp}{(2\pi)^{d-2}}\, \frac{\vec \ell_\perp^2}{[\vec \ell_\perp^2 + t(1-\alpha/z)]^3}
\nn \\ & \qquad
= \frac{\alpha_s(\mu) C_F}{2\pi}\, \theta(z)\theta(1-z) \Ga(1+\eps) (e^{\gamma_E} \mu^2)^\eps
 \Bigl[\frac{1 + (1-z)^2}{z} - \eps z\Bigr] \frac{\sin \pi\eps}{\pi\eps}\, \frac{\theta(t)}{t^{1+\eps}}\,
 \Bigl(\frac{z}{1-z}\Bigr)^\eps
 \,. \nn
\end{align}
This is the same loop integral and discontinuity as in \fig{Bone_a}. For the crossed graph with the external lines interchanged, we can again change $\ell \to -\ell$ to obtain the same expression up to $\delta(\ell^-\! - \w) \delta(\ell^+
\! + b^+) \to \delta(\ell^-\! + \w) \delta(\ell^+\! - b^+)$, which leads to a vanishing discontinuity as before. Expanding in $\eps$, we obtain
\begin{align}\label{eq:B_gqbare}
B^{\bare\one}_{g/q}(t,z)
&= \frac{\alpha_s(\mu) C_F}{2\pi}\,
   \theta(z) \biggl\{ \frac{1}{\mu^2} \cL_0\Bigl(\frac{t}{\mu^2}\Bigr) P_{gq}(z)
\nn\\ & \quad
   + \delta(t)\biggl[P_{gq}(z)\Bigl(-\frac{1}{\eps} + \ln \frac{1-z}{z} \Bigr) + \theta(1-z) z \biggr] \biggr\}
\,,\end{align}
with $P_{gq}(z)$ as in \eq{split}. The matrix element with external antiquarks gives the same result, so the mixing contributions from quarks and antiquarks are identical.

The renormalized matrix elements $B_{g/j}(t, z, \mu)$ are given in terms of the bare ones as~\cite{Stewart:2010qs}
\begin{align} \label{eq:B_ren}
B_{g/j}^\bare(t, z) = \int\! \df t'\, Z_B^g(t-t',\mu)\, B_{g/j}(t', z, \mu)
\,.\end{align}
Since $\gamma_B^g(t,\mu) =  - \mu\frac{\df}{\df\mu} Z_B^g(t,\mu)$, the NLO counterterm follows from the one-loop anomalous dimension in \eq{gaB_gen} with $\Gamma_\cusp^g(\alpha_s) = C_A \alpha_s/\pi$ and $\gamma_B^g(\alpha_s) = \beta_0 \alpha_s/(2\pi)$ (see \app{rge}),
\begin{equation} \label{eq:ZBg}
Z_B^g(t,\mu) = \delta(t) + \frac{\alpha_s(\mu)}{2\pi} \biggl\{2C_A \biggl[\frac{1}{\eps^2} \delta(t)
    - \frac{1}{\eps} \frac{1}{\mu^2}\cL_0\Bigl(\frac{t}{\mu^2}\Bigr)\biggr]
  + \frac{1}{2\eps} \beta_0\, \delta(t) \biggr\}
\,.\end{equation}
Expanding \eq{B_ren} to NLO and using the tree-level result for $B_{g/j}(t, z, \mu)$ in \eq{tree}, we find
\begin{equation} \label{eq:B_ren_1loop}
B_{g/j}^{\bare\one}(t, z) = Z^{g\one}_B(t,\mu)\, \delta_{gj}\, \delta(1-z) + B_{g/j}^\one(t,z,\mu)
\,.\end{equation}
Note that the mixing contribution $B_{g/q}$ is UV finite and not renormalized. From \eq{ZBg} and our results in \eqs{B_ggbare}{B_gqbare} we have
\begin{align} \label{eq:Bg_1loop}
B_{g/g}^\one(t,z, \mu)
  &= \frac{\alpha_s(\mu) C_A}{2\pi}\, \theta(z)
  \biggl\{  - \frac{1}{\eps}\, \delta(t) \Bigl[P_{gg}(z) + \frac{\beta_0}{2C_A}\,\delta(1-z) \Bigr]
  \nn \\ & \quad
  + \frac{2}{\mu^2} \cL_1\Bigl(\frac{t}{\mu^2}\Bigr) \delta(1-z)
  + \frac{1}{\mu^2} \cL_0\Bigl(\frac{t}{\mu^2}\Bigr) P_{gg}(z)
  \nn \\ & \quad
  + \delta(t) \Bigl[\cL_1(1-z)\frac{2(1-z+z^2)^2}{z} - P_{gg}(z) \ln z - \frac{\pi^2}{6} \delta(1-z) \Bigr] \biggr\}
\,,\nn\\
B_{g/q}^\one(t,z, \mu)
&= \frac{\alpha_s(\mu) C_F}{2\pi}\,
   \theta(z) \biggl\{ - \frac{1}{\eps}\, \delta(t) P_{gq}(z)
\nn\\ & \quad
   + \frac{1}{\mu^2} \cL_0\Bigl(\frac{t}{\mu^2}\Bigr) P_{gq}(z)
   + \delta(t)\Bigl[P_{gq}(z)\ln \frac{1-z}{z} + \theta(1-z) z \Bigr] \biggr\}
\,,\end{align}
where all $1/\eps$ divergences are now IR divergences. Subtracting the one-loop PDF matrix elements in \eq{f_g} according to \eq{B_fact_1loop}, we see that the IR divergences precisely cancel between $B_{g/j}$ and $f_{g/j}$, and the finite terms in \eq{Bg_1loop} determine the one-loop matching coefficients $\cI_{gj}(t, z, \mu)$ in \eq{Ig_results}.

To compare our expression for $\cI_{gg}^\one(t, z, \mu)$ with the result of ref.~\cite{Fleming:2006cd}, we take the moments
\begin{align} \label{eq:hC}
\hat{C}_{II}^\one(z, N)
&= M^2 \int_0^1 \! \df y\, y^N \cI_{gg}^\one[M^2(1-y),z, \mu]
= \int_0^{M^2} \! \df t\, \Bigl(1 - \frac{t}{M^2}\Bigr)^N \cI_{gg}^\one(t,z, \mu)
\nn\\
&= \frac{\alpha_s(\mu) C_A}{2\pi}\, \theta(z) \biggl[
   \ln^2\Bigl(\frac{N e^{\gamma_E} \mu^2}{M^2}\Bigr) \delta(1-z)
  - \ln\Bigl(\frac{N e^{\gamma_E} \mu^2}{M^2}\Bigr) P_{gg}(z)
  \nn \\ & \quad
  + \cL_1(1-z) \frac{2(1 - z + z^2)^2}{z} - P_{gg}(z) \ln z \biggr] + \ORd{\frac{1}{N}}
\,,\end{align}
where we used the approximation for the harmonic numbers at large $N$
\begin{equation}
H_N = \sum_{i=1}^N \frac{1}{i} = \ln N + \gamma_E + \ORd{\frac{1}{N}}
\,.\end{equation}
Our result in \eq{hC} agrees with eq.~(68) of ref.~\cite{Fleming:2006cd} except for a constant term $-\alpha_s C_A/\pi\, \delta(1-z) \pi^2/8$.

\subsection{Integrals, Discontinuities, and Plus Distributions}
\label{app:loops}

Here we list various loop integrals that are needed for the calculation in \app{calc}:
\begin{align}\label{eq:loop_int}
 \int\! \frac{\df^{d-2} \vec \ell_\perp}{(2\pi)^{d-2}}\, \frac{1}{(\vec\ell_\perp^2 + \De)^2}
  & = \frac{1}{(4\pi)^{1-\eps}} \Ga(1+\eps)\, \De^{-1-\eps}
\,, \nn \\
  \int\! \frac{\df^{d-2} \vec \ell_\perp}{(2\pi)^{d-2}}\, \frac{1}{(\vec \ell_\perp^2 + \De)^3}
  & = \frac{1}{(4\pi)^{1-\eps}} \frac{\Ga(2+\eps)}{2}\, \De^{-2-\eps}
\,, \nn \\
  \int\! \frac{\df^{d-2} \vec\ell_\perp}{(2\pi)^{d-2}}\, \frac{\vec\ell_\perp^2}{(\vec\ell_\perp^2 + \De)^3}
  & = \frac{1}{(4\pi)^{1-\eps}}(1-\eps) \frac{\Ga(1+\eps)}{2}\, \De^{-1-\eps}
\,.
\end{align}
To calculate the discontinuities of the various graphs we need the relations,
\begin{align} \label{eq:disc}
  -\theta(z) \frac{\img}{2\pi}\, \Disc_{t>0} \int_0^1 \df\al\, \De^{-1-\eps}
  &= \theta(z)\, \frac{\sin \pi\eps}{\pi\eps}\,  \frac{\theta(t)}{t^{1+\eps}}\,
  \theta(1-z) z^{1+\eps} (1-z)^{-\eps}
  \,, \nn \\
  -\theta(z) \frac{\img}{2\pi}\, \Disc_{t>0} \int_0^1 \df\al\, (1-\al) \De^{-1-\eps}
  &= \theta(z)\, \frac{\sin \pi\eps}{\pi\eps(1-\eps)}\,  \frac{\theta(t)}{t^{1+\eps}}\,
  \theta(1-z) z^{1+\eps} (1-z)^{1-\eps}
  \,, \nn \\
  -\theta(z) \frac{\img}{2\pi}\, \Disc_{t>0} \int_0^1 \df\al\, (1-\al)\, t \De^{-2-\eps}
  &= \theta(z)\, \frac{\sin \pi\eps}{\pi\eps(1+\eps)}\,  \frac{\theta(t)}{t^{1+\eps}}\,
  \theta(1-z) z^{2+\eps} (1-z)^{-\eps}
  \,.
\end{align}
The plus distributions are defined as
\begin{align} \label{eq:plusdef}
\cL_n(x)
&= \biggl[ \frac{\theta(x) \ln^n x}{x}\biggr]_+
 = \lim_{\beta \to 0} \biggl[
  \frac{\theta(x- \beta)\ln^n x}{x} +
  \delta(x- \beta) \, \frac{\ln^{n+1}\!\beta}{n+1} \biggr]
\,,\nn\\
\cL^\eta(x)
&= \biggl[ \frac{\theta(x)}{x^{1-\eta}}\biggr]_+
 = \lim_{\beta \to 0} \biggl[
  \frac{\theta(x - \beta)}{x^{1-\eta}} +
  \delta(x- \beta) \, \frac{x^\eta - 1}{\eta} \biggr]
\,,\end{align}
satisfying the boundary condition $\int_0^1 \df z\, \cL_n(z) = 0$.
We also need the distribution identity,
\begin{equation} \label{eq:plus_exp}
 \frac{1}{(1-z)^{1+\eps}} = -\frac{1}{\eps} \delta(1-z) + \cL_0(1-z) - \eps \cL_1(1-z) + \ord{\eps^2}
\,.\end{equation}

\section{Perturbative Results}
\label{app:rge}

\subsection{Hard Function}
\label{app:hardapp}

The functions $F^\zero(z)$ and $F^\one(z)$ which encode the $m_t$ dependence of the hard Wilson coefficient in \eq{CggH} are given by
\begin{align} \label{eq:Ftop}
F^\zero(z) &= \frac{3}{2z} - \frac{3}{2z}\Bigl\lvert 1 - \frac{1}{z}\Bigr\rvert
\begin{cases}
\arcsin^2(\sqrt{z})\, , & 0 < z \leq 1 \,,\\
\ln^2[-\img(\sqrt{z} + \sqrt{z-1})] \,, \quad & z > 1
\,,\end{cases}
\nn\\
F^\one(z) &=
\Bigl(5 - \frac{38}{45}\, z - \frac{1289}{4725}\, z^2 - \frac{155}{1134}\, z^3 - \frac{5385047}{65488500}\, z^4\Bigr) C_A
\nn\\ & \quad
+ \Bigl(-3 + \frac{307}{90}\, z + \frac{25813}{18900}\, z^2 + \frac{3055907}{3969000}\, z^3 +
   \frac{659504801}{1309770000}\, z^4 \Bigr) C_F + \ord{z^5}
\,.\end{align}
Here, $F^\zero(z)$ gives the well-known $m_t$ dependence of the leading-order $gg\to H$ cross section given by the virtual top-quark loop. The full analytic $m_t$ dependence of the virtual two-loop corrections to $gg\to H$ in terms of harmonic polylogarithms were obtained in refs.~\cite{Harlander:2005rq, Anastasiou:2006hc}. Since the corresponding exact expression for $F^\one(z)$ is very long, we use the results expanded in $m_H^2/m_t^2$ from ref.~\cite{Pak:2009bx}. The additional $m_t$ dependence coming from $F^\one(z)$ is small and the expansion is converging very quickly, so the expanded result is completely sufficient for practical purposes.
For completeness we also give the leading terms in $F^\two(z)$,
\begin{align} \label{eq:Ftop2}
F^\two(z) &=
\bigl(7 C_A^2 + 11 C_A C_F - 6 C_F \beta_0 \bigr) \ln(-4z-\img 0)
+ \Bigl(-\frac{419}{27} + \frac{7\pi^2}{6} + \frac{\pi^4}{72} - 44 \zeta_3 \Bigr) C_A^2
\nn \\ & \quad
+ \Bigl(-\frac{217}{2} - \frac{\pi^2}{2} + 44\zeta_3 \Bigr) C_A C_F
+ \Bigl(\frac{2255}{108} + \frac{5\pi^2}{12} + \frac{23\zeta_3}{3} \Bigr) C_A \beta_0
- \frac{5}{6} C_A T_F
\nn \\ & \quad
+ \frac{27}{2} C_F^2
+ \Bigl(\frac{41}{2} - 12 \zeta_3\Bigr) C_F \beta_0
-\frac{4}{3} C_F T_F
+ \ord{z}
\,.\end{align}
The first few higher-order terms in $z$ can be extracted from the results in refs.~\cite{Harlander:2009bw, Pak:2009bx}.

Alternatively to the one-step matching we use, one can perform a two-step matching by treating $m_H \ll m_t$ as done in refs.~\cite{Idilbi:2005er, Idilbi:2005ni, Ahrens:2008nc, Mantry:2009qz}. In this case, one first integrates out the top quark at the scale $\mu_t \simeq m_t$ and then matches from QCD onto SCET at $\mu_H \simeq m_H$. In this way one neglects power corrections of $\ord{m_H/m_t}$ but in turn the running between $\mu_t$ and $\mu_H$ allows one to sum logarithms of $m_H/m_t$. We first integrate out the top loop to get the effective Hamiltonian~\cite{Dawson:1990zj, Djouadi:1991tka}
\begin{equation} \label{eq:C1}
\mathcal{H}_\mathrm{eff} = -C_1(m_t,\mu_t)\, \frac{H}{12\pi v}\, G^{\mu\nu a}\, G_{\mu\nu}^a
 \,, \quad
 C_1(m_t,\mu_t) = \alpha_s(\mu_t) \biggl[1 + \frac{\alpha_s(\mu_t)}{4\pi} (5 C_A - 3 C_F) \biggr]
 \,,
\end{equation}
where as before $\alpha_s(\mu_t)$ is evaluated for $n_f = 5$ flavors and at higher order $C_1(m_t,\mu_t)$ contains logarithms of $m_t/\mu_t$. In the second step, we integrate out hard off-shell modes by matching onto the SCET operator in \eq{Heff},
\begin{align} \label{eq:C2}
- G_{\mu\nu}^a G^{\mu\nu a}
&= C_2(q^2, \mu_H)\, q^2 g_{\mu\nu} \cB_{\perp}^{\mu c} \cB_{\perp}^{\nu c}
 \,,\nn\\
C_2(q^2,\mu_H)
&= 1 + \frac{\alpha_s(\mu_H)}{4\pi} C_A \biggl(-\ln^2 \frac{-q^2 -\img 0}{\mu_H^2} + \frac{\pi^2}{6} \biggr)
.\end{align}
The SCET matching coefficient $C_2(q^2, \mu_H)$ corresponds to the IR finite parts of the gluon form factor. Combining the two matching steps we obtain
\begin{align}
C_{ggH}(m_t, q^2,\mu) &= C_1(m_t, \mu)\, C_2(q^2, \mu)
\nn\\
&= \alpha_s(\mu) \biggl\{1 + \frac{\alpha_s(\mu)}{4\pi} \biggl[ \Bigl(-\ln^2 \frac{-q^2-\img 0}{\mu^2}
  + 5 + \frac{\pi^2}{6} \Bigr) C_A - 3 C_F \biggr] \biggr\}
\,.\end{align}
In the second line we expanded both coefficients at the same scale $\mu$, which reproduces the $m_t\to\infty$ limit of \eq{CggH}. At NNLO, this gives the leading terms of $F^\two(z)$ in \eq{Ftop2}.
The strict $m_t\to\infty$ limit is formally necessary for a consistent two step matching. Note that in the $m_t\to\infty$ limit that is often used in the literature, the full $m_t$ dependence of the leading-order cross section in the overall factor $F^\zero(z)$ in \eq{CggH} is included, which then turns out to be a very good approximation for the total cross section~\cite{Dawson:1993qf, Spira:1995rr, Pak:2009dg, Harlander:2009my}. Considering the virtual corrections only, this way of defining the $m_t\to\infty$ limit is equivalent to the one-step matching in \eq{CggH} where one neglects the weak $m_t$ dependence of $F^\one(z)$ and using $F^\one(0)$.

\subsection{Beam Function}
\label{app:beamapp}

We obtain the $\mu_B$-dependent terms in the two-loop contribution $\cI_{ij}^\two$ in \eq{Ig} by perturbatively solving the two-loop RGE of the matching coefficients $\cI_{gj}$, given by~\cite{Stewart:2010qs}
\begin{align} \label{eq:Igj_RGE}
\mu\frac{\df}{\df\mu}\cI_{gj}(t,z,\mu)
= \sum_k \int\!\df t'\, \frac{\df z'}{z'}\, \cI_{gk}\Bigl(t - t', \frac{z}{z'},\mu\Bigr)\Bigl[\gamma_B^g(t', \mu)\, \delta_{kj} \delta(1 - z')
- \delta(t') \gamma^f_{kj}(z',\mu) \Bigr]
\,,\end{align}
where the $\gamma_B^g(t, \mu)$ is the anomalous dimensions of the beam function in \eq{gaB_gen} and $\gamma^f_{kj}(z, \mu)$ that of the PDFs. We obtain
\begin{align}
\cI_{gg}^\two(t,z,\mu_B)
&= \frac{1}{\mu_B^2} \cL_3\Bigl(\frac{t}{\mu_B^2}\Bigr) 8 C_A^2\, \delta(1-z)
+ \frac{1}{\mu_B^2} \cL_2\Bigl(\frac{t}{\mu_B^2}\Bigr)
   \bigl[ 12 C_A^2 P_{gg}(z) - 2 C_A \beta_0\, \delta(1-z) \bigr]
\nn \\ & \quad
+ \frac{1}{\mu_B^2} \cL_1\Bigl(\frac{t}{\mu_B^2}\Bigr) \biggl\{
   4 C_A^2 \Bigl[ \Bigl(\frac{4}{3} - \pi^2\Bigr) \delta(1-z) + 2 \cI_{gg}^{(1,\delta)}(z) + (P_{gg} \otimes P_{gg})(z) \Bigr]
   \nn \\ & \qquad
   + 2 C_A \beta_0  \Bigl[\frac{10}{3} \delta(1-z) - P_{gg}(z)\Bigr]
   + 8 C_F T_F n_f (P_{gq} \otimes P_{qg})(z) \biggr\}
\nn \\ & \quad
+ \frac{1}{\mu_B^2} \cL_0\Bigl(\frac{t}{\mu_B^2}\Bigr) \biggl\{
   4 C_A^2 \Bigl[\Bigl(-\frac{7}{9} + 8 \zeta_3\Bigr)\delta(1-z) - \frac{\pi^2}{3} P_{gg}(z) + (\cI_{gg}^{(1,\delta)} \otimes P_{gg})(z) \Bigr]
   \nn \\ & \qquad
   + C_A \beta_0 \Bigl[ \Bigl( -\frac{92}{9} + \frac{\pi^2}{3} \Bigr) \delta(1-z) -2\cI_{gg}^{(1,\delta)}(z) \Bigr]
   \nn \\ & \qquad
   + C_F T_F n_f \bigl[ 4\delta(1-z)  + 8(\cI_{gq}^{(1,\delta)} \otimes P_{qg})(z) \bigr] + 4 P_{gg}^\one(z) \biggr\}
   + 4 \delta(t)\, \cI_{gg}^{(2,\delta)}(z)
\,, \nn \\
\cI_{gq}^\two(t,z,\mu_B)
&= \frac{1}{\mu_B^2} \cL_2\Bigl(\frac{t}{\mu_B^2}\Bigr) 12 C_A C_F P_{gq}(z)
+ \frac{1}{\mu_B^2} \cL_1\Bigl(\frac{t}{\mu_B^2}\Bigr) \biggl\{
   4C_F^2 (P_{gq} \otimes P_{qq})(z)
   \nn \\ & \qquad
   + 4C_A C_F \bigl[ (P_{gg} \otimes P_{gq})(z) + 2\cI_{gq}^{(1,\delta)}(z) \bigr]
   - 4C_F \beta_0 P_{gq}(z) \biggr\}
\nn\\ & \quad
+ \frac{1}{\mu_B^2} \cL_0\Bigl(\frac{t}{\mu_B^2}\Bigr) \biggl\{
   4C_A C_F \Bigl[ - \frac{\pi^2}{3} P_{gq}(z) + (\cI_{gg}^{(1,\delta)}  \otimes P_{gq})(z) \Bigr]
   \\\nn & \qquad
   + 4C_F^2 (\cI_{gq}^{(1,\delta)} \otimes P_{qq})(z)
   - 4C_F \beta_0 \cI_{gq}^{(1,\delta)}(z) + 4P_{gq}^\two(z) \biggr\}
   + 4\delta(t)\, \cI_{gq}^{(2,\delta)}(z)
\,.\end{align}
The $\mu_B$-independent terms $\cI_{gj}^{(2,\delta)}(z)$ require the two-loop calculation of gluon beam function which is not yet available.
The one-loop functions $\cI_{gj}^{(1,\delta)}(z)$ entering in the above expressions were already given in \eq{Igdel_results}.
The quark and gluon splitting functions are
\begin{align} \label{eq:Pqq_def}
P_{gg}(z)
&= 2 \cL_0(1-z)z + 2\theta(1-z)\Bigl[\frac{1-z}{z} +  z(1-z)\Bigr]
\,,\nn\\
P_{gq}(z) &= \theta(1-z)\, \frac{1+(1-z)^2}{z}
\,,\nn\\
P_{qq}(z)
&= \cL_0(1-z)(1+z^2) + \frac{3}{2}\,\delta(1-z)
\,,\nn\\
P_{qg}(z) &= \theta(1-z)\bigl[(1-z)^2+ z^2\bigr]
 \,.\end{align}
The two-loop splitting functions were calculated in refs.~\cite{Furmanski:1980cm, Ellis:1996nn} and are given by~\cite{Ellis:1996nn}
\begin{align} \label{eq:Pgj_one}
P_{gg}^\one(z)
&= C_A^2 \Bigl\{\frac{1}{3} (4-\pi ^2) \cL_0(1-z)
   + (-1 + 3 \zeta_3) \delta(1-z)
   + \frac{- 253 + 294 z - 318 z^2 + 253 z^3}{18 z}
\nn\\ & \quad\quad
   + \frac{-1 +2 z - z^2+ z^3}{3 z} \pi^2
   - \frac{4}{3} (9 + 11 z^2) \ln z
   + \frac{2(1 + z -z^2)^2}{1 - z^2} \ln^2 z
\nn\\ & \quad\quad
   - 2P_{gg}(z) \ln z \ln(1-z)
   - 2P_{gg}(-z) \Bigl[\ln z \ln(1+z) + \Li_2(-z) + \frac{\pi^2}{12}  \Bigr]
   \Bigr\}
\nn\\ & \quad
  + C_A \beta_0 \Bigl[\frac{5}{3} \cL_0(1-z) +\delta(1-z) + \frac{23-29 z+19 z^2-23 z^3}{6z}+(1 + z) \ln z \Bigr]
\nn\\ & \quad
  + C_F T_F n_f
   \Bigl[-\delta(1-z) + \frac{4(1-12z + 6 z^2 + 5 z^3)}{3z} - 2(1+z) \ln^2 z -2 (3 + 5 z) \ln z \Bigr]
\,,\nn\\
P_{gq}^\one(z)
&= C_A C_F \Bigl\{
  \frac{-101 + 129 z - 51 z^2 + 44 z^3}{9z}
  - \frac{1}{3}(36 + 15z + 8 z^2) \ln z
  + 2 z \ln(1-z)
\nn\\ & \quad \quad
  + (2+z) \ln^2z
  - P_{gq}(z) \Bigl[2\ln z \ln (1-z) - \ln^2(1-z) + \frac{\pi^2}{6} \Bigr]
\nn\\ & \quad \quad
  - 2P_{gq}(-z) \Bigl[\ln z \ln(1+z) + \Li_2(-z) + \frac{\pi^2}{12}  \Bigr]
\Bigr\}
\nn\\ & \quad
   + C_F^2 \Bigl[-\frac{1}{2} (5 + 7 z)
   + \frac{1}{2}(4 + 7 z) \ln z
   + \frac{-6+ 6z -5 z^2}{z} \ln (1-z)
   - \frac{1}{2} (2 - z) \ln^2 z
\nn\\ & \quad\quad
   - P_{gq}(z) \ln^2(1-z) \Bigr]
+ C_F \beta_0 \Bigl[ \frac{2(5 -5 z + 4 z^2)}{3 z} + P_{gq}(z) \ln(1-z) \Bigr]
\,.\end{align}
The convolution of two splitting functions is defined as (the index $j$ is not summed over)
\begin{equation} \label{eq:convdef}
(P_{ij} \otimes P_{jk})(z) = \int_z^1\! \frac{\df w}{w}\, P_{ij}(w) P_{jk} \Bigl(\frac{z}{w}\Bigr)
\,,\end{equation}
and analogously for the convolution $(\cI_{ij}^{(1,\delta)} \otimes P_{jk})$. The necessary convolutions are
\begin{align} \label{eq:convres}
(P_{gq} \otimes P_{qg})(z)
&= \frac{4 + 3z - 3z^2 - 4z^3}{3 z} + 2 (1 + z) \ln z
\,, \nn \\
(P_{gq} \otimes P_{qq})(z)
&= 2 - \frac{z}{2} + (2-z) \ln z + \frac{2(2-2z+z^2)}{z} \ln(1-z)
\,, \nn \\
(P_{gg} \otimes P_{gq})(z)
&= \frac{-31+24z+3z^2+4z^3}{3z} - \frac{4(1+z+z^2)}{z} \ln z + \frac{2(2-2z+z^2)}{z} \ln(1-z)
\,, \nn \\
(P_{gg} \otimes P_{gg})(z)
&= 8 \cL_1(1-z) - \frac{2\pi^2}{3} \delta(1-z) +
  \frac{4(-11 + 9 z - 9 z^2 + 11 z^3)}{3z}
  \nn \\ & \quad
  + \frac{4(-1 - 3 z^2 + 4 z^3 - z^4)}{z(1-z)} \ln z + \frac{8(1 - 2 z + z^2 - z^3)}{z} \ln(1-z)
\,, \nn \\
(\cI_{gq}^{(1,\delta)} \otimes P_{qg})(z)
&= \frac{-13+12z+6z^2-5z^3}{9z} + \pi^2\frac{1+z}{3}
  + \frac{-4+9z^2+4z^3}{3z} \ln z
  \nn \\ & \quad
  + \frac{4+3z-3z^2-4z^3}{3z} \ln (1-z) - (1+z) \ln^2 z -2(1+z) \Li_2(z)
\,, \nn \\
(\cI_{gq}^{(1,\delta)} \otimes P_{qq})(z)
&= \frac{5 - 4 z + 2 z^2}{2z} + \pi^2 \frac{-4 + 6 z - 3 z^2}{6z}
  + \frac{-2-z}{2} \ln z + \frac{4+3z}{2} \ln (1-z)
  \nn \\ & \quad
  + \frac{z-2}{2} \ln^2 z + \frac{2(2 - 2 z + z^2)}{z} \ln(1-z) [\ln(1-z) - \ln z]
  + (z-2) \Li_2(z)
\,, \nn \\
(\cI_{gg}^{(1,\delta)} \otimes P_{gq})(z)
&= \frac{21 - 26 z + 5 z^2}{6z} + \pi^2 \frac{-2-6z-3z^2}{6z}+
  \frac{9 - 30 z - 9 z^2 - 4 z^3}{3z} \ln z
  \nn \\ & \quad
  + \frac{-31 + 24 z + 3 z^2 + 4 z^3}{3z} \ln(1-z)
  + \frac{2(1+z+z^2)}{z} \ln^2 z
  \nn \\ & \quad
  + \frac{2 -2z+z^2}{z} \ln(1-z) [\ln(1-z) - 2 \ln z] + (8+2z) \Li_2(z)
\,, \nn \\
(\cI_{gg}^{(1,\delta)} \otimes P_{gg})(z)
&= 6\cL_2(1-z)- \pi^2 \cL_0(1-z) + 4 \zeta_3 \delta(1-z) +
  \frac{(1-z)(67-2z+67z^2)}{9z}
  \nn \\ & \quad
   + \pi^2 \frac{-3 + 2 z - 7 z^2 + 3 z^3}{3z} + \frac{2(11-21z+6z^2-22z^3)}{3z} \ln z
  \nn \\ & \quad
   + \frac{4(-11+9z-9z^2+11z^3)}{3z} \ln(1-z) + \frac{2(1+3z^2-4z^3+z^4)}{z(1-z)} \ln^2 z
  \nn \\ & \quad
   + \frac{8(-1+2z-3z^2+2z^3-z^4)}{z(1-z)} \ln z \ln(1-z)
  \nn \\ & \quad
   + \frac{6(1-2z+z^2-z^3)}{z} \ln^2(1-z) +
  8(1+z) \Li_2(z)
  \,.
\end{align}

\subsection{Renormalization Group Evolution}
\label{app:rgeapp}

The hard Wilson coefficient satisfies the RGE
\begin{align} \label{eq:C_RGE}
\mu \frac{\df}{\df\mu} C_{ggH}(m_t, q^2, \mu) &= \gamma_H^g(q^2, \mu)\, C_{ggH}(m_t, q^2, \mu)
\,,\nn\\
\gamma_H^g(q^2, \mu) &=
\Gamma_\cusp^g[\alpha_s(\mu)]\, \ln\frac{- q^2 - \img 0}{\mu^2} + \gamma_H^g[\alpha_s(\mu)]
\,,\end{align}
whose solution gives \eq{Hrun}. The beam function RGE is given in \eq{Bg_RGE}. The soft function
satisfies an analogous RGE,
\begin{align} \label{eq:S_RGE}
\mu\frac{\df}{\df\mu} S_B^{gg}(k, \mu)
&= \int\! \df k'\, \gamma_S^g(k - k', \mu)\, S_B^{gg}(k', \mu)
\,,\nn\\
\gamma_S^g(k, \mu)
&= 4\,\Gamma_\cusp^g[\alpha_s(\mu)]\, \frac{1}{\mu} \cL_0\Bigl(\frac{k}{\mu}\Bigr) +
\gamma_S^g[\alpha_s(\mu)]\, \delta(k)
\,,\end{align}
whose solution yields \eq{SBrun}. The functions $K_\Gamma^i(\mu_0, \mu)$, $\eta_\Gamma^i(\mu_0, \mu)$, $K_\gamma(\mu_0, \mu)$ required for the RGE solutions of the hard, beam, and soft functions in \sec{calc} are defined as
\begin{align} \label{eq:Keta_def}
K_\Gamma^i(\mu_0, \mu)
& = \int_{\alpha_s(\mu_0)}^{\alpha_s(\mu)}\!\frac{\df\alpha_s}{\beta(\alpha_s)}\,
\Gamma_\cusp^i(\alpha_s) \int_{\alpha_s(\mu_0)}^{\alpha_s} \frac{\df \alpha_s'}{\beta(\alpha_s')}
\,,\qquad
\eta_\Gamma^i(\mu_0, \mu)
= \int_{\alpha_s(\mu_0)}^{\alpha_s(\mu)}\!\frac{\df\alpha_s}{\beta(\alpha_s)}\, \Gamma_\cusp^i(\alpha_s)
\,,\nn \\
K_\gamma(\mu_0, \mu)
& = \int_{\alpha_s(\mu_0)}^{\alpha_s(\mu)}\!\frac{\df\alpha_s}{\beta(\alpha_s)}\, \gamma(\alpha_s)
\,.\end{align}
Expanding the beta function and anomalous dimensions in powers of $\alpha_s$,
\begin{align}
\beta(\alpha_s) &=
- 2 \alpha_s \sum_{n=0}^\infty \beta_n\Bigl(\frac{\alpha_s}{4\pi}\Bigr)^{n+1}
\,, \quad
\Gamma^i_\cusp(\alpha_s) = \sum_{n=0}^\infty \Gamma^i_n \Bigl(\frac{\alpha_s}{4\pi}\Bigr)^{n+1}
\,, \quad
\gamma(\alpha_s) = \sum_{n=0}^\infty \gamma_n \Bigl(\frac{\alpha_s}{4\pi}\Bigr)^{n+1}
\,,\end{align}
their explicit expressions at NNLL are (suppressing the superscript $i$ on $K_\Ga^i$, $\eta_\Ga^i$ and $\Ga^i_n$),
\begin{align} \label{eq:Keta}
K_\Gamma(\mu_0, \mu) &= -\frac{\Gamma_0}{4\beta_0^2}\,
\biggl\{ \frac{4\pi}{\alpha_s(\mu_0)}\, \Bigl(1 - \frac{1}{r} - \ln r\Bigr)
   + \biggl(\frac{\Gamma_1 }{\Gamma_0 } - \frac{\beta_1}{\beta_0}\biggr) (1-r+\ln r)
   + \frac{\beta_1}{2\beta_0} \ln^2 r
\nn\\ & \hspace{10ex}
+ \frac{\alpha_s(\mu_0)}{4\pi}\, \biggl[
  \biggl(\frac{\beta_1^2}{\beta_0^2} - \frac{\beta_2}{\beta_0} \biggr) \Bigl(\frac{1 - r^2}{2} + \ln r\Bigr)
  + \biggl(\frac{\beta_1\Gamma_1 }{\beta_0 \Gamma_0 } - \frac{\beta_1^2}{\beta_0^2} \biggr) (1- r+ r\ln r)
\nn\\ & \hspace{10ex}
  - \biggl(\frac{\Gamma_2 }{\Gamma_0} - \frac{\beta_1\Gamma_1}{\beta_0\Gamma_0} \biggr) \frac{(1- r)^2}{2}
     \biggr] \biggr\}
\,, \nn\\
\eta_\Gamma(\mu_0, \mu) &=
 - \frac{\Gamma_0}{2\beta_0}\, \biggl[ \ln r
 + \frac{\alpha_s(\mu_0)}{4\pi}\, \biggl(\frac{\Gamma_1 }{\Gamma_0 }
 - \frac{\beta_1}{\beta_0}\biggr)(r-1)
\nn \\ & \hspace{10ex}
 + \frac{\alpha_s^2(\mu_0)}{16\pi^2} \biggl(
    \frac{\Gamma_2 }{\Gamma_0 } - \frac{\beta_1\Gamma_1 }{\beta_0 \Gamma_0 }
      + \frac{\beta_1^2}{\beta_0^2} -\frac{\beta_2}{\beta_0} \biggr) \frac{r^2-1}{2}
    \biggr]
\,, \nn\\
K_\gamma(\mu_0, \mu) &=
 - \frac{\gamma_0}{2\beta_0}\, \biggl[ \ln r
 + \frac{\alpha_s(\mu_0)}{4\pi}\, \biggl(\frac{\gamma_1 }{\gamma_0 }
 - \frac{\beta_1}{\beta_0}\biggr)(r-1) \biggr]
\,.\end{align}
Here, $r = \alpha_s(\mu)/\alpha_s(\mu_0)$ and the running coupling at the scale $\mu$ is given in terms of that at the reference scale $\mu_0$ by the three-loop expression
\begin{equation} \label{eq:alphas}
\frac{1}{\alpha_s(\mu)} = \frac{X}{\alpha_s(\mu_0)}
  +\frac{\beta_1}{4\pi\beta_0}  \ln X
  + \frac{\alpha_s(\mu_0)}{16\pi^2} \biggr[
  \frac{\beta_2}{\beta_0} \Bigl(1-\frac{1}{X}\Bigr)
  + \frac{\beta_1^2}{\beta_0^2} \Bigl( \frac{\ln X}{X} +\frac{1}{X} -1\Bigr) \biggl]
\,,\end{equation}
where $X\equiv 1+\alpha_s(\mu_0)\beta_0 \ln(\mu/\mu_0)/(2\pi)$.
At NNLL, we require the three-loop $\beta$ function and gluon cusp anomalous dimension and the two-loop non-cusp anomalous dimensions.
The coefficients of the \MSbar $\beta$ function to three loops are~\cite{Tarasov:1980au, Larin:1993tp}
\begin{align} \label{eq:betaexp}
\beta_0 &= \frac{11}{3}\,C_A -\frac{4}{3}\,T_F\,n_f
\,,\nn\\
\beta_1 &= \frac{34}{3}\,C_A^2  - \Bigl(\frac{20}{3}\,C_A\, + 4 C_F\Bigr)\, T_F\,n_f
\,, \nn\\
\beta_2 &=
\frac{2857}{54}\,C_A^3 + \Bigl(C_F^2 - \frac{205}{18}\,C_F C_A
 - \frac{1415}{54}\,C_A^2 \Bigr)\, 2T_F\,n_f
 + \Bigl(\frac{11}{9}\, C_F + \frac{79}{54}\, C_A \Bigr)\, 4T_F^2\,n_f^2
\,.\end{align}
The coefficients of the gluon cusp anomalous dimension in \MSbar are~\cite{Vogt:2004mw}
\begin{align} \label{eq:Gacuspexp}
\Gamma^g_0 &= 4C_A
\,,\nn\\
\Gamma^g_1 &= 4C_A \Bigl[\Bigl( \frac{67}{9} -\frac{\pi^2}{3} \Bigr)\,C_A  -
   \frac{20}{9}\,T_F\, n_f \Bigr]
\,,\nn\\
\Gamma^g_2 &= 4C_A \Bigl[
\Bigl(\frac{245}{6} -\frac{134 \pi^2}{27} + \frac{11 \pi ^4}{45}
  + \frac{22 \zeta_3}{3}\Bigr)C_A^2
  + \Bigl(- \frac{418}{27} + \frac{40 \pi^2}{27}  - \frac{56 \zeta_3}{3} \Bigr)C_A\, T_F\,n_f
\nn\\ & \hspace{8ex}
  + \Bigl(- \frac{55}{3} + 16 \zeta_3 \Bigr) C_F\, T_F\,n_f
  - \frac{16}{27}\,T_F^2\, n_f^2 \Bigr]
\,,\end{align}
and are equal to $C_A/C_F \Gamma^q_{0,1,2}$. The anomalous dimension coefficients for the hard Wilson coefficient to three loops are~\cite{Moch:2005tm, Idilbi:2005ni, Idilbi:2006dg}
\begin{align}
\gamma_{H\,0}^g &= -2 \beta_0
\,,\nn\\
\gamma_{H\,1}^g
&= \Bigl(-\frac{118}{9} + 4\zeta_3\Bigr)C_A^2 +
\Bigl(-\frac{38}{9}+\frac{\pi^2}{3} \Bigr) C_A\, \beta_0 - 2 \beta_1
\,,\nn\\
\gamma_{H\,2}^g
&= \Bigl(-\frac{60875}{162} + \frac{634\pi^2}{81} +\frac{8\pi^4}{5}+\frac{1972\zeta_3}{9}
- \frac{40\pi^2 \zeta_3}{9} - 32\zeta_5 \Bigr)C_A^3
\nn \\ & \quad +
\Bigl(\frac{7649}{54}+\frac{134\pi^2}{81} - \frac{61\pi^4}{45}
 - \frac{500\zeta_3}{9}\Bigr) C_A^2\, \beta_0 +
\Bigl(\frac{466}{81}+\frac{5\pi^2}{9}-\frac{28 \zeta_3}{3}\Bigr) C_A\, \beta_0^2
\nn \\ & \quad +
\Bigl(-\frac{1819}{54} + \frac{\pi^2}{3} + \frac{4\pi^4}{45} + \frac{152\zeta_3}{9}\Bigr) C_A\, \beta_1
-2 \beta_2
\,.\end{align}
The non-cusp anomalous dimension for the gluon beam function is equal to that of the gluon jet function which is given by $\gamma_J^g(\alpha_s) = - 2\gamma_H^g(\alpha_s) - \gamma_f^g(\alpha_s)$. Here, $\gamma_f^g(\alpha_s)$ is the coefficient of the $\delta(1-z)$ in the gluon PDF anomalous dimension which is known to three loops~\cite{Vogt:2004mw}. The resulting coefficients to three loops are
\begin{align} \label{eq:gammaBgexp}
\gamma_{B\,0}^g &= 2 \beta_0
\,,\nn\\
\gamma_{B\,1}^g
&= \Bigl(\frac{182}{9} - 32\zeta_3\Bigr)C_A^2 +
\Bigl(\frac{94}{9}-\frac{2\pi^2}{3}\Bigr) C_A\, \beta_0 + 2\beta_1
\,,\nn\\
\gamma_{B\,2}^g
&= \Bigl(\frac{49373}{81} - \frac{944 \pi^2}{81} - \frac{16\pi^4}{5} - \frac{4520 \zeta_3}{9}
+ \frac{128\pi^2 \zeta_3}{9} + 224 \zeta_5 \Bigr)C_A^3
\nn \\ & \quad +
\Bigl(-\frac{6173}{27} - \frac{376 \pi^2}{81} + \frac{13\pi^4}{5}
+ \frac{280\zeta_3}{9}\Bigr) C_A^2\, \beta_0 +
\Bigl(-\frac{986}{81}-\frac{10\pi^2}{9}+\frac{56 \zeta_3}{3}\Bigr) C_A\, \beta_0^2
\nn \\ & \quad +
\Bigl(\frac{1765}{27} - \frac{2\pi^2}{3} - \frac{8\pi^4}{45} - \frac{304\zeta_3}{9}\Bigr) C_A\, \beta_1
+ 2 \beta_2
\,. \end{align}
At NNLL we only need $\gamma_H^g$ and $\gamma_B^g$ at two loops. The three-loop coefficients are given for completeness.
The result in \eq{gammaBgexp} agrees with that given in ref.~\cite{Becher:2009th} for the gluon jet function.\footnote{Apart from a typo in ref.~\cite{Becher:2009th} where one of the terms in the $C_A^2 n_f$ contribution is missing a $\pi^2$.}
The non-cusp anomalous dimension of the gluon beam-thrust soft function is given by $\gamma_S^{g}(\alpha_s) = -2\gamma_H^g(\alpha_s) - 2\gamma_B^g(\alpha_s)$.

\subsection{Singular Fixed-Order NLO and NNLO Coefficients}
\label{app:singular}

In this Appendix we give our results for the perturbative coefficients $C_{ij}$ entering in the fixed-order cross section in \eq{sigmaFO}. The coefficients for quarks and antiquarks are related as follows:
\begin{align}
C_{g \bar q}(z_a,z_b,\tau, Y, \mu)
&= C_{gq}(z_a,z_b,\tau, Y, \mu)
\,, \quad
C_{q g}(z_a, z_b,\tau, Y, \mu)
= C_{\bar qg}(z_a, z_b,\tau, Y, \mu)
\,, \nn \\
C_{q g}(z_a, z_b,\tau, Y, \mu)
&= C_{gq}(z_b, z_a, \tau, -Y, \mu)
\,, \nn \\
C_{q \bar q'}^\sing(z_a,z_b,\tau, Y, \mu)
&= C_{\bar q q'}^\sing(z_a,z_b,\tau, Y, \mu)
= C_{\bar q \bar q'}^\sing(z_a,z_b,\tau, Y, \mu)
= C_{qq'}^\sing(z_a,z_b,\tau, Y, \mu)
\,.\end{align}
The first line follows from charge-conjugation invariance and the second line from parity invariance of QCD. The last line is only true because we limit ourselves to contributions from gluon beam functions. Consequently, we only need to consider $C_{gg}$, $C_{gq}$, and $C_{qq'}$. The small contributions involving quark beam functions that we neglect correspond to $q\bar q\to H$ production, where the $q\bar q$ pair couples to the Higgs either directly or indirectly through a two-loop $q\bar q\to gg\to H$ diagram.

As in \eq{Cijdecomp} we split the coefficients into singular and nonsingular parts,
where the singular coefficients are further decomposed as in \eq{Cijsing},
\begin{equation}
C_{ij}^\sing(z_a,z_b,\tau, Y,\mu)
= C_{ij}^{-1} (z_a,z_b,Y,\mu)\,\delta(\tau)
+ \sum_{k\geq 0} C_{ij}^k (z_a,z_b,Y,\mu)\, \cL_k(\tau)
\,.\end{equation}
Each of the coefficients has an expansion in $\alpha_s$, which we write as ($x = \sing$ or $x = k \geq -1$)
\begin{equation}
C_{ij}^x = C_{ij}^{x\zero} + \frac{\alpha_s(\mu)}{2\pi}\, C_{ij}^{x\one} + \frac{\alpha_s^2(\mu)}{(2\pi)^2}\, C_{ij}^{x\two} + \dotsb
\,,\end{equation}
corresponding to the LO, NLO, and NNLO contributions. (The overall $\alpha_s^2$ of the Higgs cross section is factored out in \eq{sigmaFO}.)
At leading order, the only nonzero coefficient is
\begin{equation} \label{eq:Cggzero}
C_{gg}^{\sing\zero}(z_a,z_b,\tau, Y, \mu) = \delta(\tau)\, \delta(1-z_a)\, \delta(1-z_b)
\,.\end{equation}
The singular NLO terms are fully contained in the resummed result at NNLL. Hence, the simplest way to obtain them is to set $\mu_H = \mu_B = \mu_S = \mu$ in \eq{TauBcm_run}, which eliminates the evolution factors, and expand it in $\alpha_s(\mu)$. Using the NLO results for the hard, beam, and soft functions, we find
\begin{align}
C_{gg}^{\sing\one}(z_a,z_b,\tau, Y,\mu)
&= C_A\, \delta(1-z_a)\, \theta(z_b) \biggl\{-2 \cL_1(\tau)\, \delta(1-z_b) + \cL_0(\tau)\, P_{gg}(z_b)
   \nn \\ & \quad
   + \delta(\tau) \biggl[
   \Bigl(\frac{2\pi^2}{3} + Y^2 + \frac{F^\one}{2C_A}\Bigr)\delta(1-z_b)
   - (Y + 2L_\mu) P_{gg}(z_b)
   + \cI_{gg}^{(1,\delta)}(z_b) \biggr] \biggr\}
   \nn \\ & \quad
   + (z_a \lra z_b, Y \to -Y)
   \,, \nn \\
 C_{gq}^{\sing\one}(z_a,z_b,\tau, Y,\mu)
 &= C_F\, \delta(1-z_a)\, \theta(z_b) \bigl\{ \cL_0(\tau) P_{gq}(z_b)
 \nn \\ & \quad
 + \delta(\tau) \bigl[\cI_{gq}^{(1,\delta)}(z_b) - (Y + 2L_\mu) P_{gq}(z_b) \bigr] \bigr\}
\,.\end{align}
Here $L_\mu \equiv \ln(\mu/m_H)$ and $F^\one \equiv F^\one[m_H^2 / (4m_t^2)]$. The $L_\mu$ terms in the NLO coefficients precisely cancel the $\mu$ dependence in the leading-order cross section coming from \eq{Cggzero}.

The NNLO coefficients $C_{ij}^{k\two}$ for $k \geq 0$ are reproduced by the resummed result at NNLL. Once we include the $\mu$-dependent NNLO terms in the hard, beam, and soft functions (which are determined from the NNLL resummation), we can obtain the $C_{ij}^{k\two}$ as before by setting $\mu_H = \mu_B = \mu_S = \mu$ in \eq{TauBcm_run} and expanding to $\ord{\alpha_s^2}$. For $C_{ij}^{-1\two}$ we are only able to obtain some parts analytically. We write it as [see \eq{Cm1terms}]
\begin{equation}
C_{ij}^{-1\two}(z_a,z_b,Y,\mu)
= c_{ij}^\pi(z_a,z_b,Y) + c_{ij}^\mu(z_a,z_b,Y,\mu) + c_{ij}^\res(z_a,z_b,Y)
\,.\end{equation}
The first contribution denotes the $\pi^2$ terms, which we get by taking $\mu_H = \mu = m_H$ in the resummed result (as opposed to taking $\mu_H = -\img m_H$, which resums them into the hard evolution factor). The second contribution contains the $\mu$-dependent terms proportional to $L_\mu = \ln(\mu/m_H)$, which we are able to obtain by requiring that they cancel the $\mu$ dependence of the singular NLO result.
Obtaining an analytic expression for the remaining piece, $c_{ij}^\res(z_a,z_b,Y)$, requires the complete two-loop hard, beam, and soft functions. Its contribution to the cross section after convolution with the PDFs and integrated over $Y$ is extracted numerically from the fixed-order NNLO cross section as discussed in \subsec{nonsingular}.

For the nonzero coefficients of $C_{gg}^{\sing\two}$ we find
\begin{align}
C_{gg}^{3\two}(z_a,z_b,Y,\mu) &= 8 C_A^2\, \delta(1-z_a) \delta(1-z_b)
\,, \nn \\
C_{gg}^{2\two}(z_a,z_b,Y,\mu)
&= \delta(1-z_a) \Bigl\{ -6 C_A^2 P_{gg}(z_b)
   + \frac{3}{2} C_A \beta_0\, \delta(1-z_b) \Bigr\}
   + (z_a \lra z_b)
\,, \nn \\
C_{gg}^{1\two}(z_a,z_b,Y,\mu)
&= \delta(1-z_a) \Bigl\{ C_A^2 \Bigl[ \Bigl(-\frac{4}{3} - \frac{11}{3} \pi^2 -4 Y^2\Bigr) \delta(1-z_b)
   + 4(Y + 2L_\mu) P_{gg}(z_b)
   \nn \\ & \quad\quad
   + (P_{gg} \otimes P_{gg})(z_b)
   -4 \cI_{gg}^{(1,\delta)}(z_b) \Bigr]
   - 2C_A F^\one \delta(1-z_b)
   \nn \\ & \quad
   - C_A \beta_0 \Bigl[\Bigl(\frac{5}{3} + 2L_\mu \Bigr) \delta(1-z_b) + \frac{1}{2} P_{gg}(z_b) \Bigr]
   + 2C_F T_F n_f (P_{gq} \otimes P_{qg})(z_b)\Bigr\}
   \nn \\ & \quad
   + C_A^2 P_{gg}(z_a) P_{gg}(z_b)
   + (z_a \lra z_b, Y \to -Y)
\,, \nn \\
C_{gg}^{0\two}(z_a,z_b,Y,\mu)
&= \delta(1-z_a) \Bigl\{C_A^2 \Bigl[
   (1 + 5 \zeta_3) \delta(1-z_b)
   + 2(\pi^2 + Y^2) P_{gg}(z_b)
   \nn \\ & \quad\quad
   - (Y+2L_\mu) (P_{gg} \otimes P_{gg})(z_b)
   + (\cI_{gg}^{(1,\delta)} \otimes P_{gg})(z_b) \Bigr]
   \nn \\ & \quad
   - \frac{1}{2} C_A \beta_0 \Bigl[\Bigl(2 + \frac{\pi^2}{6} + Y^2\Bigr) \delta(1-z_b)
   - (Y + 2L_\mu) P_{gg}(z_b)
   + \cI_{gg}^{(1,\delta)}(z_b) \Bigr]
   \nn \\ & \quad
   + 2C_F T_F n_f \Bigl[\frac{1}{2}\delta(1-z_b)
   - (Y + 2L_\mu) (P_{gq} \otimes P_{qg}) (z_b)
   + (\cI_{gq}^{(1,\delta)} \otimes P_{qg})(z_b) \Bigr]
   \nn \\ & \quad
   + C_A F^\one P_{gg}(z_b) + P_{gg}^\one(z_b)
   \Bigr\}
   \nn \\ & \quad
  + C_A^2\, \Bigl[\cI_{gg}^{(1,\delta)}(z_a) - 2L_\mu P_{gg}(z_a) \Bigr] P_{gg}(z_b)
    + (z_a \lra z_b, Y \to -Y)
\,.\end{align}
The two-loop splitting function, $P_{gg}^\one(z)$ is given in \eq{Pgj_one}. The convolutions $(P_{ij} \otimes P_{jk})(z)$ and $(\cI_{ij}^{(1,\delta)} \otimes P_{jk})(z)$ are defined in \eq{convdef} and their analytical expressions are given in \eq{convres}. For the $\mu$-dependent and $\pi^2$ terms of $C_{gg}^{-1\two}$ we get
\begin{align} \label{eq:c_mu_pi2}
c_{gg}^\mu(z_a,z_b,Y,\mu)
&= -2  L_\mu \delta(1-z_a) \Bigl\{
   C_A^2 \Bigl[
   \Bigl(\frac{4\pi^2}{3} + 2Y^2\Bigr)P_{gg}(z_b)
   \\ & \qquad
   - (Y + L_\mu) (P_{gg} \otimes P_{gg}) (z_b)
   + \cI_{gg}^{(1,\delta)} \otimes P_{gg} (z_b) \Bigr]
   \nn \\ & \quad
  - \frac{1}{2} C_A \bt_0 \Bigl[  \Bigl(\frac{2\pi^2}{3} + Y^2\Bigr) \delta(1-z_b) - (Y + L_\mu) P_{gg}(z_b)
   + \cI_{gg}^{(1,\delta)}(z_b) \Bigr]
   \nn \\ & \quad
   +2 C_F T_F n_f\, \Bigl[ - (Y+L_\mu)(P_{gq} \otimes P_{qg})(z_b) + \cI_{gq}^{(1,\delta)} \otimes P_{qg}(z_b) \Bigr]
   \nn \\ & \quad
   - \frac{1}{4} \bigl(\bt_1 + F^\one \bt_0\bigr) \delta(1-z_b) + C_A F^\one P_{gg}(z_b) + P_{gg}^\one(z_b)
   \Bigr\}
   \nn \\ & \quad
   -2 C_A^2\, L_\mu \Bigl[\cI_{gg}^{(1,\delta)}(z_a) - L_\mu P_{gg}(z_a)\Bigr] P_{gg}(z_b)
   + (z_a \lra z_b, Y \to -Y)
\,,\nn\\
c_{gg}^\pi(z_a,z_b,Y)
&= \pi^2 \delta(1-z_a) \Bigl\{ C_A^2 \Bigr[
 \Bigl(\frac{1}{3} + \frac{\pi^2}{3} + Y^2\Bigr)\delta(1-z_b)
  - Y P_{gg}(z_b) + \cI_{gg}^{(1,\delta)}(z_b) \Bigr]
  \nn \\ & \quad
  + \frac{5}{12} C_A \beta_0\, \delta(1-z_b) + \frac{1}{2} C_A F^\one\delta(1-z_b) \Bigr\}
  + (z_a \lra z_b, Y \to -Y)
\nn\,.\end{align}
For $C_{gq}^{\sing\two}$ we find
\begin{align}
C_{gq}^{3\two}(z_a,z_b,Y,\mu)
&= 0
\,,\\
C_{gq}^{2\two}(z_a,z_b,Y,\mu)
&= -6 C_A C_F\, \delta(1-z_a) P_{gq}(z_b)
\,, \nn \\
C_{gq}^{1\two}(z_a,z_b,Y,\mu)
&= \delta(1-z_a)
   \Bigl\{C_A C_F \Bigl[ 4(Y+2L_\mu) P_{gq}(z_b)
   + (P_{gg} \otimes P_{gq})(z_b) -4 \cI_{gq}^{(1,\delta)}(z_b) \Bigr]
   \nn \\ & \quad
   + C_F^2 (P_{gq} \otimes P_{qq})(z_b)
   - C_F \beta_0 P_{gq}(z_b) \Bigr\}
   + 2C_A C_F P_{gg}(z_a) P_{gq}(z_b)
\,,\nn \\
C_{gq}^{0\two}(z_a,z_b,Y,\mu)
&= \delta(1-z_a) \Bigl\{
   C_A C_F \Bigl[2(\pi^2 + Y^2) P_{gq}(z_b)
   - (Y + 2L_\mu) (P_{gg} \otimes P_{gq}) (z_b)
   \nn \\ & \qquad
    + (\cI_{gg}^{(1,\delta)} \otimes P_{gq})(z_b) \Bigr]
   \nn \\ & \quad
   + C_F^2 \Bigl[- (Y + 2L_\mu) (P_{gq}\otimes P_{qq})(z_b)
   + (\cI_{gq}^{(1,\delta)} \otimes P_{qq})(z_b) \Bigr]
   \nn \\ & \quad
    + C_F \beta_0 \Bigl[(Y + 2L_\mu) P_{gq}(z_b) - \cI_{gq}^{(1,\delta)}(z_b) \Bigr]
    + C_F F^\one P_{gq}(z_b)
   + P_{gq}^\one(z_b) \Bigr\}
   \nn \\ & \quad
   + C_A C_F \Bigl[P_{gg}(z_a)\,\cI_{gq}^{(1,\delta)} (z_b) +\cI_{gg}^{(1,\delta)}(z_a) P_{gq}(z_b) -4 L_\mu\, P_{gg}(z_a) P_{gq}(z_b) \Bigr]
\nn\,.\end{align}
The two-loop splitting function $P_{gq}^\one(z)$ is given in \eq{Pgj_one}. For $C_{gq}^{-1\two}$, the $\mu$-dependent and $\pi^2$ terms are
\begin{align}
c_{gq}^\mu(z_a,z_b,Y,\mu)
&= -2 L_\mu \delta(1-z_a) \Bigl\{
   C_A C_F \Bigl[\Bigl(\frac{4\pi^2}{3} + 2Y^2 \Bigr)P_{gq}(z_b)
   - (Y + L_\mu) (P_{gg} \otimes P_{gq})(z_b)
   \nn \\ & \quad
   + (\cI_{gg}^{(1,\delta)} \otimes P_{gq})(z_b)\Bigr]
   + C_F^2 \Bigl[- (Y + L_\mu) (P_{gq} \otimes P_{qq})(z_b) + (\cI_{gq}^{(1,\delta)} \otimes P_{qq})(z_b)\Bigr]
   \nn \\ & \quad
   + C_F \bt_0 \Bigl[ (Y+L_\mu) P_{gq}(z_b) - \cI_{gq}^{(1,\delta)}(z_b)\Bigr]
   + C_F F^\one P_{gq}(z_b) + P_{gq}^\one(z_b)
   \Bigr\}
   \nn \\ & \quad
   - 2C_A C_F L_\mu \Bigl[ P_{gg}(z_a)\, \cI_{gq}^{(1,\delta)}(z_b) + \cI_{gg}^{(1,\delta)}(z_a)  P_{gq}(z_b) - 2 L_\mu\, P_{gg}(z_a) P_{gq}(z_b)  \Bigr]
\,,\nn\\
c_{gq}^\pi(z_a,z_b,Y)
&= \pi^2 C_A C_F\,\delta(1-z_a) \Bigl[- Y P_{gq}(z_b) + \cI_{gq}^{(1,\delta)}(z_b) \Bigr]
\,.\end{align}
Finally, for $C_{qq'}^{\sing\two}$ we find
\begin{align}
C_{qq'}^{3\two}(z_a,z_b,Y,\mu) &= C_{qq'}^{2\two}(z_a,z_b,Y,\mu) = 0
\,, \nn \\
C_{qq'}^{1\two}(z_a,z_b,Y,\mu) &= 2C_F^2\, P_{gq}(z_a) P_{gq'}(z_b)
\,, \nn \\
C_{qq'}^{0\two}(z_a,z_b,Y,\mu)
& = C_F^2 \Bigl[ P_{gq}(z_a)\,\cI_{gq'}^{(1,\delta)}(z_b) + \cI_{gq}^{(1,\delta)}(z_a)P_{gq'}(z_b) -4 L_\mu\, P_{gq}(z_a) P_{gq'}(z_b) \Bigr]
\,.\end{align}
There is no contribution to $C_{qq'}^{-1\two}$ from $\pi^2$ summation, $c_{qq'}^\pi = 0$. The $\mu$-dependent terms are
\begin{equation}
c_{qq'}^\mu(z_a,z_b,Y,\mu)
= - 2 C_F^2 L_\mu \Bigl[ P_{gq}(z_a)\, \cI_{gq'}^{(1,\delta)}(z_b) + \cI_{gq}^{(1,\delta)}(z_a)  P_{gq'}(z_b) - 2 L_\mu\, P_{gq}(z_a) P_{gq'}(z_b)  \Bigr]
\,.\end{equation}

\bibliographystyle{../jhep}
\bibliography{../pp}

\providecommand{\href}[2]{#2}\begingroup\raggedright\begin{thebibliography}{10%
0}

\bibitem{Aaltonen:2010yv}
{\bf CDF and D0} Collaboration, T.~Aaltonen {\em et.~al.}, {\it {Combination of
  Tevatron searches for the standard model Higgs boson in the $W^+W^-$ decay
  mode}},  {\em Phys. Rev. Lett.} {\bf 104} (2010) 061802,
  [\href{http://arXiv.org/abs/1001.4162}{{\tt arXiv:1001.4162}}].

\bibitem{Aad:2009wy}
{\bf The ATLAS} Collaboration, G.~Aad {\em et.~al.}, {\it {Expected Performance
  of the ATLAS Experiment - Detector, Trigger and Physics}},
  \href{http://arXiv.org/abs/0901.0512}{{\tt arXiv:0901.0512}}.

\bibitem{CMSnoteWW}
{\bf CMS} Collaboration, {\it {Search Strategy for a Standard Model Higgs Boson
  Decaying to Two $W$ Bosons in the Fully Leptonic Final State}}, .
  CMS-PAS-HIG-08-006.

\bibitem{Dittmar:1996ss}
M.~Dittmar and H.~K. Dreiner, {\it {How to find a Higgs boson with a mass
  between $155$ GeV -- $180$ GeV at the LHC}},  {\em Phys. Rev. D} {\bf 55}
  (1997) 167--172, [\href{http://arXiv.org/abs/hep-ph/9608317}{{\tt
  hep-ph/9608317}}].

\bibitem{Aaltonen:2010cm}
{\bf The CDF} Collaboration, T.~Aaltonen {\em et.~al.}, {\it {Inclusive Search
  for Standard Model Higgs Boson Production in the $WW$ Decay Channel using the
  CDF II Detector}},  {\em Phys. Rev. Lett.} {\bf 104} (2010) 061803,
  [\href{http://arXiv.org/abs/1001.4468}{{\tt arXiv:1001.4468}}].

\bibitem{Abazov:2010ct}
{\bf The D0} Collaboration, V.~M. Abazov {\em et.~al.}, {\it {Search for Higgs
  boson production in dilepton and missing energy final states with $\sim
  5.4\mathrm{fb^{-1}}$ of $p\bar{p}$ collisions at $\sqrt s =1.96$ TeV}},  {\em
  Phys. Rev. Lett.} {\bf 104} (2010) 061804,
  [\href{http://arXiv.org/abs/1001.4481}{{\tt arXiv:1001.4481}}].

\bibitem{:2010ar}
{\bf CDF and D0} Collaboration, {\it {Combined CDF and D0 Upper Limits on
  Standard Model Higgs-Boson Production with up to $6.7$ fb$^{-1}$ of Data}},
  \href{http://arXiv.org/abs/1007.4587}{{\tt arXiv:1007.4587}}.

\bibitem{Anastasiou:2009bt}
C.~Anastasiou, G.~Dissertori, M.~Grazzini, F.~St\"ockli, and B.~R. Webber, {\it
  {Perturbative QCD effects and the search for a $H\to WW \to \ell\nu\ell\nu$
  signal at the Tevatron}},  {\em JHEP} {\bf 08} (2009) 099,
  [\href{http://arXiv.org/abs/0905.3529}{{\tt arXiv:0905.3529}}].

\bibitem{Baglio:2010um}
J.~Baglio and A.~Djouadi, {\it {Predictions for Higgs production at the
  Tevatron and the associated uncertainties}},  {\em JHEP} {\bf 10} (2010) 064,
  [\href{http://arXiv.org/abs/1003.4266}{{\tt arXiv:1003.4266}}].

\bibitem{Demartin:2010er}
F.~Demartin, S.~Forte, E.~Mariani, J.~Rojo, and A.~Vicini, {\it {The impact of
  PDF and alphas uncertainties on Higgs Production in gluon fusion at hadron
  colliders}},  {\em Phys. Rev. D} {\bf 82} (2010) 014002,
  [\href{http://arXiv.org/abs/1004.0962}{{\tt arXiv:1004.0962}}].

\bibitem{Baglio:2010yf}
J.~Baglio and A.~Djouadi, {\it {Addendum to: Predictions for Higgs production
  at the Tevatron and the associated uncertainties}},
  \href{http://arXiv.org/abs/1009.1363}{{\tt arXiv:1009.1363}}.

\bibitem{Dawson:1990zj}
S.~Dawson, {\it {Radiative corrections to Higgs boson production}},  {\em Nucl.
  Phys. B} {\bf 359} (1991) 283--300.

\bibitem{Djouadi:1991tka}
A.~Djouadi, M.~Spira, and P.~M. Zerwas, {\it {Production of Higgs bosons in
  proton colliders: QCD corrections}},  {\em Phys. Lett. B} {\bf 264} (1991)
  440--446.

\bibitem{Spira:1995rr}
M.~Spira, A.~Djouadi, D.~Graudenz, and P.~M. Zerwas, {\it {Higgs boson
  production at the LHC}},  {\em Nucl. Phys. B} {\bf 453} (1995) 17--82,
  [\href{http://arXiv.org/abs/hep-ph/9504378}{{\tt hep-ph/9504378}}].

\bibitem{Harlander:2002wh}
R.~V. Harlander and W.~B. Kilgore, {\it {Next-to-next-to-leading order Higgs
  production at hadron colliders}},  {\em Phys. Rev. Lett.} {\bf 88} (2002)
  201801, [\href{http://arXiv.org/abs/hep-ph/0201206}{{\tt hep-ph/0201206}}].

\bibitem{Anastasiou:2002yz}
C.~Anastasiou and K.~Melnikov, {\it {Higgs boson production at hadron colliders
  in NNLO QCD}},  {\em Nucl. Phys. B} {\bf 646} (2002) 220--256,
  [\href{http://arXiv.org/abs/hep-ph/0207004}{{\tt hep-ph/0207004}}].

\bibitem{Ravindran:2003um}
V.~Ravindran, J.~Smith, and W.~L. van Neerven, {\it {NNLO corrections to the
  total cross section for Higgs boson production in hadron hadron collisions}},
   {\em Nucl. Phys. B} {\bf 665} (2003) 325--366,
  [\href{http://arXiv.org/abs/hep-ph/0302135}{{\tt hep-ph/0302135}}].

\bibitem{Pak:2009dg}
A.~Pak, M.~Rogal, and M.~Steinhauser, {\it {Finite top quark mass effects in
  NNLO Higgs boson production at LHC}},  {\em JHEP} {\bf 02} (2010) 025,
  [\href{http://arXiv.org/abs/0911.4662}{{\tt arXiv:0911.4662}}].

\bibitem{Harlander:2009my}
R.~V. Harlander, H.~Mantler, S.~Marzani, and K.~J. Ozeren, {\it {Higgs
  production in gluon fusion at next-to-next-to-leading order QCD for finite
  top mass}},  {\em Eur. Phys. J. C} {\bf 66} (2010) 359--372,
  [\href{http://arXiv.org/abs/0912.2104}{{\tt arXiv:0912.2104}}].

\bibitem{Aglietti:2004nj}
U.~Aglietti, R.~Bonciani, G.~Degrassi, and A.~Vicini, {\it {Two-loop light
  fermion contribution to Higgs production and decays}},  {\em Phys. Lett. B}
  {\bf 595} (2004) 432--441, [\href{http://arXiv.org/abs/hep-ph/0404071}{{\tt
  hep-ph/0404071}}].

\bibitem{Actis:2008ug}
S.~Actis, G.~Passarino, C.~Sturm, and S.~Uccirati, {\it {NLO Electroweak
  Corrections to Higgs Boson Production at Hadron Colliders}},  {\em Phys.
  Lett. B} {\bf 670} (2008) 12--17, [\href{http://arXiv.org/abs/0809.1301}{{\tt
  arXiv:0809.1301}}].

\bibitem{Anastasiou:2008tj}
C.~Anastasiou, R.~Boughezal, and F.~Petriello, {\it {Mixed QCD-electroweak
  corrections to Higgs boson production in gluon fusion}},  {\em JHEP} {\bf 04}
  (2009) 003, [\href{http://arXiv.org/abs/0811.3458}{{\tt arXiv:0811.3458}}].

\bibitem{Djouadi:2005gi}
A.~Djouadi, {\it {The Anatomy of electro-weak symmetry breaking. I: The Higgs
  boson in the standard model}},  {\em Phys. Rept.} {\bf 457} (2008) 1--216,
  [\href{http://arXiv.org/abs/hep-ph/0503172}{{\tt hep-ph/0503172}}].

\bibitem{Boughezal:2009fw}
R.~Boughezal, {\it {Theoretical Status of Higgs Production at Hadron Colliders
  in the SM}},  \href{http://arXiv.org/abs/0908.3641}{{\tt arXiv:0908.3641}}.

\bibitem{Anastasiou:2008ik}
C.~Anastasiou, G.~Dissertori, F.~St\"ockli, and B.~R. Webber, {\it {QCD
  radiation effects on the $H\to WW \to \ell\nu\ell\nu$ signal at the LHC}},
  {\em JHEP} {\bf 03} (2008) 017, [\href{http://arXiv.org/abs/0801.2682}{{\tt
  arXiv:0801.2682}}].

\bibitem{Catani:2001cr}
S.~Catani, D.~de~Florian, and M.~Grazzini, {\it {Direct Higgs production and
  jet veto at the Tevatron and the LHC in NNLO QCD}},  {\em JHEP} {\bf 01}
  (2002) 015, [\href{http://arXiv.org/abs/hep-ph/0111164}{{\tt
  hep-ph/0111164}}].

\bibitem{Anastasiou:2004xq}
C.~Anastasiou, K.~Melnikov, and F.~Petriello, {\it {Higgs boson production at
  hadron colliders: Differential cross sections through next-to-next-to-leading
  order}},  {\em Phys. Rev. Lett.} {\bf 93} (2004) 262002,
  [\href{http://arXiv.org/abs/hep-ph/0409088}{{\tt hep-ph/0409088}}].

\bibitem{Davatz:2006ut}
G.~Davatz {\em et.~al.}, {\it {Combining Monte Carlo generators with
  next-to-next-to-leading order calculations: Event reweighting for Higgs boson
  production at the LHC}},  {\em JHEP} {\bf 07} (2006) 037,
  [\href{http://arXiv.org/abs/hep-ph/0604077}{{\tt hep-ph/0604077}}].

\bibitem{Anastasiou:2007mz}
C.~Anastasiou, G.~Dissertori, and F.~St\"ockli, {\it {NNLO QCD predictions for
  the $H\to WW \to \ell\nu\ell\nu$ signal at the LHC}},  {\em JHEP} {\bf 09}
  (2007) 018, [\href{http://arXiv.org/abs/0707.2373}{{\tt arXiv:0707.2373}}].

\bibitem{Grazzini:2008tf}
M.~Grazzini, {\it {NNLO predictions for the Higgs boson signal in the $H\to WW
  \to \ell\nu\ell\nu$ and $H\to ZZ\to 4\ell$ decay channels}},  {\em JHEP} {\bf
  02} (2008) 043, [\href{http://arXiv.org/abs/0801.3232}{{\tt
  arXiv:0801.3232}}].

\bibitem{Berger:2010nc}
E.~L. Berger, Q.-H. Cao, C.~B. Jackson, T.~Liu, and G.~Shaughnessy, {\it {Higgs
  Boson Search Sensitivity in the $H \to WW$ Dilepton Decay Mode at $\sqrt s =
  7$ and $10$ TeV}},  {\em Phys. Rev. D} {\bf 82} (2010) 053003,
  [\href{http://arXiv.org/abs/1003.3875}{{\tt arXiv:1003.3875}}].

\bibitem{Frixione:2002ik}
S.~Frixione and B.~R. Webber, {\it {Matching NLO QCD computations and parton
  shower simulations}},  {\em JHEP} {\bf 06} (2002) 029,
  [\href{http://arXiv.org/abs/hep-ph/0204244}{{\tt hep-ph/0204244}}].

\bibitem{Frixione:2003ei}
S.~Frixione, P.~Nason, and B.~R. Webber, {\it {Matching NLO QCD and parton
  showers in heavy flavour production}},  {\em JHEP} {\bf 08} (2003) 007,
  [\href{http://arXiv.org/abs/hep-ph/0305252}{{\tt hep-ph/0305252}}].

\bibitem{Alioli:2008tz}
S.~Alioli, P.~Nason, C.~Oleari, and E.~Re, {\it {NLO Higgs boson production via
  gluon fusion matched with shower in POWHEG}},  {\em JHEP} {\bf 04} (2009)
  002, [\href{http://arXiv.org/abs/0812.0578}{{\tt arXiv:0812.0578}}].

\bibitem{Hamilton:2009za}
K.~Hamilton, P.~Richardson, and J.~Tully, {\it {A Positive-Weight
  Next-to-Leading Order Monte Carlo Simulation for Higgs Boson Production}},
  {\em JHEP} {\bf 04} (2009) 116, [\href{http://arXiv.org/abs/0903.4345}{{\tt
  arXiv:0903.4345}}].

\bibitem{Sjostrand:2006za}
T.~Sj{\"o}strand, S.~Mrenna, and P.~Skands, {\it Pythia 6.4 physics and
  manual},  {\em JHEP} {\bf 05} (2006) 026,
  [\href{http://arXiv.org/abs/hep-ph/0603175}{{\tt hep-ph/0603175}}].

\bibitem{Sjostrand:2007gs}
T.~Sj{\"o}strand, S.~Mrenna, and P.~Skands, {\it {A Brief Introduction to
  PYTHIA 8.1}},  {\em Comput. Phys. Commun.} {\bf 178} (2008) 852--867,
  [\href{http://arXiv.org/abs/0710.3820}{{\tt arXiv:0710.3820}}].

\bibitem{Corcella:2000bw}
G.~Corcella {\em et.~al.}, {\it Herwig 6: An event generator for hadron
  emission reactions with interfering gluons (including supersymmetric
  processes)},  {\em JHEP} {\bf 01} (2001) 010,
  [\href{http://arXiv.org/abs/hep-ph/0011363}{{\tt hep-ph/0011363}}].

\bibitem{Corcella:2002jc}
G.~Corcella {\em et.~al.}, {\it {HERWIG 6.5 release note}},
  \href{http://arXiv.org/abs/hep-ph/0210213}{{\tt hep-ph/0210213}}.

\bibitem{deFlorian:1999zd}
D.~de~Florian, M.~Grazzini, and Z.~Kunszt, {\it {Higgs production with large
  transverse momentum in hadronic collisions at next-to-leading order}},  {\em
  Phys. Rev. Lett.} {\bf 82} (1999) 5209--5212,
  [\href{http://arXiv.org/abs/hep-ph/9902483}{{\tt hep-ph/9902483}}].

\bibitem{Ravindran:2002dc}
V.~Ravindran, J.~Smith, and W.~L. Van~Neerven, {\it {Next-to-leading order QCD
  corrections to differential distributions of Higgs boson production in hadron
  hadron collisions}},  {\em Nucl. Phys. B} {\bf 634} (2002) 247--290,
  [\href{http://arXiv.org/abs/hep-ph/0201114}{{\tt hep-ph/0201114}}].

\bibitem{Glosser:2002gm}
C.~J. Glosser and C.~R. Schmidt, {\it {Next-to-leading corrections to the Higgs
  boson transverse momentum spectrum in gluon fusion}},  {\em JHEP} {\bf 12}
  (2002) 016, [\href{http://arXiv.org/abs/hep-ph/0209248}{{\tt
  hep-ph/0209248}}].

\bibitem{Anastasiou:2005qj}
C.~Anastasiou, K.~Melnikov, and F.~Petriello, {\it {Fully differential Higgs
  boson production and the di-photon signal through next-to-next-to-leading
  order}},  {\em Nucl. Phys. B} {\bf 724} (2005) 197--246,
  [\href{http://arXiv.org/abs/hep-ph/0501130}{{\tt hep-ph/0501130}}].

\bibitem{Collins:1984kg}
J.~C. Collins, D.~E. Soper, and G.~Sterman, {\it {Transverse Momentum
  Distribution in Drell-Yan Pair and $W$ and $Z$ Boson Production}},  {\em
  Nucl. Phys. B} {\bf 250} (1985) 199.

\bibitem{Balazs:2000wv}
C.~Balazs and C.~P. Yuan, {\it {Higgs boson production at the LHC with soft
  gluon effects}},  {\em Phys. Lett. B} {\bf 478} (2000) 192--198,
  [\href{http://arXiv.org/abs/hep-ph/0001103}{{\tt hep-ph/0001103}}].

\bibitem{Berger:2002ut}
E.~L. Berger and J.-w. Qiu, {\it {Differential cross section for Higgs boson
  production including all-orders soft gluon resummation}},  {\em Phys. Rev. D}
  {\bf 67} (2003) 034026, [\href{http://arXiv.org/abs/hep-ph/0210135}{{\tt
  hep-ph/0210135}}].

\bibitem{Bozzi:2003jy}
G.~Bozzi, S.~Catani, D.~de~Florian, and M.~Grazzini, {\it {The $q_T$ spectrum
  of the Higgs boson at the LHC in QCD perturbation theory}},  {\em Phys. Lett.
  B} {\bf 564} (2003) 65--72, [\href{http://arXiv.org/abs/hep-ph/0302104}{{\tt
  hep-ph/0302104}}].

\bibitem{Kulesza:2003wn}
A.~Kulesza, G.~Sterman, and W.~Vogelsang, {\it {Joint resummation for Higgs
  production}},  {\em Phys. Rev. D} {\bf 69} (2004) 014012,
  [\href{http://arXiv.org/abs/hep-ph/0309264}{{\tt hep-ph/0309264}}].

\bibitem{Idilbi:2005er}
A.~Idilbi, X.-d. Ji, and F.~Yuan, {\it {Transverse momentum distribution
  through soft-gluon resummation in effective field theory}},  {\em Phys. Lett.
  B} {\bf 625} (2005) 253--263,
  [\href{http://arXiv.org/abs/hep-ph/0507196}{{\tt hep-ph/0507196}}].

\bibitem{Bozzi:2005wk}
G.~Bozzi, S.~Catani, D.~de~Florian, and M.~Grazzini, {\it {Transverse-momentum
  resummation and the spectrum of the Higgs boson at the LHC}},  {\em Nucl.
  Phys. B} {\bf 737} (2006) 73--120,
  [\href{http://arXiv.org/abs/hep-ph/0508068}{{\tt hep-ph/0508068}}].

\bibitem{Mantry:2009qz}
S.~Mantry and F.~Petriello, {\it {Factorization and Resummation of Higgs Boson
  Differential Distributions in Soft-Collinear Effective Theory}},  {\em Phys.
  Rev. D} {\bf 81} (2010) 093007, [\href{http://arXiv.org/abs/0911.4135}{{\tt
  arXiv:0911.4135}}].

\bibitem{Laenen:2000ij}
E.~Laenen, G.~Sterman, and W.~Vogelsang, {\it {Recoil and threshold corrections
  in short distance cross-sections}},  {\em Phys. Rev. D} {\bf 63} (2001)
  114018, [\href{http://arXiv.org/abs/hep-ph/0010080}{{\tt hep-ph/0010080}}].

\bibitem{deFlorian:2009hc}
D.~de~Florian and M.~Grazzini, {\it {Higgs production through gluon fusion:
  updated cross sections at the Tevatron and the LHC}},  {\em Phys. Lett. B}
  {\bf 674} (2009) 291--294, [\href{http://arXiv.org/abs/0901.2427}{{\tt
  arXiv:0901.2427}}].

\bibitem{Davatz:2004zg}
G.~Davatz, G.~Dissertori, M.~Dittmar, M.~Grazzini, and F.~Pauss, {\it
  {Effective K-factors for $g g \to H \to W W \to \ell\nu\ell\nu$ at the LHC}},
   {\em JHEP} {\bf 05} (2004) 009,
  [\href{http://arXiv.org/abs/hep-ph/0402218}{{\tt hep-ph/0402218}}].

\bibitem{Papaefstathiou:2010bw}
A.~Papaefstathiou, J.~M. Smillie, and B.~R. Webber, {\it {Resummation of
  transverse energy in vector boson and Higgs boson production at hadron
  colliders}},  {\em JHEP} {\bf 04} (2010) 084,
  [\href{http://arXiv.org/abs/1002.4375}{{\tt arXiv:1002.4375}}].

\bibitem{Stewart:2009yx}
I.~W. Stewart, F.~J. Tackmann, and W.~J. Waalewijn, {\it {Factorization at the
  LHC: From PDFs to Initial State Jets}},  {\em Phys. Rev. D} {\bf 81} (2010)
  094035, [\href{http://arXiv.org/abs/0910.0467}{{\tt arXiv:0910.0467}}].

\bibitem{Catani:2003zt}
S.~Catani, D.~de~Florian, M.~Grazzini, and P.~Nason, {\it {Soft-gluon
  resummation for Higgs boson production at hadron colliders}},  {\em JHEP}
  {\bf 07} (2003) 028, [\href{http://arXiv.org/abs/hep-ph/0306211}{{\tt
  hep-ph/0306211}}].

\bibitem{Moch:2005ky}
S.~Moch and A.~Vogt, {\it {Higher-order soft corrections to lepton pair and
  Higgs boson production}},  {\em Phys. Lett. B} {\bf 631} (2005) 48--57,
  [\href{http://arXiv.org/abs/hep-ph/0508265}{{\tt hep-ph/0508265}}].

\bibitem{Idilbi:2005ni}
A.~Idilbi, X.-d. Ji, J.-P. Ma, and F.~Yuan, {\it {Threshold resummation for
  Higgs production in effective field theory}},  {\em Phys. Rev. D} {\bf 73}
  (2006) 077501, [\href{http://arXiv.org/abs/hep-ph/0509294}{{\tt
  hep-ph/0509294}}].

\bibitem{Laenen:2005uz}
E.~Laenen and L.~Magnea, {\it {Threshold resummation for electroweak
  annihilation from DIS data}},  {\em Phys. Lett. B} {\bf 632} (2006) 270--276,
  [\href{http://arXiv.org/abs/hep-ph/0508284}{{\tt hep-ph/0508284}}].

\bibitem{Ravindran:2006cg}
V.~Ravindran, {\it {Higher-order threshold effects to inclusive processes in
  QCD}},  {\em Nucl. Phys. B} {\bf 752} (2006) 173--196,
  [\href{http://arXiv.org/abs/hep-ph/0603041}{{\tt hep-ph/0603041}}].

\bibitem{Ahrens:2010rs}
V.~Ahrens, T.~Becher, M.~Neubert, and L.~L. Yang, {\it {Updated Predictions for
  Higgs Production at the Tevatron and the LHC}},  {\em Phys. Lett. B} {\bf
  698} (2011) 271--274, [\href{http://arXiv.org/abs/1008.3162}{{\tt
  arXiv:1008.3162}}].

\bibitem{Stewart:2010tn}
I.~W. Stewart, F.~J. Tackmann, and W.~J. Waalewijn, {\it {N-Jettiness: An
  Inclusive Event Shape to Veto Jets}},  {\em Phys. Rev. Lett.} {\bf 105}
  (2010) 092002, [\href{http://arXiv.org/abs/1004.2489}{{\tt
  arXiv:1004.2489}}].

\bibitem{Martin:2009iq}
A.~D. Martin, W.~J. Stirling, R.~S. Thorne, and G.~Watt, {\it {Parton
  distributions for the LHC}},  {\em Eur. Phys. J. C} {\bf 63} (2009) 189--285,
  [\href{http://arXiv.org/abs/0901.0002}{{\tt arXiv:0901.0002}}].

\bibitem{Cacciari:2008gp}
M.~Cacciari, G.~P. Salam, and G.~Soyez, {\it {The anti-$k_t$ jet clustering
  algorithm}},  {\em JHEP} {\bf 04} (2008) 063,
  [\href{http://arXiv.org/abs/0802.1189}{{\tt arXiv:0802.1189}}].

\bibitem{fastjet}
M.~Cacciari, G.~P. Salam, and G.~Soyez. \href{http://fastjet.fr/}{{\tt
  http://fastjet.fr/}}.

\bibitem{Kidonakis:2010dk}
N.~Kidonakis, {\it {Next-to-next-to-leading soft-gluon corrections for the top
  quark cross section and transverse momentum distribution}},  {\em Phys. Rev.
  D} {\bf 82} (2010) 114030, [\href{http://arXiv.org/abs/1009.4935}{{\tt
  arXiv:1009.4935}}].

\bibitem{Stewart:2010qs}
I.~W. Stewart, F.~J. Tackmann, and W.~J. Waalewijn, {\it {The Quark Beam
  Function at NNLL}},  {\em JHEP} {\bf 09} (2010) 005,
  [\href{http://arXiv.org/abs/1002.2213}{{\tt arXiv:1002.2213}}].

\bibitem{Fleming:2006cd}
S.~Fleming, A.~K. Leibovich, and T.~Mehen, {\it {Resummation of Large Endpoint
  Corrections to Color-Octet $J/\psi$ Photoproduction}},  {\em Phys. Rev. D}
  {\bf 74} (2006) 114004, [\href{http://arXiv.org/abs/hep-ph/0607121}{{\tt
  hep-ph/0607121}}].

\bibitem{Bauer:2000ew}
C.~W. Bauer, S.~Fleming, and M.~E. Luke, {\it {Summing Sudakov logarithms in $B
  \to X_s\gamma$ in effective field theory}},  {\em Phys. Rev. D} {\bf 63}
  (2000) 014006, [\href{http://arXiv.org/abs/hep-ph/0005275}{{\tt
  hep-ph/0005275}}].

\bibitem{Bauer:2000yr}
C.~W. Bauer, S.~Fleming, D.~Pirjol, and I.~W. Stewart, {\it An effective field
  theory for collinear and soft gluons: Heavy to light decays},  {\em Phys.
  Rev. D} {\bf 63} (2001) 114020,
  [\href{http://arXiv.org/abs/hep-ph/0011336}{{\tt hep-ph/0011336}}].

\bibitem{Bauer:2001ct}
C.~W. Bauer and I.~W. Stewart, {\it Invariant operators in collinear effective
  theory},  {\em Phys. Lett. B} {\bf 516} (2001) 134--142,
  [\href{http://arXiv.org/abs/hep-ph/0107001}{{\tt hep-ph/0107001}}].

\bibitem{Bauer:2001yt}
C.~W. Bauer, D.~Pirjol, and I.~W. Stewart, {\it Soft-collinear factorization in
  effective field theory},  {\em Phys. Rev. D} {\bf 65} (2002) 054022,
  [\href{http://arXiv.org/abs/hep-ph/0109045}{{\tt hep-ph/0109045}}].

\bibitem{Bauer:2002nz}
C.~W. Bauer, S.~Fleming, D.~Pirjol, I.~Z. Rothstein, and I.~W. Stewart, {\it
  Hard scattering factorization from effective field theory},  {\em Phys. Rev.
  D} {\bf 66} (2002) 014017, [\href{http://arXiv.org/abs/hep-ph/0202088}{{\tt
  hep-ph/0202088}}].

\bibitem{Ligeti:2008ac}
Z.~Ligeti, I.~W. Stewart, and F.~J. Tackmann, {\it {Treating the b quark
  distribution function with reliable uncertainties}},  {\em Phys. Rev. D} {\bf
  78} (2008) 114014, [\href{http://arXiv.org/abs/0807.1926}{{\tt
  arXiv:0807.1926}}].

\bibitem{Harlander:2005rq}
R.~Harlander and P.~Kant, {\it {Higgs production and decay: Analytic results at
  next-to-leading order QCD}},  {\em JHEP} {\bf 12} (2005) 015,
  [\href{http://arXiv.org/abs/hep-ph/0509189}{{\tt hep-ph/0509189}}].

\bibitem{Anastasiou:2006hc}
C.~Anastasiou, S.~Beerli, S.~Bucherer, A.~Daleo, and Z.~Kunszt, {\it {Two-loop
  amplitudes and master integrals for the production of a Higgs boson via a
  massive quark and a scalar-quark loop}},  {\em JHEP} {\bf 01} (2007) 082,
  [\href{http://arXiv.org/abs/hep-ph/0611236}{{\tt hep-ph/0611236}}].

\bibitem{Harlander:2009bw}
R.~V. Harlander and K.~J. Ozeren, {\it {Top mass effects in Higgs production at
  next-to-next-to-leading order QCD: virtual corrections}},  {\em Phys. Lett.
  B} {\bf 679} (2009) 467--472, [\href{http://arXiv.org/abs/0907.2997}{{\tt
  arXiv:0907.2997}}].

\bibitem{Pak:2009bx}
A.~Pak, M.~Rogal, and M.~Steinhauser, {\it {Virtual three-loop corrections to
  Higgs boson production in gluon fusion for finite top quark mass}},  {\em
  Phys. Lett. B} {\bf 679} (2009) 473--477,
  [\href{http://arXiv.org/abs/0907.2998}{{\tt arXiv:0907.2998}}].

\bibitem{Parisi:1979xd}
G.~Parisi, {\it {Summing Large Perturbative Corrections in QCD}},  {\em Phys.
  Lett. B} {\bf 90} (1980) 295.

\bibitem{Sterman:1986aj}
G.~Sterman, {\it {Summation of Large Corrections to Short Distance Hadronic
  Cross-Sections}},  {\em Nucl. Phys. B} {\bf 281} (1987) 310.

\bibitem{Magnea:1990zb}
L.~Magnea and G.~Sterman, {\it {Analytic continuation of the Sudakov
  form-factor in QCD}},  {\em Phys. Rev. D} {\bf 42} (1990) 4222--4227.

\bibitem{Eynck:2003fn}
T.~O. Eynck, E.~Laenen, and L.~Magnea, {\it {Exponentiation of the Drell-Yan
  cross section near partonic threshold in the DIS and MS-bar schemes}},  {\em
  JHEP} {\bf 06} (2003) 057, [\href{http://arXiv.org/abs/hep-ph/0305179}{{\tt
  hep-ph/0305179}}].

\bibitem{Ahrens:2008qu}
V.~Ahrens, T.~Becher, M.~Neubert, and L.~L. Yang, {\it {Origin of the Large
  Perturbative Corrections to Higgs Production at Hadron Colliders}},  {\em
  Phys. Rev. D} {\bf 79} (2009) 033013,
  [\href{http://arXiv.org/abs/0808.3008}{{\tt arXiv:0808.3008}}].

\bibitem{Ahrens:2008nc}
V.~Ahrens, T.~Becher, M.~Neubert, and L.~L. Yang, {\it {Renormalization-Group
  Improved Prediction for Higgs Production at Hadron Colliders}},  {\em Eur.
  Phys. J. C} {\bf 62} (2009) 333--353,
  [\href{http://arXiv.org/abs/0809.4283}{{\tt arXiv:0809.4283}}].

\bibitem{Chiu:2008vv}
J.-y. Chiu, R.~Kelley, and A.~V. Manohar, {\it {Electroweak Corrections using
  Effective Field Theory: Applications to the LHC}},  {\em Phys. Rev. D} {\bf
  78} (2008) 073006, [\href{http://arXiv.org/abs/0806.1240}{{\tt
  arXiv:0806.1240}}].

\bibitem{Chiu:2009mg}
J.-y. Chiu, A.~Fuhrer, R.~Kelley, and A.~V. Manohar, {\it {Factorization
  Structure of Gauge Theory Amplitudes and Application to Hard Scattering
  Processes at the LHC}},  {\em Phys. Rev. D} {\bf 80} (2009) 094013,
  [\href{http://arXiv.org/abs/0909.0012}{{\tt arXiv:0909.0012}}].

\bibitem{Mantry:2010mk}
S.~Mantry and F.~Petriello, {\it {Transverse Momentum Distributions from
  Effective Field Theory with Numerical Results}},  {\em Phys. Rev. D} {\bf 83}
  (2011) 053007, [\href{http://arXiv.org/abs/1007.3773}{{\tt
  arXiv:1007.3773}}].

\bibitem{Schwartz:2007ib}
M.~D. Schwartz, {\it {Resummation and NLO Matching of Event Shapes with
  Effective Field Theory}},  {\em Phys. Rev. D} {\bf 77} (2008) 014026,
  [\href{http://arXiv.org/abs/0709.2709}{{\tt arXiv:0709.2709}}].

\bibitem{Fleming:2007xt}
S.~Fleming, A.~H. Hoang, S.~Mantry, and I.~W. Stewart, {\it {Top Jets in the
  Peak Region: Factorization Analysis with NLL Resummation}},  {\em Phys. Rev.
  D} {\bf 77} (2008) 114003, [\href{http://arXiv.org/abs/0711.2079}{{\tt
  arXiv:0711.2079}}].

\bibitem{Hoang:2007vb}
A.~H. Hoang and I.~W. Stewart, {\it {Designing Gapped Soft Functions for Jet
  Production}},  {\em Phys. Lett. B} {\bf 660} (2008) 483--493,
  [\href{http://arXiv.org/abs/0709.3519}{{\tt arXiv:0709.3519}}].

\bibitem{Abbate:2010xh}
R.~Abbate, M.~Fickinger, A.~H. Hoang, V.~Mateu, and I.~W. Stewart, {\it {Thrust
  at N3LL with Power Corrections and a Precision Global Fit for alphas(mZ)}},
  \href{http://arXiv.org/abs/1006.3080}{{\tt arXiv:1006.3080}}.

\bibitem{Martin:2009bu}
A.~D. Martin, W.~J. Stirling, R.~S. Thorne, and G.~Watt, {\it {Uncertainties on
  $\alpha_s$ in global PDF analyses and implications for predicted hadronic
  cross sections}},  {\em Eur. Phys. J. C} {\bf 64} (2009) 653--680,
  [\href{http://arXiv.org/abs/0905.3531}{{\tt arXiv:0905.3531}}].

\bibitem{Stewart:2010pd}
I.~W. Stewart, F.~J. Tackmann, and W.~J. Waalewijn, {\it {The Beam Thrust Cross
  Section for Drell-Yan at NNLL Order}},  {\em Phys. Rev. Lett.} {\bf 106}
  (2011) 032001, [\href{http://arXiv.org/abs/1005.4060}{{\tt
  arXiv:1005.4060}}].

\bibitem{Manohar:2006nz}
A.~V. Manohar and I.~W. Stewart, {\it {The zero-bin and mode factorization in
  quantum field theory}},  {\em Phys. Rev. D} {\bf 76} (2007) 074002,
  [\href{http://arXiv.org/abs/hep-ph/0605001}{{\tt hep-ph/0605001}}].

\bibitem{Altarelli:1977zs}
G.~Altarelli and G.~Parisi, {\it {Asymptotic Freedom in Parton Language}},
  {\em Nucl. Phys. B} {\bf 126} (1977) 298.

\bibitem{Dawson:1993qf}
S.~Dawson and R.~Kauffman, {\it {QCD corrections to Higgs boson production:
  nonleading terms in the heavy quark limit}},  {\em Phys. Rev. D} {\bf 49}
  (1994) 2298--2309, [\href{http://arXiv.org/abs/hep-ph/9310281}{{\tt
  hep-ph/9310281}}].

\bibitem{Furmanski:1980cm}
W.~Furmanski and R.~Petronzio, {\it {Singlet Parton Densities Beyond Leading
  Order}},  {\em Phys. Lett. B} {\bf 97} (1980) 437.

\bibitem{Ellis:1996nn}
R.~K. Ellis and W.~Vogelsang, {\it {The evolution of parton distributions
  beyond leading order: the singlet case}},
  \href{http://arXiv.org/abs/hep-ph/9602356}{{\tt hep-ph/9602356}}.

\bibitem{Tarasov:1980au}
O.~V. Tarasov, A.~A. Vladimirov, and A.~Y. Zharkov, {\it {The Gell-Mann-Low
  Function of QCD in the Three Loop Approximation}},  {\em Phys. Lett. B} {\bf
  93} (1980) 429--432.

\bibitem{Larin:1993tp}
S.~A. Larin and J.~A.~M. Vermaseren, {\it {The three-loop QCD $\beta$ function
  and anomalous dimensions}},  {\em Phys. Lett. B} {\bf 303} (1993) 334--336,
  [\href{http://arXiv.org/abs/hep-ph/9302208}{{\tt hep-ph/9302208}}].

\bibitem{Vogt:2004mw}
A.~Vogt, S.~Moch, and J.~A.~M. Vermaseren, {\it {The three-loop splitting
  functions in QCD: The singlet case}},  {\em Nucl. Phys. B} {\bf 691} (2004)
  129--181, [\href{http://arXiv.org/abs/hep-ph/0404111}{{\tt hep-ph/0404111}}].

\bibitem{Moch:2005tm}
S.~Moch, J.~A.~M. Vermaseren, and A.~Vogt, {\it {Three-loop results for quark
  and gluon form factors}},  {\em Phys. Lett. B} {\bf 625} (2005) 245--252,
  [\href{http://arXiv.org/abs/hep-ph/0508055}{{\tt hep-ph/0508055}}].

\bibitem{Idilbi:2006dg}
A.~Idilbi, X.~dong Ji, and F.~Yuan, {\it {Resummation of Threshold Logarithms
  in Effective Field Theory For DIS, Drell-Yan and Higgs Production}},  {\em
  Nucl. Phys. B} {\bf 753} (2006) 42--68,
  [\href{http://arXiv.org/abs/hep-ph/0605068}{{\tt hep-ph/0605068}}].

\bibitem{Becher:2009th}
T.~Becher and M.~D. Schwartz, {\it {Direct photon production with effective
  field theory}},  {\em JHEP} {\bf 02} (2010) 040,
  [\href{http://arXiv.org/abs/0911.0681}{{\tt arXiv:0911.0681}}].

\end{thebibliography}\endgroup

\end{document}